\documentclass[12pt,letterpaper,oneside,openany]{book}
\pdfoutput=1

\usepackage[top=1in, bottom=1in, left=1in, right=1in]{geometry}

\usepackage{sectsty}
\chapterfont{\centering} 
\usepackage{setspace}
\usepackage{appendix}

\usepackage[titles]{tocloft}
\newlength\oldchwidth
\newlength\newchwidth
\newcommand{\labelchapters}{
    \renewcommand{\cftchappresnum}{Chapter~}
    \renewcommand{\cftchapaftersnum}{:~}
    \setlength{\cftchapnumwidth}{\oldchwidth}
    \addtolength\cftchapnumwidth\newchwidth
}
\newcommand{\listappendicesname}{List of Appendices}
\newlistof{appendices}{loa}{\listappendicesname}
\newlength\newappwidth
\newcommand{\labelappendices}{
    \renewcommand{\cftchappresnum}{Appendix~}
    \renewcommand{\cftchapaftersnum}{:~}
    \setlength{\cftchapnumwidth}{\oldchwidth}
    \addtolength\cftchapnumwidth\newappwidth
}

\usepackage{etoolbox} 

\usepackage{graphicx}
\usepackage{amsmath}
\usepackage{amsfonts}

\newcommand\fft[2]{{\frac{#1}{#2}}}
\newcommand\ft[2]{{\textstyle\frac{#1}{#2}}}
\newcommand\nn{\nonumber}
\begin{document}

\frontmatter

\pagenumbering{gobble} 

\newgeometry{top=2.5in}
\thispagestyle{empty}

{\begin{center}
\begin{doublespace}
{\Large \bf High-temperature asymptotics of the 4d superconformal index}\\[.5cm] 
by\\
Arash Arabi Ardehali\\[.5cm] 
\end{doublespace}
\begin{singlespace}
A dissertation submitted in partial fulfillment\\
of the requirements for the degree of\\
Doctor of Philosophy\\
(Physics)\\ 
in the University of Michigan\\
2016 
\end{singlespace}
\end{center}

\vspace*{5cm}
\begin{singlespace}
\noindent Doctoral Committee:\\[4mm] 
\hspace*{1cm} Professor James T. Liu, Chair\\
\hspace*{1cm} Professor Igor Kriz\\
\hspace*{1cm} Professor Finn Larsen\\
\hspace*{1cm} Professor Roberto D. Merlin\\
\hspace*{1cm} Professor Leopoldo A. Pando Zayas
\end{singlespace}}
\restoregeometry
\clearpage



\pagenumbering{roman} 
\setcounter{page}{2}  

\phantomsection \addcontentsline{toc}{chapter}{Abstract}
\newgeometry{top=2in}

\begin{center}
\textbf{\huge{Abstract}}\\[1.5cm]
\end{center}

\noindent This dissertation contains a study of certain
four-dimensional superconformal field theories (4d SCFTs). Any 4d
SCFT has a spectrum of local operators. Some of these operators sit
in short representations of the 4d $\mathcal{N}=1$ superconformal
group SU($2,2|1$), and can be quantified using a partition function
known as the 4d superconformal index. The superconformal index
$\mathcal{I}(b,\beta)$ is a function of two positive real
parameters: the squashing parameter $b$, and the inverse temperature
$\beta$. Our study in the present dissertation is focused on the
temperature- (or $\beta$-) dependence of the superconformal index of
4d SCFTs.

The superconformal index of a typical Lagrangian 4d SCFT is given by
a special function known as an elliptic hypergeometric integral
(EHI). The high-temperature limit of the index corresponds to the
hyperbolic limit of the EHI. The hyperbolic limit of certain special
EHIs has been analyzed by Eric~Rains around 2006; extending Rains's
techniques, we discover a surprisingly rich structure in the
high-temperature limit of a (rather large) class of EHIs that arise
as the superconformal index of unitary Lagrangian 4d SCFTs with
\emph{non-chiral matter content}. Our result has implications for
$\mathcal{N}=1$ dualities, the AdS/CFT correspondence, and
supersymmetric gauge dynamics on $R^3\times S^1$.

We also investigate the high-temperature asymptotics of the
large-$N$ limit of the superconformal index of a class of
holographic 4d SCFTs (described by toric quiver gauge theories with
SU($N$) nodes). We show that from this study a rather general
solution to the problem of holographic Weyl anomaly in
AdS$_5$/CFT$_4$ at the subleading order (in the $1/N$ expansion)
emerges.\\

Most of this dissertation is based on published works by Jim Liu,
Phil Szepietowski, and the author. We include here a few previously
unpublished results as well, one of which is the high-temperature
asymptotics of the superconformal index of puncture-less SU($2$)
class-$\mathcal{S}$ theories.

\restoregeometry
\clearpage

\phantomsection \addcontentsline{toc}{chapter}{Acknowledgements}
\newgeometry{top=2in}
\begin{center}
\textbf{\Large{Acknowledgements}}\\[2cm]
\end{center}

I have been very privileged to have Jim Liu as my thesis advisor.
The work described in this dissertation is the result of a long and
enjoyable collaboration with him and Phil Szepietowski. I would like
to express my deepest gratitude to both of them, for the wonderful
and formative time I spent with them.

I am thankful to Finn Larsen and Leo Pando Zayas, with whom I had
shorter collaborations, which were nevertheless memorably
delightful, and very educational. I am indebted to Finn and Leo also
for their continual encouragement and support during my phd.

While carrying out the analysis reported in Chapter~3 of this
dissertation, I have benefited from ideas and suggestions of Peter
Miller and Eric Rains; I am grateful to both of them for helpful
discussions on the subject.

I would like to thank Ratindranath Akhoury, Henriette Elvang, Finn
Larsen, Roberto Merlin, Leo Pando Zayas, and Len Sander for the
numerous enjoyable scientific conversations we have had, and also
for the physics that I have learned from them in their unforgettably
nice classes.

I am grateful to Lydia Bieri and Igor Kriz for their inspiring and
memorable courses on geometry and topology, through which I learned
to appreciate the exciting depth of modern mathematical ideas.

Thanks to all of the graduate students in the physics department at
University of Michigan for the fun time we had together. I wish to
acknowledge in particular the fruitful scientific interactions with
Anthony Charles, Mahdis Ghodrati, Marios Hadjiantonis, Jack Kearney,
Gino Knodel, Pedro Lisbao, Alejandro Lopez, Tim Olson, Uttam Paudel,
Vimal Rathee, Sam Roland, and Bob Zheng.

Finally, I would especially like to thank all of my friends and
family for their unending and unconditional support.
\restoregeometry
\clearpage


\phantomsection
\labelchapters 
\tableofcontents \clearpage


\phantomsection \addcontentsline{toc}{chapter}{List of Figures}
\listoffigures \clearpage

\phantomsection
\labelappendices 
\addcontentsline{toc}{chapter}{List of Appendices} \listofappendices
\clearpage



\mainmatter

\addtocontents{toc}{\protect\contentsline{chapter}{\vspace*{-6mm}}{}{}} 

\chapter{Introduction}

Quantum Field Theory (QFT) is, among other things, the theoretical
framework for understanding fundamental particles and their
interactions. The particles in a given QFT\footnote{A specific
\emph{quantum system} may be described by a specific \emph{model} in
quantum field theory; with an abuse of terminology, we will refer to
different ``model''s in the QFT framework as different ``QFT''s.
Also, we will have in mind only conventional QFT models, consisting
of fields with spin $\le 1$.} are divided into \emph{bosons}
carrying spin zero or one, and \emph{fermions} carrying spin
one-half; matter---in the form common on Earth---typically consists
of the fermions, and the bosons mediate interactions between the
matter particles.

Despite theoretical successes in the regime where the QFT particles
interact weakly and \emph{perturbation theory} accurately describes
a wide range of observed phenomena, lack of progress on
long-standing strong-interaction problems in QFT (such as the
problem of quark confinement) indicates that more powerful
\emph{non-perturbative} techniques are needed.

A promising arena wherein to uncover non-perturbative structures in
QFT is the realm of \emph{Supersymmetric} (SUSY) QFTs. These are
theories enjoying a powerful symmetry that, roughly speaking,
exchanges their fermions and bosons. Many of the properties of
supersymmetric theories are under better analytic control, sometimes
even in the strong-interaction regime, thanks to their large
symmetry group.

A subset of SUSY QFTs are yet much more symmetric, and it is natural
to start with them in the quest for non-perturbative understanding
of QFTs. These are Conformal SUSY QFTs, also known as SuperConformal
Field Theories (SCFTs). An SCFT has a Hilbert space that is
invariant not only under supersymmetry, but also under the action of
the conformal group, which in four space-time dimensions can be
described\footnote{For brevity of exposition, we will not
distinguish here between the symmetry group and its universal
cover.} as SU($2,2$); with minimal supersymmetry (i.e. four Poincare
and four conformal supercharges) added, the conformal group extends
to the $\mathcal{N}=1$ superconformal group SU($2,2|1$).

Besides serving as toy models of the richer non-conformal SUSY QFTs,
SCFTs play a conceptually important role in the renormalization
group (RG) approach to SUSY QFT: one can often think of
non-conformal SUSY QFTs as describing ``flows'' between some SCFT in
the ultraviolet (UV), and some SCFT in the infrared (IR) regime of
energies. Thus, SCFTs can serve as signposts on the landscape of
SUSY QFTs.\\

The main piece of data of an SCFT is the spectrum---of the various
quantum numbers---of the states in its Hilbert space. The states of
a conformal field theory are in one-to-one correspondence with the
local operators in the theory \cite{Polchinski:ST}. Therefore an
important objective when studying a given SCFT is to classify and
count the local operators/states of the theory.

Since the Hilbert space of an SCFT is invariant under the 4d
superconformal group, the states/operators in it are labeled by
quantum numbers associated to the generators of the maximal compact
bosonic subgroup of SU($2,2|1$). A subset of the states/operators
(known as BPS states/operators), which have specific relations
between their various quantum numbers, sit in short representations
of the superconformal algebra. This sector is expected to be
protected against ``smooth deformations'' arising from RG flows or
interactions, because the quantum numbers of the short
representations do not undergo smooth changes\footnote{This is true
modulo multiplet recombination. The possibilities for recombination
are limited though, by the fact that the R-charges of local
operators (in a Lagrangian 4d SCFT) are algebraic numbers; c.f.
\cite{Intriligator:2003jj}.}; the states/operators in this sector
are hence under better analytic control thanks to this
``topological'' structure that the superconformal symmetry induces
on the Hilbert space. The topological sector consisting of the short
representations is sometimes referred to as the \emph{BPS sector} of
the SCFT. The \emph{superconformal index}
\cite{Romelsberger:2005eg,Kinney:2005ej} is a particular partition
function which efficiently quantifies this controllable sector of an
SCFT. As a partition function, it depends on an inverse-temperature
parameter\footnote{The index also depends on a squashing parameter
$b$ which for simplicity we suppress (i.e. set to unity) in the
present chapter.} $\beta$ used to weigh various states with
Boltzmann-type factors. Investigating the $\beta$-dependence of the
superconformal index---or \emph{the index}---of various interesting
SCFTs is the main goal of this dissertation. More precisely, we
would like to understand how the index behaves as the temperature is
taken to infinity---or $\beta$ is taken to zero. (The
low-temperature ($\beta\to\infty$) asymptotics of the index is
rather trivial; see \cite{Ardehali:2015c}.)

It is worth emphasizing that the ``temperature'' parameter in the
index does not admit a thermal interpretation: in a path-integral
picture, the index is computed as the partition function on
$S^3\times S^1$ with \emph{periodic} boundary conditions around the
$S^1$ \cite{Assel:2014}, while the more familiar thermal partition
functions are computed with fermions having anti-periodic boundary
conditions around the Euclidean time circle. Nevertheless, the
Boltzmann-type factors entering the definition of the index [see for
instance Eq.~\ref{eq:indexDefInt} below] suggest this stretch of
terminology, and thus we will keep referring to $\beta$ as `inverse
temperature'.\\

The reader without prior familiarity with superconformal indices may
wonder if investigating the index, which is only a certain measure
of a certain sector of an SCFT, is a worthwhile endeavor. The
following two remarkable applications of the index respond to this
question in the positive.
\begin{itemize}
\item Application to supersymmetric duality: the so-called IR dualities
in SUSY QFT imply that two differently formulated SCFTs (e.g. two
SCFTs with different field contents) are exactly equivalent, even at
the \emph{non-perturbative} (!) level; the superconformal index can
serve as a probe of this equivalence, since a proposal for duality
of two formulations may be valid only if the indices computed using
the different formulations are equal. The great power of the index
in probing supersymmetric dualities was demonstrated in the seminal
paper of Dolan and Osborn \cite{Dolan:2008} around 2008.

\item Application to holography: according to the AdS/CFT
correspondence, the states encoded in (or ``counted by'') the
superconformal index of a holographic SCFT (such as the
$\mathcal{N}=4$ SYM) correspond to states of a quantum gravity
theory in Anti-de~Sitter (AdS) space. In particular, it is expected
that the index will help the microscopic counting of high-energy
quantum gravity states, such as Giant Gravitons
\cite{Bourdier:2015}.
\end{itemize}

We will see that understanding the high-temperature asymptotics of
the index not only leads to advances in both of the directions
itemized above, but also opens up new prospects for understanding
the \emph{non-perturbative} (!) low-energy dynamics of 4d
supersymmetric gauge theories compactified on a circle. (More
precisely, the high-temperature asymptotics of the index of an SCFT
formulated in the UV as a gauge theory on $R^4$, seems to encode
information on the Coulomb branch dynamics of the gauge theory on
$R^3\times S^1$; see subsection~\ref{sec:crossed}.)\\

In the remaining parts of this chapter, we first introduce the 4d
superconformal index, along with its famous precedent, the quantum
mechanical Witten index, more elaborately in
section~\ref{sec:WittenIntro}. Then we proceed in
section~\ref{sec:finiteNintro} to highlight our main result: the
high-temperature asymptotics of the superconformal index of
finite-rank Lagrangian unitary non-chiral 4d SCFTs. In
section~\ref{sec:LargeNintro} we discuss the high-temperature
asymptotics of the \emph{large-rank limit} of the indices of a class
of holographic SCFTs, and explain how our results address (for the
class of theories under study) the computation of the Holographic
Weyl Anomaly at the subleading order in the $1/N$ expansion.\\

\section{From Witten index to the superconformal
index}\label{sec:WittenIntro}

Consider a unitary quantum mechanical system enjoying
\emph{supersymmetry}. That is to say, there exists a fermionic
operator $Q$, referred to as the \emph{supercharge} operator, acting
on the Hilbert space of the system, and satisfying
\begin{equation}
\{Q,Q^{\dagger}\}=H\quad\quad\text{and}\quad\quad
Q^2=0,\label{eq:Qdef}
\end{equation}
with $H$ the Hamiltonian operator.

Existence of $Q$ implies that states of nonzero energy are paired in
the system: it can be easily checked using (\ref{eq:Qdef}) that
$Q+Q^{\dagger}$ provides a one-to-one mapping from the set of
bosonic states $|\mathrm{b}\rangle$ with $H|\mathrm{b}\rangle\neq
0$, to the set of fermionic states $|\mathrm{f}\rangle$ with
$H|\mathrm{f}\rangle\neq0$.

But the zero-energy states \emph{are not} necessarily paired:
\begin{equation}
\begin{split}
0=\langle \mathrm{b}| H |\mathrm{b}\rangle=\ & \langle \mathrm{b}|
\{Q,Q^\dagger\}|\mathrm{b}\rangle=\langle
\mathrm{b}| Q\ Q^\dagger|\mathrm{b}\rangle+\langle \mathrm{b}| Q^\dagger\ Q|\mathrm{b}\rangle\\
&\Rightarrow (Q+Q^\dagger)|\mathrm{b}\rangle=0,
\end{split}
\end{equation}
with a similar argument applying to the fermionic zero-energy
states. Since the states with zero energy are annihilated by $Q$ and
$Q^\dagger$, we say that the zero-energy states are
``supersymmetric''. Since (\ref{eq:Qdef}) implies that $H$
necessarily has non-negative eigenvalues, we can further say that
the zero-energy states are \emph{supersymmetric ground states} of
the theory.

Therefore in unitary supersymmetric quantum mechanics, the
\textbf{Witten index} \cite{Witten:1982}
\begin{equation}
\begin{split}
\mathcal{I}^W:=&\sum_i (-1)^F
e^{-\hat{\beta}E_i}\\
=&n^b_{\mathrm{z.e.}}-n^f_{\mathrm{z.e.}}
\end{split}
\end{equation}
(with $F$ the fermion-number operator, evaluating to $1$ on
fermionic states, and to $0$ on bosonic states), receives
contributions only from unpaired zero-energy (or supersymmetric
ground-) states; the index is thus obviously independent of
$\hat{\beta}$.

Less obviously, $\mathcal{I}^W$ is also independent\footnote{Modulo
subtleties (see \cite{Witten:1982}) that are not relevant for our
discussion.} of the \emph{interaction strength} in the system! The
reason is that any continuous deformation of the system, such as
that induced by RG flows or by variation of the interaction
couplings, should cause the supersymmetric ground-states to acquire
nonzero energy only in pairs; similarly, any nonzero-energy state
that as a result of the continuous deformation becomes a
supersymmetric ground-state, should be accompanied by a partner
state all along. This independence of the index from continuous
deformations gives it a topological character---hence the
well-deserved title ``index''.

Since $\mathcal{I}^W$ is independent of the couplings, it can be
computed even in strongly interacting theories. It can thus provide
information that would otherwise be inaccessible through (the more
conventional) perturbative means. The original work of Witten used
this index to probe the ground-state(s) of non-abelian
supersymmetric gauge theories. The idea was that a \emph{nonzero
index} would mean that either $n^b_{\mathrm{z.e.}}\neq0$ or
$n^f_{\mathrm{z.e.}}\neq0$, and thus it would imply the existence of
supersymmetric ground-states, and hence the \emph{absence of
spontaneous supersymmetry breaking}. It is worth emphasizing the
remarkable fact that a weak-coupling calculation of $\mathcal{I}^W$
can yield nontrivial information about the (strongly-interacting)
ground-state of a non-abelian gauge theory!\\

The above discussion in the context of supersymmetric quantum
mechanics can now help us to extend the concept of an index to
unitary 4d SCFTs.

In any 4d SCFT there exists a supercharge operator\footnote{In fact
any 4d SCFT has at least four such operators. The $Q$ that we
consider here is one [it doesn't matter which] of the two that
transform inside a $(0,1/2)$ representation of the (complexified)
Lorentz group.} $Q$, such that
\begin{equation}
\{Q,Q^{\dagger}\}=H-2J_2^z-\frac{3}{2}R\quad\quad\text{and}\quad\quad
Q^2=0,
\end{equation}
with $H$ the Hamiltonian in the radial quantization, $J_2^z$ the
third generator of the right-handed\footnote{Had we chosen a $Q$
operator transforming inside a $(1/2,0)$ representation of the
Lorentz group, $J_2^z$ would be replaced with $J_1^z$.} Lorentz
SU($2$), and $R$ the generator of the U($1$)$_R$ inside the
$\mathcal{N}=1$ superconformal group SU($2,2|1$).

Therefore, in analogy with the above quantum mechanical discussion,
we can define the following Witten index for unitary 4d SCFTs:
\begin{equation}
\mathcal{I}^W:=\mathrm{Tr}(-1)^{F}e^{-\hat{\beta}(E-2j_2-\frac{3}{2}r)},\label{eq:unrefnd}
\end{equation}
with $E,j_2,r$ the eigenvalues of $H,J_2^z,R$, and with the trace
taken over the Hilbert space in the radial quantization. Similarly
to the quantum mechanical case above, the dependence on
$\hat{\beta}$ drops out, since only states with vanishing
$E-2j_2-\frac{3}{2}r$ (hence sitting in short representations of
SU($2,2|1$)) have a chance of surviving bose-fermi cancelations.

It turns out that the combination $H-R/2$ commutes with the
supercharge $Q$ used above. We can hence refine $\mathcal{I}^W$ with
a fugacity $e^{-\beta}$ for the combination $E-r/2$, without ruining
the cancelations underlying its topological character. We refer to
this refined Witten index as the \textbf{4d superconformal index}
\cite{Romelsberger:2005eg,Kinney:2005ej}:
\begin{equation}
 \mathcal{I}
(\beta):=\mathrm{Tr}(-1)^{F}e^{-\beta(E-\frac{r}{2})}e^{-\hat{\beta}(E-2j_2-\frac{3}{2}r)}.
\end{equation}
In fact without the refinement with $\beta$, the index---as defined
in (\ref{eq:unrefnd})---is divergent in interesting SCFTs; thus the
Boltzmann-type factor $e^{-\beta(E-\frac{r}{2})}$ is actually
necessary as a regulator. Since the superconformal index does not
depend on $\hat{\beta}$, we can write
\begin{equation}
 \mathcal{I}
(\beta)=\sum(-1)^{F}e^{-\beta(E-\frac{r}{2})},\label{eq:indexDefInt}
\end{equation}
with the sum taken either over the local operators in the SCFT, or
equivalently (via the CFT state/operator correspondence) over the
states in the radial quantization.

The index (\ref{eq:indexDefInt}) can be computed in closed form for
a wide variety of interesting SCFTs. For instance the index of a
free chiral multiplet is given by an \emph{elliptic gamma function}
(see appendix~\ref{sec:app1} for the definition of the elliptic
gamma):
\begin{equation}
 \mathcal{I}_{\chi}
(\beta)=\Gamma(e^{-2\beta/3};e^{-\beta},e^{-\beta}).\label{eq:indexFreeChi}
\end{equation}
In theories with several decoupled chiral multiplets, the index
would be given by a product of the corresponding elliptic gamma
functions. In gauge theories with several chiral multiplets, the
index would be given by a product of several chiral-multiplet gamma
functions and several vector-multiplet gamma function, integrated
(roughly speaking) over the gauge group so as to project the result
onto the gauge-singlet sector. We will explain this more carefully
in Chapter~\ref{chap:2}.\\

\section{High-temperature asymptotics of the index of finite-$N$ gauge
theories}\label{sec:finiteNintro}

We now summarize the rich structure we find in the high-temperature
limit of the superconformal index. Throughout this dissertation we
focus on the index of unitary 4d SCFTs that admit a gauge theory
description with \emph{non-chiral matter content}; the SU($N$)
$\mathcal{N}=4$ SYM and the SU($N$) SQCD fixed points are the
examples that we ask the reader to keep in mind while reading the
somewhat abstract discussion below. We emphasize that in the present
section we are considering gauge theories with a finite rank; the
large-$N$ limit of superconformal indices will be discussed in the
next section of the present chapter.\\

The index of a \emph{Lagrangian SCFT} [by that we mean an SCFT
admitting a gauge theory description] is given by an
\textbf{elliptic hypergeometric integral} (EHI) \cite{Dolan:2008}.
This is an expression of the form $\int f(\beta;x_1,\dots,x_{r_G})\
\mathrm{d}^{r_G}x$, with $r_G$ the rank of the gauge group $G$ of
the Lagrangian SCFT, $\beta \ (>0)$ the inverse temperature, and
$x_i\ (\in [-1/2,1/2])$ the integration variables. The function $f$
is a complicated special function of its arguments, given explicitly
as a product of several elliptic gamma functions; moreover, when the
SCFT is non-chiral, $f$ is real and positive semi-definite. The
integral over $-1/2<x_i<1/2$ roughly projects onto the gauge-singlet
sector; colloquially speaking, it washes out the contribution of
non-gauge-invariant operators to the index\footnote{This is
analogous to how the zeroth (or the ``singlet'') Fourier component
of a periodic real function is obtained by integrating that
function.}.

The high-temperature ($\beta\to0$) limit of the index corresponds to
the \textbf{hyperbolic limit} of the EHI. This limit has been
rigorously analyzed by Eric~Rains \cite{Rains:2009} (around 2006) in
certain special EHIs. We put the EHIs studied by Rains in the wider
context of the EHIs arising from non-chiral unitary Lagrangian 4d
SCFTs. In this generalized framework, the methods of Rains can be
extended to uncover a surprisingly rich structure. We find (using,
in particular, appropriate uniform estimates (derived in
appendix~\ref{sec:app2}) for the elliptic gamma function) that in
the $\beta\to 0$ limit $\mathcal{I}(\beta)$ simplifies as
\begin{equation}
\begin{split}
\mathcal{I}(\beta)=\int f(\beta;\mathbf{x})\
\mathrm{d}^{r_G}x\overset{\beta\to 0}{\longrightarrow}\int
e^{-(\mathcal{E}_0^{DK}(\beta)+V^{\mathrm{eff}}(x_1,\dots,x_{r_G};\beta))}\
\mathrm{d}^{r_G}x,\label{eq:IsimpIntro}
\end{split}
\end{equation}
with
\begin{equation}
\begin{split}
\mathcal{E}^{DK}_0(\beta)=-\frac{16\pi^2}{3\beta}(c-a),\label{eq:EdkEquivIntro}
\end{split}
\end{equation}
where $c$ and $a$ are the central charges\footnote{The central
charges ($\in\mathbb{R}^{>0}$) are measures of the number of degrees
of freedom in the SCFT. For an SCFT described by an SU($N$) gauge
theory, $c$ and $a$ are typically of order $N^2$ at large $N$; for
example, for the SU($N$) $\mathcal{N}=4$ SYM we have
$c=a=(N^2-1)/4$. See Chapter~3 for a precise expression for $c-a$ in
terms of the matter content.} of the SCFT. We have given a
superscript \emph{DK} to $\mathcal{E}_0$, because a proposal of
Di~Pietro and Komargodski \cite{DiPietro:2014} implies the
high-temperature asymptotics $\mathcal{I}(\beta)\approx
e^{-\mathcal{E}^{DK}_0(\beta)}$ (see \cite{Aharony:2013a} for an
earlier hint of this asymptotic formula).

We observe from (\ref{eq:IsimpIntro}) that an effective potential
$V^{\mathrm{eff}}(\mathbf{x};\beta)$ dictates the high-temperature
asymptotics of $\mathcal{I}(\beta)$. It turns out that
\begin{equation}
V^{\mathrm{eff}}(\mathbf{x};\beta)=\frac{4\pi^2}{\beta}L_h(\mathbf{x}),\label{eq:VeffLh}
\end{equation}
with $L_h$ a continuous, real, piecewise linear function of the
$x_i$, which is determined by the matter content of the SCFT
(examples can be found in the Figures~\ref{fig:A1}, \ref{fig:SO5},
and \ref{fig:ISS} below). We will refer to $L_h$ as the \emph{Rains
function} of the SCFT. The relations (\ref{eq:IsimpIntro}) and
(\ref{eq:VeffLh}) imply that the index localizes in the $\beta\to0$
limit to the locus of minima of $L_h$. We thus find
\begin{equation}
\begin{split}
\mathcal{I}(\beta)\approx
e^{-(\mathcal{E}_0^{DK}(\beta)+V^{\mathrm{eff}}_{\mathrm{min}}(\beta))}.\label{eq:IsimpIntro2}
\end{split}
\end{equation}
Taking the logarithm of the two sides, we can write this as [the
subleading term and the error estimate will be justified in Chapter
3]
\begin{equation}
\ln \mathcal{I}(\beta)= \frac{16\pi^2}{3\beta}(c-a-\frac{3}{4}L_{h\
min})+\mathrm{dim}\mathfrak{h}_{qu}\ln(\frac{2\pi}{\beta})+O(\beta^0),\label{eq:dkSCFTsCorrected}
\end{equation}
with $L_{h\ \mathrm{min}}$ (which we will prove to be $\le 0$) the
minimum of the Rains function over $-1/2\le x_i\le 1/2$, and
$\mathrm{dim}\mathfrak{h}_{qu}$ the dimension of the locus of minima
of $L_h$.\\

The minimization problem for $L_h(\mathbf{x})$ can often be
analytically solved on a case by case basis (as in
\cite{Rains:2009}) using certain generalized triangle inequalities
(GTIs); for most SCFTs of interest to us, the required GTI is
obtained as a corollary of Rains's GTI, which can be found in
appendix~\ref{sec:app3}.\\

Note that the leading piece in (\ref{eq:dkSCFTsCorrected}) takes the
same form as the Di~Pietro-Komargodski formula
$\ln\mathcal{I}(\beta)\approx
-\mathcal{E}^{DK}_0(\beta)=\frac{16\pi^2}{3\beta}(c-a)$, but with
the ``shifted $c-a$'' defined as
\begin{equation}
(c-a)_{\mathrm{shifted}}:= c-a-\frac{3}{4}L_{h\ \mathrm{min}}.
\end{equation}
This last relation appears to be analogous to the equation
\begin{equation}
c_{\mathrm{eff}}= c-24h_{\mathrm{min}},
\end{equation}
frequently discussed in the context of non-unitary 2d CFTs (see e.g.
\cite{Kutasov:1991ks}).\\


One application of the result (\ref{eq:dkSCFTsCorrected}) is to
supersymmetric dualities. Dual SCFTs must have identical partition
functions. Comparison of the indices provides one of the strongest
tests of any proposed duality between $\mathcal{N}=1$ SCFTs
\cite{Dolan:2008,Spirido:2009}. The full comparison of the
multiple-integrals computing superconformal indices is, however,
extremely challenging, except for the few cases (corresponding to
various SQCD-type theories
\cite{Dolan:2008,Spirido:2009,Spirido:2011or}) already established
in the mathematics literature (e.g. in the celebrated work of Rains
\cite{Rains:2005} (from around 2005) on ``transformations'' of
elliptic hypergeometric integrals). Rather, known dualities are
frequently used to conjecture new identities between multi-variable
integrals of elliptic hypergeometric type
\cite{Dolan:2008,Spirido:2009,Spirido:2011or,Kutasov:2014}.

We propose comparison of the high-temperature asymptotics of the
indices. Since dual SCFTs have equal central charges, the relation
(\ref{eq:dkSCFTsCorrected}) implies that dual SCFTs must also have
equal $L_{h\ \mathrm{min}}$ and $\mathrm{dim}\mathfrak{h}_{qu}$;
these new non-trivial tests of supersymmetric dualities were checked
in \cite{Ardehali:2015c} for several specific cases, validating
well-known duality conjectures. A few examples of the applications
of these tests can be found also in subsection~\ref{sec:dualApp}
below. We emphasize that these tests are independent of `t~Hooft
anomaly matchings (see \cite{Ardehali:2015c} for a more detailed
discussion).\\

Another application of (\ref{eq:dkSCFTsCorrected}) is to holography.
For the specific case of the SU($N$) $\mathcal{N}=4$ SYM, we have
$c-a=L_{h\ \mathrm{min}}=0$ and $\mathrm{dim}\mathfrak{h}_{qu}=N-1$.
This means that the asymptotic growth of the index of this SCFT is
power-law. Producing this power-law asymptotics from the holographic
dual seems to require the state counting of supersymmetric (more
precisely, $1/16$-BPS) giant gravitons \cite{Bourdier:2015}; this
appears to be a very interesting objective within reach of current
technology. See subsection~\ref{sec:holoApp} for a more detailed
discussion.

\subsection{Relation to previous work}

Only two previous works attacked the problem of the high-temperature
asymptotics of the 4d superconformal index of rather general
Lagrangian SCFTs. (Other papers have considered this problem in
special theories; see subsection~1.2 of \cite{Ardehali:2015c} for
references to such papers.)

In the 2013 work of Aharony et. al. \cite{Aharony:2013a} an EHI-type
expression for the index was considered. Then, assuming that at high
temperatures the integrand of the EHI is localized around the unit
element of the gauge group, the relation $\mathcal{I}(\beta)\approx
e^{-\mathcal{E}^{DK}_0(\beta)}Z_{S^3}$ (or more precisely, an
equivariant generalization thereof) was arrived at; $Z_{S^3}$ stands
for the three-sphere partition function of the dimensionally reduced
daughter of the 4d SCFT (see subsection~\ref{sec:S3} below for a
matrix-integral expression for $Z_{S^3}$). The authors of
\cite{Aharony:2013a} pointed out, however, that the result can not
be trusted in general, as $Z_{S^3}$ may be divergent (as the cut-off
of the matrix-integral computing it is taken to infinity) due to an
unlifted Coulomb branch in the 3d theory; see \cite{Aharony:2013b}
for an explicit discussion of unlifted Coulomb branches.

In the 2014 work of Di~Pietro and Komargodski \cite{DiPietro:2014}
no explicit form for the index was assumed. But it was assumed that
the 4d SCFT is Lagrangian, and that $Z_{S^3}$ is at most power-law
divergent with respect to the cut-off ($\propto 1/\beta$) of the
high-temperature effective field theory describing the massless
sector of the circle-compactified theory living on $S^3$. It was
then intuitively argued that such power-law divergences would modify
the asymptotics $\mathcal{I}(\beta)\approx
e^{-\mathcal{E}^{DK}_0(\beta)}$ only at the (generically) subleading
order in a small-$\beta$ expansion, such that
$\mathcal{I}(\beta)\approx
(\frac{1}{\beta})^{n_m}e^{-\mathcal{E}^{DK}_0(\beta)}$, with $n_m$
related to the number of unlifted moduli.\\

In the present dissertation (following \cite{Ardehali:2015c}) we
show that Rains's rigorous machinery in \cite{Rains:2009} can be
adapted for a definitive general analysis of the high-temperature
asymptotics of the superconformal indices of non-chiral unitary 4d
Lagrangian SCFTs. We derive results that clarify the following
points:
\begin{itemize}
\item $[$explicit study of various examples leads to the
conjecture that$]$ in theories where $Z_{S^3}$ is power-law
divergent, the (generically) subleading power-law asymptotics of
$\mathcal{I}(\beta)$ can be most nicely associated with a ``Coulomb
branch'' picture \emph{in the crossed channel} (see
subsection~\ref{sec:crossed});
\item in some of the most interesting SCFTs (more specifically, in
certain interacting $\mathcal{N}=1$ SCFTs with $c<a$), $Z_{S^3}$ is
\emph{exponentially divergent}, and as a result even the leading
asymptotics $\mathcal{I}(\beta)\approx
e^{-\mathcal{E}^{DK}_0(\beta)}$ receives a modification, with the
correct asymptotics reading $\mathcal{I}(\beta)\approx
e^{-(\mathcal{E}^{DK}_0(\beta)+V^{\mathrm{eff}}_{\mathrm{min}})}$
(see section~\ref{sec:general}, and subsections~\ref{sec:iss} and
\ref{sec:BCI}).\\
\end{itemize}

\section{Taking the large-$N$ limit of the index
first}\label{sec:LargeNintro}

In holography, or more specifically in the AdS/CFT correspondence,
the large-$N$ limit of gauge theories plays an important role. We
will focus on a certain class of holographic SCFTs when discussing
the large-$N$ limit; these are SCFTs arising from toric quiver
gauges theories. One of their important features is that they are
dual to IIB string theory on AdS$_5\times$SE$_5$, with SE$_5$ a
\emph{toric} Sasaki-Einstein 5-manifold.

Taking the large-$N$ limit of the index of these theories one
obtains the \emph{multi-trace index} of the SCFT
\cite{Gadde:2010holo}; this is the index of the multi-trace
operators of the SCFT in the planar limit. This index is
holographically dual to the multi-particle index of the gravity
side; the multi-particle index receives contributions from
multi-particle Kaluza-Klein (KK) states in the bulk. The
multi-particle index can be related through simple combinatorial
procedures (namely via \emph{plethystic} exponentials/logarithms
\cite{Benvenuti:2006}) to the single-particle index of the gravity
theory, which receives contributions only from the bulk
single-particle KK states.\\

In a series of papers written by Jim~Liu, Phil~Szepietowski, and the
author, it was discovered that the high-temperature asymptotics of
the single-particle index encodes the bulk KK fields' contribution
to the \emph{subleading holographic Weyl anomaly}
\cite{Ardehali:2014,Ardehali:2015a,Ardehali:2015b}. The problem of
holographic Weyl anomaly is to reproduce the central charges $a$ and
$c$ of a holographic SCFT from its gravitational dual\footnote{The
expression ``Weyl anomaly'' is used because the central charges
determine, among other things, the anomalous behavior of the SCFT
partition function under Weyl re-scalings of the spacetime metric;
see e.g. \cite{Henningson:1998gx}.}. In a large-$N$ expansion, the
leading ($O(N^2)$) piece of the central charges can be
holographically obtained using Einstein gravity on the AdS side;
this was done around 1998 \cite{Henningson:1998gx}. Obtaining the
subleading ($O(1)$) piece of the central charges from the gravity
side was more challenging, until the relation with the
superconformal index was understood
\cite{Ardehali:2014,Ardehali:2015a,Ardehali:2015b}.

The holographic connection between the subleading central charges
and the single-particle index is derived roughly as follows. First
of all, long multiplets of SU($2,2|1$) in the bulk KK spectrum do
not contribute to either the single-particle index or the
holographic central charges. Next, for short multiplets,
\emph{irrespective of the type} (which could be chiral, anti-chiral,
conserved, semi-long I, or semi-long II), the holographic
contribution to the central charges takes a simple form, determined
by the high-temperature asymptotics of the contribution of the
multiplet to the single-particle index. Summing up the contributions
of all the KK particles in the bulk, one concludes that the
high-temperature asymptotics of the single-particle index is related
to the subleading holographic Weyl anomaly. This relation will be
discussed further in Chapter 4; there we will explain how the
relation leads to a solution to the problem of Holographic Weyl
Anomaly in toric quiver SCFTs.\\

\section{Overview of the publications this dissertation is based on}

\begin{itemize}
\item A.~A. Ardehali, J.~T.~Liu, and P.~Szepietowski,
 \emph{$c-a$ from the $\mathcal{N}=1$ superconformal index,}
  JHEP {\bf 1412}, 145 (2014) [arXiv:1407.6024 [hep-th]]. (Listed as reference
  \cite{Ardehali:2014}.)\\
  This work established a holographic relation between the difference of
  the central charges (i.e. $c-a$) and the single-particle index, in the context of
  4d SCFTs dual to IIB theory on AdS$_5\times$SE$_5$ (with SE$_5$ a Sasaki-Einstein
  5-manifold). The relation was then checked
  explicitly for toric quiver SCFTs (with SU($N$) nodes) without
  adjoint matter and with a smooth dual SE$_5$; this successful check can be
  considered a test of AdS/CFT at the subleading order (in $1/N$) for an infinite class
  of holographic SCFTs.

  The paper also conjectured the
  holographically derived
  relation between $c-a$ and the index to hold for all (not necessarily holographic)
  4d SCFTs; this conjecture was ruled out later in \cite{Ardehali:2015c}.

\item A.~A.~Ardehali, J.~T.~Liu, and P.~Szepietowski,
 \emph{Central charges from the $\mathcal{N}=1$ superconformal index,} Phys. Rev. Lett. {\bf 114}, 091603 (2015)
  [arXiv:1411.5028 [hep-th]]. (Listed as reference
  \cite{Ardehali:2015a}.)\\
  This work extended the holographic result of the previous paper to
  expressions for the $O(N^0)$ pieces of $a$ and $c$ separately. (Note that for SCFTs dual to
  AdS$_5\times$SE$_5$ we always have $c-a=O(N^0)$.) The relations were
  then explicitly checked for toric quiver SCFTs (with SU($N$) nodes) without
  adjoint matter and dual to smooth SE$_5$; this check constitutes a very strong
  and general test of AdS/CFT at the subleading order in the $1/N$ expansion.

  The
  paper also presented general conjectures for extracting the central
  charges of any (finite rank, not necessarily holographic) 4d SCFT from its index, but
  those conjectures were later ruled out in \cite{Ardehali:2015c}.

\item A.~A.~Ardehali, J.~T.~Liu, and P.~Szepietowski,
 \emph{High-temperature expansion of supersymmetric partition functions,} JHEP {\bf
 1507}, 113 (2015)
 [arXiv:1502.07737 [hep-th]]. (Listed as reference
  \cite{Ardehali:2015b}.)\\
 This work generalized the above-mentioned AdS/CFT matching of the subleading
 central charges to all toric quivers with SU($N$) nodes
 (even to quivers with adjoint matter fields and/or with singular
 dual SE$_5$; there were two extra assumptions made though, as explained in Chapter~4 below).

This paper contains also the first correct calculation of the SUSY
Casimir energy in the literature; it thereby clarified the
connection between the 4d superconformal index, and its
corresponding SUSY partition function computed by path-integration
over $S_b^3\times S_\beta^1$ (with $S_b^3$ the unit three-sphere
with squashing parameter $b$, and $\beta$ the circumference of the
circle). This result appeared shortly afterwards also in the
independent work of Assel et. al. \cite{Assel:2015s}.

 The paper \cite{Ardehali:2015b} also proposed a conjecture for the high-temperature asymptotics
 of the indices of general (finite-rank) 4d SCFTs; that conjecture was ruled out
 later in \cite{Ardehali:2015c}.

\item A.~A.~Ardehali,
 \emph{High-temperature asymptotics of supersymetric partition functions,}
  [arXiv:1512.03376 [hep-th]]. (Listed as reference
  \cite{Ardehali:2015c}.)\\
  This paper extended Rains's analysis \cite{Rains:2009} to study the high- (and low-) temperature asymptotics
  of the index of Lagrangian SCFTs with a semi-simple gauge group
  (under some extra simplifying assumptions spelled out
  at the beginning of the Discussion section in \cite{Ardehali:2015c}).\\

\end{itemize}

\vfill

\section{Novel results}

There are three previously unpublished results in the present dissertation.\\

The first is an improved derivation (compared to the original one in
\cite{Ardehali:2015c}) of the asymptotics of the indices of
\emph{non-chiral} SCFTs. This derivation is given in Chapter 3, and
leads to Eq.~(\ref{eq:LagIndexSimp6noTheta}), which is our main
result. The original derivation (reported in \cite{Ardehali:2015c})
of Eq.~(\ref{eq:LagIndexSimp6noTheta}) was based on the physically
expected---but mathematically unjustified---assumption that certain
cancelations do not occur in the high-temperature limit of the EHIs
arising from SUSY gauge theories (see the comments below Eq.~(3.15)
of \cite{Ardehali:2015c}).\\

The second previously unpublished result is the asymptotics, shown
in (\ref{eq:AsyClassS}), of the index of the puncture-less SU($2$)
class-$\mathcal{S}$ theories of genus $g\ge 2$; the result is
interesting: these $\mathcal{N}=2$ SCFTs satisfy the
Di~Pietro-Komargodski formula, even though they famously have the
unusual balance $c<a$ between their central charges. These theories
are thus to be contrasted with the $\mathcal{N}=1$ SCFTs with $c<a$
discussed in subsections~\ref{sec:iss} and \ref{sec:BCI}, which do
not satisfy the Di~Pietro-Komargodski formula.\\

The third novel result is the relation between the high-temperature
asymptotics of the single-trace and multi-trace indices, shown in
(\ref{eq:largeNAsy}), and its corollary in Eq.~(\ref{eq:AkAnsatz}).
Part of the relation (\ref{eq:AkAnsatz}) was given as an ansatz in
\cite{Ardehali:2015b}; we not only prove that ansatz in
appendix~\ref{sec:app4}, but also derive a piece of it that was left
undetermined in \cite{Ardehali:2015b}.

\chapter{The 4d superconformal index}\label{chap:2}

The goal of this chapter is to write down---and to explain---the
explicit expression for the elliptic hypergeometric integral (EHI)
whose high-temperature asymptotics we will analyze (under certain
simplifying conditions) in the next chapter. This expression can be
found in Eq.~(\ref{eq:LagEquivIndex}) below.

In the physical context, the EHI in Eq.~(\ref{eq:LagEquivIndex}) may
arise as the superconformal index of a 4d Lagrangian SCFT.
Elaborating on the physical context is the purpose of the following
two sections. The reader not interested in---or already familiar
with---this physical context can skip directly to the third section
below (i.e. section \ref{sec:EHI}) where the EHI of our interest is
spelled out.\\

\section{Background: The building blocks of a unitary 4d Lagrangian
SCFT}\label{sec:BG}

The Hilbert space of a 4d SCFT is invariant under the action of the
4d $\mathcal{N}=1$ superconformal group SU($2,2|1$). The generators
of this group constitute the 4d superconformal algebra.

The bosonic part of the 4d superconformal algebra consists of the 4d
conformal algebra and a U($1$) automorphism referred to as the
U($1$)$_R$. We denote the charge of a state under $H$ (the generator
of dilations, which in the radial quantization becomes the
Hamiltonian) by $E$, the charge under $R$ (the generator of the
U($1$)$_R$) by $r$, and the charges under $J^z_1$ and $J^z_2$ (the
Cartan generators of the left and right SU($2$) spins of the Lorentz
group) by $j_1$ and $j_2$. All these charges are real numbers, $j_1$
and $j_2$ are half-integers, and unitarity implies $E\ge0$.

The fermionic part the 4d superconformal algebra consists of the
supercharges $Q_\alpha$, $\bar{Q}_{\dot{\alpha}}$, and their
conformal partners $S^\alpha$, $\bar{S}^{\dot{\alpha}}$.
Importantly, we have
$\{Q_\alpha,\bar{Q}_{\dot{\alpha}}\}=2P_{\alpha\dot{\alpha}}$, and
$\{S^\alpha,\bar{S}^{\dot{\alpha}}\}=2K^{\alpha\dot{\alpha}}$, with
$P$ and $K$ respectively the generators of translations and special
conformal transformations.\\

A computationally efficient description of a Lagrangian SCFT is
provided, however, not through the Hilbert space perspective, but by
the field content of the gauge theory that flows to it. The field
content of a supersymmetric gauge theory is organized inside
supermultiplets. Focusing on interacting unitary 4d Lagrangian SCFTs
with fields of spin $\le 1$, we are left with two possible
supermultiplets: chiral multiplets and vector multiplets.

A chiral multiplet consists of a complex scalar and a Weyl fermion,
whereas a vector multiplet consists of a vector boson and a Weyl
fermion. (Note that since we are interested in SCFTs, we are
restricting our attention to QFTs with massless field content.)

The scalar inside a chiral multiplet $\chi$ has an R-charge that we
denote by $r_\chi$; the R-charge of the supersymmetric partner (the
Weyl fermion in the same multiplet) is $r_\chi-1$. On the other
hand, a vector boson has zero R-charge, and its superpartner (the
Weyl fermion in the same multiplet, referred to as the gaugino) has
R-charge 1.

The interaction of massless vector bosons is described by a gauge
theory. This means, among other things, that the vector boson field
transforms in the adjoint representation of a gauge group $G$ (which
we take to be a compact matrix Lie group with a semi-simple
algebra). A chiral multiplet $\chi$ in the theory may transform in a
representation $\mathcal{R}_\chi$ of the gauge group $G$.\\

With the above background in mind, and for our purposes below,
\textbf{we take the following as the defining data of a unitary 4d
Lagrangian SCFT}: $i)$ a \emph{gauge group} $G$, which we take to be
a compact semi-simple matrix Lie group of rank $r_G$, denote its
typical root vector by $\alpha:=(\alpha_1,\dots,\alpha_{r_G})$, and
denote the set of all the roots by $\Delta_G$; $ii)$ a finite number
of \emph{chiral multiplets} $\chi_j:=\{\mathcal{R}_j,r_j\}$, with
$j=1,\dots,n_\chi$, where $\mathcal{R}_j$ is a finite-dimensional
irreducible representation of $G$, whose typical weight vector we
denote by $\rho^j:=(\rho^j_1,\dots,\rho^j_{r_G})$, and the set of
all the weights of $\mathcal{R}_j$ we denote by $\Delta_{j}$, while
$r_j\ (\in]0,2[)$ is the \emph{R-charge} of the chiral multiplet
$\chi_j$.

We further demand that the following \emph{anomaly cancelation}
conditions be satisfied by the $\mathcal{R}_{j}$ and the $r_{j}$:
\begin{equation}
\sum_j\sum_{\rho^j\in\Delta_{j}}\rho^j_l\rho^j_m\rho^j_n=0,\quad
(\mathrm{for\ all\ }l,m,n)
\end{equation}
\begin{equation}
\sum_j\sum_{\rho^j\in\Delta_j}\rho^j_l=0, \quad (\mathrm{for\ all\ }
l)
\end{equation}
\begin{equation}
\sum_j(r_j-1)\sum_{\rho^j\in\Delta_j}\rho^j_l\rho^j_m+\sum_{\alpha\in\Delta_G}\alpha_{l}\alpha_{m}=0,
\quad (\mathrm{for\ all\ }l,m)
\end{equation}
\begin{equation}
\sum_j(r_j-1)^2\sum_{\rho^j\in\Delta_j}\rho^j_l=0 \quad
(\mathrm{for\ all\ }l).
\end{equation}
These relations correspond respectively to cancelation of the
following anomalies: $i)$ the gauge$^3$ anomaly; $ii)$ the
gauge-gravitational-gravitational anomaly; $iii)$ the
U(1)$_R$-gauge-gauge anomaly; and $iv)$ the
gauge-U(1)$_R$-U(1)$_R$ anomaly.\\

Note that it would be more appropriate to say that the above data
defines `a SUSY gauge theory with a U($1$) R-symmetry', and not
necessarily an SCFT. In particular, the above conditions on the data
do not guarantee that the $r_\chi$ are the \emph{superconformal}
R-charges of the chiral multiplets in the IR fixed point of the SUSY
gauge theory defined by the above data; for instance, the SU($N_c$)
SQCD with R-charge assignment $r_\chi=1-N_c/N_f$, for $N_f>N_c$ but
outside the conformal window, \emph{does} satisfy the above
conditions, even though its IR fixed point is free, with emergent
accidental symmetries mixing with its U($1$)$_R$ in the infrared.
Therefore we keep in mind that only a subset of the SUSY gauge
theories defined by the above data lead to SCFTs with the chiral
multiplets in the IR having superconformal R-charges $r_\chi$. On
the other hand, any SUSY gauge theory with U($1$) R-symmetry---as
defined by the above data---can be assigned an EHI via
Eq.~(\ref{eq:LagEquivIndex}) below; for non-conformal theories the
resulting EHI can be thought of as arising from path-integration
(c.f. \cite{Assel:2014}), rather than from a ``superconformal''
index calculation.\\

\section{Definition of the index}

The superconformal index is defined as
\begin{equation}
\mathcal
I(b,\beta)=\mathrm{Tr}\left[(-1)^Fe^{-\hat\beta(E-2j_2-\fft32r)}
p^{j_1+j_2+\fft12r}q^{-j_1+j_2+\fft12r}\right],
\label{eq:equivIndexDef}
\end{equation}
with $p=e^{-b\beta}$ and $q=e^{-b^{-1}\beta}$; we take $b,\beta>0$,
and refer to $b$ as the \emph{squashing parameter}, and $\beta$ as
the \emph{inverse temperature} (the reason for these names will
become clear shortly); the special case with $b=1$ corresponds to
the index introduced in Chapter 1. The trace in the above relation
is over the Hilbert space of the theory on $S^3\times \mathbb{R}$,
with $S^3$ the round unit three-sphere, and $\mathbb{R}$ the time
direction. The index is independent of $\hat\beta$ because it only
receives uncanceled contributions from states with
$E-2j_2-\fft32r=0$. In a superconformal theory, these states
correspond to operators that sit in short representations of the
superconformal algebra. The index of an SCFT thus encodes exact
(non-perturbative) information about the operator spectrum of the
underlying theory.\\

The exponents of $p$ and $q$ correspond to operators that commute
with the supercharge used in the definition of the index: the
expression $E-2j_2-\fft32r$ is $Q$-exact for a particular
supercharge $Q$, and the combinations $J^z_1+J^z_2+R/2$ and
$-J^z_1+J^z_1+R/2$ both commute with that $Q$. Therefore $p$ and $q$
refine the Witten index $\mathcal
I^W=\mathrm{Tr}[(-1)^Fe^{-\hat\beta(E-2j_2-\fft32r)}]$ without
ruining the cancelations underlying its topological character. In
fact without refinement with $p$ and $q$, the index $\mathcal I^W$
is often divergent, and thus $p$ and $q$ are necessary as
regulators.\\

\section{Evaluation of the index}\label{sec:EHI}

There are two ways to compute the index of a Lagrangian SCFT. The
Hamiltonian route goes through the so-called Romelsberger
prescription \cite{Romelsberger:2007pre}. The Lagrangian route uses
the supersymmetric localization of the path-integral on $S_b^3\times
S^1_\beta$, where $S^3_b$ is the unit three-sphere with squashing
parameter $b>0$, and $\beta>0$ is the circumference of the circle
\cite{Assel:2014}.

Originally, the indices of Lagrangian SCFTs were computed using the
Romelsberger prescription; see for instance the work of Dolan and
Osborn from 2008 \cite{Dolan:2008}. Later on, supersymmetric
localization caught up, and not only reproduced the correct
expression for the index, but also gave an extra Casimir-type factor
which is of physical significance; see \cite{Nawata:2011} for the
localization computation for the $\mathcal{N}=4$ theory, and the
2014 paper of Assel et. al. \cite{Assel:2014} for the result for the
case with more general matter content (the correct evaluation of the
Casimir-type factor was done later in \cite{Ardehali:2015b} and
\cite{Assel:2015s}).

In the present section we evaluate the index of a general unitary
Lagrangian 4d SCFT (defined as in section~\ref{sec:BG}) using the
Romelsberger prescription (see
\cite{Romelsberger:2007pre,Dolan:2008}). According to the
prescription, one starts with adding up the \emph{single-letter}
indices of various multiplets, and then plethystically exponentiates
the result. To project onto the gauge-singlet sector though, one
should $i)$ make the single-letter indices character-valued, and
$ii)$ integrate the result of the plethystic exponentiation against
the Haar measure of the gauge group.\\

A chiral multiplet $\chi=(\phi_r,\psi_{r-1})$, along with its
CP-conjugate multiplet $\bar{\chi}=(\bar\phi_{-r},\bar\psi_{-r+1})$,
contributes
\begin{equation}
i_\chi(z;p,q)=\sum_{\rho^{\chi}
\in\Delta_\chi}\frac{(pq)^{r_\chi/2}z^{\rho^\chi}-(pq)^{1-r_\chi/2}z^{-\rho^\chi}}{(1-p)(1-q)},\label{eq:iChi}
\end{equation}
to the total single-letter index. Recall that the set $\Delta_\chi$
consists of as many weights $\rho^{\chi}$ as the dimension of the
representation $\mathcal{R}_\chi$. Also, our symbolic notation
$z^{\rho^{\chi}}$ should be understood as
$z_1^{\rho^{\chi}_1}\times\dots\times z_{r_G}^{\rho^{\chi}_{r_G}}$,
where $\rho^{\chi}\equiv (\rho^{\chi}_1,\dots,\rho^{\chi}_{r_G})$,
with $r_G$ the rank of the gauge group.

The first term in the numerator of (\ref{eq:iChi}) is the
contribution $(pq)^{r_\chi/2}$ that $\phi_r$ makes to the index,
multiplied by the character $\sum_{\rho^{\chi}
\in\Delta_\chi}z^{\rho^{\chi}}$ of the representation
$\mathcal{R}_\chi$ of $G$ under which $\chi$ transforms. The second
term in the numerator of (\ref{eq:iChi}) is the contribution
$(pq)^{1-r_\chi/2}$ of $\bar{\psi}_{-r+1}$ to the index, multiplied
by the character of the representation $\bar{\mathcal{R}}_\chi$ of
$G$ under which $\bar{\chi}$ transforms. The denominator of
(\ref{eq:iChi}) comes from summing up the geometric series arising
from adding the contributions of the conformal descendants of
$\phi_r$ and $\bar{\psi}_{-r+1}$ (see section 2 of \cite{Dolan:2008}
for the details).

The plethystic exponential of $i_\chi(z;p,q)$ is given by a product
of several elliptic gamma functions:
\begin{equation}
\mathcal{I}_\chi(z;p,q):=\exp(\sum_{n=1}^\infty\frac{i_\chi(z^n;p^n,q^n)}{n})=\prod_{\rho^{\chi}
\in\Delta_\chi}\Gamma((pq)^{r_\chi/2} z^{\rho^{\chi}}).
\end{equation}
The elliptic gamma function $\Gamma(\ast)$ is a special function
explained in appendix~\ref{sec:app1}.\\

The vector multiplets in the theory contribute to the total
single-letter index as
\begin{equation}
\begin{split}
i_v(z;p,q)&=\left(-\frac{p}{(1-p)(1-q)}-\frac{q}{1-q}+\frac{pq}{(1-p)(1-q)}\right)[r_G+\sum_{\alpha_+}(z^{\alpha_+}+z^{-\alpha_+})]\\
&=\frac{2pq-p-q}{(1-p)(1-q)}[r_G+\sum_{\alpha_+}(z^{\alpha_+}+z^{-\alpha_+})].\label{eq:iVec}
\end{split}
\end{equation}
The $\alpha_+$ are the positive roots of $G$. By $z^{\alpha_+}$ we
mean $z_1^{\alpha_1}\times\dots\times z_{r_G}^{\alpha_{r_G}}$, where
$\alpha_+\equiv(\alpha_1,\dots,\alpha_{r_G})$.

Inside the brackets on the RHS of the first line of (\ref{eq:iVec})
we have the character of the adjoint representation of $G$. Inside
the parentheses on the RHS of the first line of (\ref{eq:iVec}) we
have respectively the contribution of the first gaugino, the second
gaugino, and the gauge field, along with their conformal
descendants; the $p$-descendants of the second gaugino are not taken
into account because the equation of motion relates them to the
$q$-descendants of the first gaugino (see section 2 of
\cite{Dolan:2008} for the details).

The plethystic exponential of $i_v(z;p,q)$ yields a product of
Pochhammer symbols and elliptic gamma functions:
\begin{equation}
\mathcal{I}_v(z;p,q):=\exp(\sum_{n=1}^\infty\frac{i_v(z^n;p^n,q^n)}{n})=
\frac{(p;p)^{r_G}(q;q)^{r_G}}{\prod_{\alpha_+}(1-z^{+\alpha_+})(1-z^{-\alpha_+})\Gamma(
z^{\pm\alpha_+})}.
\end{equation}
The \emph{Pochhammer symbol} $(\ast;\ast)$ is a special function
explained in appendix~\ref{sec:app1}.\\

Multiplying the contribution of the various chiral multiplets
$\prod_\chi\mathcal{I}_\chi(z;p,q)$ by the contribution of the
vector multiplet(s) $\mathcal{I}_v(z;p,q)$ we obtain [alternatively
we could have summed up the character-valued single-letter indices
of various multiplets, and then plethystically exponentiated the
result]
\begin{equation}
\begin{split}
\mathcal{I}(z;p,q)=(p;p)^{r_G}(q;q)^{r_G} \frac{\prod_\chi
\prod_{\rho^{\chi} \in\Delta_\chi}\Gamma((pq)^{r_\chi/2}
z^{\rho^{\chi}})}{\prod_{\alpha_+}(1-z^{+\alpha_+})(1-z^{-\alpha_+})\Gamma(
z^{\pm\alpha_+})}.\label{eq:LagEquivIndex}
\end{split}
\end{equation}
The above index receives contributions from non-gauge-invariant
operators. By integrating it against the Haar measure of the gauge
group
\begin{equation}
\mathrm{d}\mu=\frac{1}{|W|}\mathrm{d}^{r_G}x
\prod_{\alpha_+}(1-z^{+\alpha_+})(1-z^{-\alpha_+}),
\end{equation}
we arrive at the contribution of only the gauge-singlet sector. On
the RHS of the above relation, $|W|$ is the order of the Weyl group
of $G$, and $z_j=e^{2\pi i x_j}$.

The end result is the following elliptic hypergeometric integral
[for comparison with \cite{Rains:2009} note that $\omega_{1\
\mathrm{there}}=ib_{\mathrm{here}}$, $\omega_{2\
\mathrm{there}}=ib_{\mathrm{here}}^{-1}$, and
$v_{\mathrm{there}}=\frac{\beta_{\mathrm{here}}}{2\pi}$]:
\begin{equation}
\boxed{\begin{split}
\mathcal{I}(b,\beta)=\frac{(p;p)^{r_G}(q;q)^{r_G}}{|W|}\int
\mathrm{d}^{r_G}x \frac{\prod_\chi \prod_{\rho^{\chi}
\in\Delta_\chi}\Gamma((pq)^{r_\chi/2}
z^{\rho^{\chi}})}{\prod_{\alpha_+}\Gamma(
z^{\pm\alpha_+})}.\label{eq:LagEquivIndex}
\end{split}}
\end{equation}
The integral is over the unit hypercube $x_j\in [-1/2,1/2]$ in the
Cartan subalgebra (or alternatively, over the maximal torus of $G$
in the space of $z_j$).\\

Since the expression in Eq.~(\ref{eq:LagEquivIndex}) might seem a
bit complicated, let us specialize it to a very simple case: the
SU($2$) SQCD with three flavors. The gauge group SU($2$) has rank
$r_G=1$. The Weyl group of SU($N$) is the permutation group of $N$
elements, so it has order $N!$, which for SU($2$) becomes $2$. We
have three chiral quark multiplets with
$\rho^{\chi_1}_1,\rho^{\chi_2}_1,\rho^{\chi_3}_1=\pm 1$, and three
chiral anti-quark multiplets with
$\rho^{\chi_{4}}_1,\rho^{\chi_{5}}_1,\rho^{\chi_{6}}_1=\mp1$ (each
of the chiral multiplets has two weights ($\pm 1$), because they sit
in two-dimensional representations of the gauge group). All the
chiral multiplets have R-charge $r_\chi=1/3$. Finally, the group
SU($2$) has two roots, corresponding to the raising and lowering
operators of the 3d angular momentum, and the positive root (the
raising operator) has $\alpha_+=2$. All in all, we get for this
simple example
\begin{equation}
\begin{split}
\mathcal{I}_{N_c=2,N_f=3}(b,\beta)=\frac{(p;p)(q;q)}{2}\int_{-1/2}^{1/2}
\mathrm{d}x \frac{\Gamma^6((pq)^{1/6} z^{\pm 1} )}{\Gamma( z^{\pm
2})}.\label{eq:LagIndexSU2SQCD}
\end{split}
\end{equation}

Many explicit expressions for the index $\mathcal{I}(b,\beta)$ of
specific 4d SCFTs can be found in
\cite{Spirido:2009,Spirido:2011or,Ardehali:2015c}. A few specific
examples will be spelled out in the next chapter as well.

\subsection*{Miscellaneous remarks}

To further clarify the notation we are using for the roots and
weights, we add that with our notation the three-dimensional
representation of SU($3$) has weights
$(\rho_1,\rho_2)=(1,0),(0,1),(-1,-1)$, and the positive roots of
SU($3$) are $\alpha_+=(1,-1),(2,1),(1,2)$.\\

If a Lagrangian 4d SCFT has emergent accidental symmetries mixing
with its ultraviolet U($1$)$_R$ to give the superconformal
U($1$)$_R$ in the infrared, the Romelsberger prescription can not be
applied to it. For such SCFTs, the EHI in (\ref{eq:LagEquivIndex})
can be interpreted as arising from path-integration of the UV gauge
theory using the ultraviolet U($1$)$_R$, but the EHI would not
coincide with the superconformal index of the IR SCFT. In this
dissertation we do not discuss the superconformal index of such
SCFTs.\\

The EHIs studied by Rains in \cite{Rains:2009} correspond to the
Sp($2N$) and SU($N$) supersymmetric quantum chromodynamics theories
\cite{Dolan:2008}. (Note that for simplicity we are focusing on the
special case where all the $u_r$ and $v_r$ in \cite{Rains:2009} are
equal.)

For a mathematically oriented introduction to the EHIs studied in
\cite{Rains:2009} see \cite{Rains:talk}.


\chapter{High-temperature asymptotics of the index of non-chiral
theories}\label{sec:finiteNchap}

We now focus on \emph{non-chiral} SCFTs: those in which nonzero
$\rho^\chi$ come in pairs with opposite signs. With this
restriction, the hyperbolic limit of the EHI shown in
(\ref{eq:LagEquivIndex}) can be analyzed completely reliably, as
described below.

\section{General analysis}\label{sec:general}

\subsection{Step 1: simplifying the EHI to an ordinary integral}

The high-temperature asymptotics of the index
(\ref{eq:LagEquivIndex}) of a non-chiral SCFT is found as follows.
Using (\ref{eq:PochAsy}), the Pochhammer symbols in the prefactor of
(\ref{eq:LagEquivIndex}) can be immediately replaced with their
asymptotic expressions. We have
\begin{equation}
(p;p)^{r_G}(q;q)^{r_G}\simeq e^{-\pi^2(b+b^{-1})r_G/6\beta}\times
\left(\frac{2\pi}{\beta}\right)^{r_G}\times
e^{\beta(b+b^{-1})r_G/24},\quad\quad(\text{as
$\beta\to0$})\label{eq:prePochAsyLead}
\end{equation}
with the symbol $\simeq$ as defined in appendix~\ref{sec:app2}.

The asymptotics of the integrand of (\ref{eq:LagEquivIndex}) can be
obtained from the estimates in (\ref{eq:GammaAsyPosNegZ2}). With the
aid of (\ref{eq:prePochAsyLead}) and (\ref{eq:GammaAsyPosNegZ2}) we
find the $\beta\to0$ asymptotics of $\mathcal{I}$
as\footnote{Compared to the expression in (3.9) of
\cite{Ardehali:2015c}, the RHS of (\ref{eq:LagIndexSimp1}) lacks a
phase $i\Theta$ in the exponent because (as explained in
\cite{Ardehali:2015c}) in non-chiral theories $\Theta=0$. Also, the
RHS of (\ref{eq:LagIndexSimp1}) has the extra factors $1/|W|$,
$W_0(b)$, $e^{\beta E_{\mathrm{susy}}(b)}$, and
$W(\mathbf{x};b,\beta)$ which were absent in \cite{Ardehali:2015c};
these arise here because in the analysis below we are using
estimates that are stronger than the estimates used in
\cite{Ardehali:2015c}.}
\begin{equation}
\begin{split}
\mathcal I(b,\beta)\simeq \frac{1}{|W|}
\left(\frac{2\pi}{\beta}\right)^{r_G}e^{-\mathcal{E}^{DK}_0(b,\beta)}W_0(b)e^{\beta
E_{\mathrm{susy}}(b)} \int_{\mathfrak{h}_{cl}} \mathrm{d}^{r_G}x\
e^{-V^{\mathrm{eff}}(\mathbf{x};b,\beta)}W(\mathbf{x};b,\beta),\label{eq:LagIndexSimp1}
\end{split}
\end{equation}
with $\mathfrak{h}_{cl}$---which in the path-integral picture can be
interpreted \cite{Ardehali:2015c} as the ``classical'' moduli-space
of the holonomies around $S^1_\beta$---denoting the unit hypercube
$x_i\in [-1/2,1/2]$, and with
\begin{equation}
\begin{split}
\mathcal{E}^{DK}_0(b,\beta)=\frac{\pi^2}{3\beta}(\frac{b+b^{-1}}{2})\mathrm{Tr}R,\label{eq:EdkEquiv}
\end{split}
\end{equation}
\begin{equation}
\begin{split}
V^{\mathrm{eff}}(\mathbf{x};b,\beta)=\frac{4\pi^2}{\beta}(\frac{b+b^{-1}}{2})L_h(\mathbf{x}),\label{eq:VeffEquiv1}
\end{split}
\end{equation}
\begin{equation}
\begin{split}
E_{\mathrm{susy}}(b)=\frac{1}{6}(\frac{b+b^{-1}}{2})^3\mathrm{Tr}R^3-(\frac{b+b^{-1}}{2})(\frac{b^2+b^{-2}}{24})\mathrm{Tr}R.\label{eq:EcBequiv}
\end{split}
\end{equation}
The `t~Hooft anomalies in the above relations are given by
\begin{equation}
\begin{split}
\mathrm{Tr}R&:=\mathrm{dim}G+\sum_\chi(r_\chi-1)\mathrm{dim}\mathcal{R}_\chi=-16(c-a),\\
\mathrm{Tr}R^3&:=\mathrm{dim}G+\sum_\chi(r_\chi-1)^3\mathrm{dim}\mathcal{R}_\chi=\frac{16}{9}(5a-3c).
\end{split}
\end{equation}
We have also defined $W_0(b)$, and the \emph{real} functions
$L_h(\mathbf{x})$ and $W(\mathbf{x};b,\beta)$ via
\begin{equation}
\begin{split}
L_h(\mathbf{x}):= \frac{1}{2}\sum_{\chi}(1-r_\chi)\sum_{\rho^{\chi}
\in\Delta_\chi}\vartheta(\langle\rho^{\chi}\cdot
\mathbf{x}\rangle)-\sum_{\alpha_+}\vartheta(\langle\alpha_+\cdot
\mathbf{x}\rangle),\label{eq:LhDef}
\end{split}
\end{equation}
\begin{equation}
\begin{split}
W_0(b)=\prod_\chi\prod_{\rho^\chi=0}\Gamma_h(r_\chi\omega),\label{eq:W0Def}
\end{split}
\end{equation}
\begin{equation}
\begin{split}
W(\mathbf{x};b,\beta)=\prod_\chi
\prod_{\rho^\chi_+}\frac{\psi_b(-\frac{2\pi
i}{\beta}\{\langle\rho^\chi_+\cdot\mathbf{x}\rangle\}+(r_\chi-1)\frac{b+b^{-1}}{2})}{\psi_b(-\frac{2\pi
i}{\beta}\{\langle\rho^\chi_+\cdot\mathbf{x}\rangle\}-(r_\chi-1)\frac{b+b^{-1}}{2})}
\prod_{\alpha_+}\frac{\psi_b(-\frac{2\pi
i}{\beta}\{\langle\alpha_+\cdot\mathbf{x}\rangle\}+\frac{b+b^{-1}}{2})}{\psi_b(-\frac{2\pi
i}{\beta}\{\langle\alpha_+\cdot\mathbf{x}\rangle\}-\frac{b+b^{-1}}{2})}.\label{eq:W+Def}
\end{split}
\end{equation}
The function $\vartheta(x)$ in (\ref{eq:LhDef}) is defined as
$\vartheta(x):=\{x\}(1-\{x\})$ (with the fractional part function
defined as $\{x\}:=x-\lfloor x\rfloor$). In (\ref{eq:W0Def}), the
second product is over the zero weights of $\mathcal{R}_\chi$ (the
adjoint representation, for instance, has $r_G$ such weights), and
$\omega$ is defined as $\omega:=i(b+b^{-1})/2$. The $\rho^\chi_+$ in
(\ref{eq:W+Def}) denote the positive weights of $\mathcal{R}_\chi$.
The \emph{non-compact quantum dilogarithm} $\psi_b(\ast)$ and the
\emph{hyperbolic gamma} $\Gamma_h(\ast)$ are special functions
explained in appendix~\ref{sec:app1}.

That $L_h(\mathbf{x})$ is real should be obvious from the definition
of $\vartheta(x)$; that $W(\mathbf{x};b,\beta)$ is real follows from
(\ref{eq:hyperbolicGammaConj}) and (\ref{eq:hyperbolicGammaPsi}).\\

Note that in (\ref{eq:LagIndexSimp1}) we are claiming that \emph{the
matrix-integral is approximated well with the integral of its
approximate integrand}. This is true because the estimates we have
used inside the integrand are $i)$ uniform, and $ii)$ accurate up to
exponentially small corrections of the type $e^{-1/\beta}$; these
two strong conditions---on the integrand estimates---were not
satisfied in the treatment of \cite{Ardehali:2015c}.

Now, from (\ref{eq:hyperbolicGammaConj}) it follows that $W_0(b)$ is
a real number; it is moreover nonzero and finite, as we are assuming
$r_\chi\in]0,2[$ (the zeros and poles of the hyperbolic gamma
function are described in appendix~\ref{sec:app1}). We would thus
make an $O(\beta^0)$ error in the asymptotics of
$\ln\mathcal{I}(b,\beta)$ by setting $W_0(b)$, along with $|W|$ and
$e^{\beta E_{\mathrm{susy}}(b)}$, to unity. In other words,
\begin{equation}
\begin{split}
\mathcal I(b,\beta)\approx
\left(\frac{2\pi}{\beta}\right)^{r_G}e^{-\mathcal{E}^{DK}_0(b,\beta)}
\int_{\mathfrak{h}_{cl}} \mathrm{d}^{r_G}x\
e^{-V^{\mathrm{eff}}(\mathbf{x};b,\beta)}W(\mathbf{x};b,\beta),\label{eq:LagIndexSimpO1}
\end{split}
\end{equation}
with an $O(\beta^0)$ error upon taking the logarithm of the two
sides.

We are hence left with the asymptotic analysis of the integral
$\int_{\mathfrak{h}_{cl}} e^{-V}W$. From here, standard methods of
asymptotic analysis can be employed.

\subsection{Step 2: asymptotic analysis of the simplified integral}

Before continuing our asymptotic analysis further, we note that the
star of our show, the real function $L_h$ which determines the
effective potential\footnote{Somewhat surprisingly, $L_h$ also
appears in the $n\to1$ limit of the zero-point energy associated to
nonzero spatial holonomies on $S^1\times S^3/\mathbb{Z}_n$; c.f.
Eq.~(29) of the arXiv preprint of \cite{Benini:2011et} (with $\nu,a$
in there set to zero). It might be possible to clarify this
coincidence by analytically continuing the results of
\cite{Benini:2011et} (see also \cite{Razamat:2013bw,Nieri:2015}) to
non-integer $n$, and then using modular properties of the
generalized elliptic gamma functions employed in that work.}
$V^{\mathrm{eff}}(\mathbf{x};b,\beta)$, is \emph{piecewise linear};
the quadratic terms in it cancel because of the ABJ
U($1$)$_R$-gauge-gauge anomaly cancelation:
\begin{equation}
\frac{\partial^2 L_h(\mathbf{x})}{\partial x_i\partial
x_j}=\sum_\chi(r_\chi-1)\sum_{\rho^\chi\in\Delta_\chi}\rho^\chi_i\rho^\chi_j+\sum_{\alpha}\alpha_{i}\alpha_{j}=0.
\end{equation}
Also, $L_h$ is continuous, is even under $\mathbf{x}\to
-\mathbf{x}$, and vanishes at $\mathbf{x}=0$; these properties
follow from the properties of the function $\vartheta(x)$ defined
above. We refer to $L_h(\mathbf{x})$ as the \emph{Rains function} of
the SCFT. This function has been analyzed by Rains \cite{Rains:2009}
in the special cases of the elliptic hypergeometric integrals
associated to SU($N$) and Sp($N$) SQCD theories.\\

Writing $V^{\mathrm{eff}}$ in terms of the Rains function $L_h$,
(\ref{eq:LagIndexSimpO1}) simplifies to
\begin{equation}
\begin{split}
\mathcal{I}(b,\beta)\approx \left(\frac{2\pi}{\beta}\right)^{r_G}
e^{-\mathcal{E}^{DK}_0(b,\beta)}\int_{\mathfrak{h}_{cl}}
\mathrm{d}^{r_G}x\
e^{-\frac{4\pi^2}{\beta}(\frac{b+b^{-1}}{2})L_h(\mathbf{x})}W(\mathbf{x};b,\beta).\label{eq:LagIndexSimp1noTheta}
\end{split}
\end{equation}

It will be useful for us to know that $W(\mathbf{x};b,\beta)$ is a
positive semi-definite function of $\mathbf{x}$; this follows from
(\ref{eq:hyperbolicGammaConj}) and (\ref{eq:hyperbolicGammaPsi}).

To analyze the integral in (\ref{eq:LagIndexSimp1noTheta}), first
note that the integrand is not smooth over $\mathfrak{h}_{cl}$. We
hence break $\mathfrak{h}_{cl}$ into sets on which $L_h$ is linear.
These sets can be obtained as follows. Define
\begin{equation}
\begin{split}
\mathcal{S}_g:=\bigcup_{\alpha_+}\{\mathbf{x}\in
\mathfrak{h}_{cl}|\langle\alpha_+\cdot
\mathbf{x}\rangle\in&\mathbb{Z}\},\quad\quad\mathcal{S}_\chi:=\bigcup_{\rho^\chi_+}\{\mathbf{x}\in\mathfrak{h}_{cl}|\langle\rho^{\chi}_+\cdot
\mathbf{x}\rangle\in\mathbb{Z}\},\\
&\mathcal{S}:=\bigcup_{\chi}\mathcal{S}_\chi\cup
\mathcal{S}_g.\label{eq:SSsDef}
\end{split}
\end{equation}
It should be clear that everywhere in $\mathfrak{h}_{cl}$, except on
$\mathcal{S}$, the function $L_h$ is guaranteed to be linear---and
therefore smooth.

The set $\mathcal{S}$ consists of a union of codimension one affine
hyperplanes inside the space of the $x_i$. These hyperplanes chop
$\mathfrak{h}_{cl}$ into (finitely many, convex) polytopes
$\mathcal{P}_n$. The integral in (\ref{eq:LagIndexSimp1noTheta})
then decomposes to
\begin{equation}
\begin{split}
\mathcal{I}(b,\beta)\approx e^{-\mathcal{E}^{DK}_0(b,\beta)}\sum_n
\left(\frac{2\pi}{\beta}\right)^{r_G} \int_{\mathcal{P}_n}
\mathrm{d}^{r_G}x\
e^{-\frac{4\pi^2}{\beta}(\frac{b+b^{-1}}{2})L_h(\mathbf{x})}W(\mathbf{x};b,\beta).\label{eq:LagIndexSimp2noTheta}
\end{split}
\end{equation}

Let $\mathcal{S}^{(\beta)}_g$ denote the set of all points in
$\mathfrak{h}_{cl}$ that are at a distance less than $N_0\beta$ from
$\mathcal{S}_g$, with some fixed $N_0>0$. We divide $\mathcal{P}_n$
into $i)$ $\mathcal{P}_n\cap\mathcal{S}^{(\beta)}_g$, and $ii)$ the
rest of $\mathcal{P}_n$, which we denote by $\mathcal{P}'_n$. Now,
by taking $N_0$ to be large enough, we can push $\mathcal{P}'_n$
away from the zeros of $\psi_b$, and thus make
$w_i<W(\mathbf{x};b,\beta)<w_s$ over $\mathcal{P}'_n$ (with some
$0<w_i$ and some $w_s<\infty$). Therefore the contribution that the
$n$th summand in (\ref{eq:LagIndexSimp2noTheta}) receives from
$\mathcal{P}'_n$ is well approximated (with an $O(\beta^0)$ error
upon taking the logs) by
\begin{equation}
\begin{split}
J_n:=\left(\frac{2\pi}{\beta}\right)^{r_G}\int_{\mathcal{P}'_n}
\mathrm{d}^{r_G}x\
e^{-\frac{4\pi^2}{\beta}(\frac{b+b^{-1}}{2})L_h(\mathbf{x})}.\label{eq:ithIntegralSimplPp}
\end{split}
\end{equation}

Let's further replace $\mathcal{P}'_n$ in
(\ref{eq:ithIntegralSimplPp}) with $\mathcal{P}_n$; we will shortly
see that this replacement introduces a negligible error. We would
hence like to estimate
\begin{equation}
\begin{split}
I_n:=\left(\frac{2\pi}{\beta}\right)^{r_G}\int_{\mathcal{P}_n}
\mathrm{d}^{r_G}x\
e^{-\frac{4\pi^2}{\beta}(\frac{b+b^{-1}}{2})L_h(\mathbf{x})}.\label{eq:ithIntegralSimplP}
\end{split}
\end{equation}

Since $L_h$ is linear on each $\mathcal{P}_n$, its minimum over
$\mathcal{P}_n$ is guaranteed to be realized on
$\partial\mathcal{P}_n$. Let us assume that this minimum occurs on
the $k$th $j$-face of $\mathcal{P}_n$, which we denote by
$j_n$-$\mathcal{F}^k_n$. We denote the value of $L_h$ on this
$j$-face by $L_{h\ \mathrm{min}}^n$. Equipped with this notation, we
can write (\ref{eq:ithIntegralSimplP}) as
\begin{equation}
\begin{split}
I_n= \left(\frac{2\pi}{\beta}\right)^{r_G}
e^{-\frac{4\pi^2}{\beta}(\frac{b+b^{-1}}{2})L_{h\
\mathrm{min}}^n}\int_{\mathcal{P}_n} \mathrm{d}^{r_G}x\
e^{-\frac{4\pi^2}{\beta}(\frac{b+b^{-1}}{2})\Delta
L^n_h(\mathbf{x})},\label{eq:LagIndexSimp3noTheta}
\end{split}
\end{equation}
where $\Delta L^n_h(\mathbf{x}):=L_h(\mathbf{x})-L_{h\
\mathrm{min}}^n$ is a linear function on $\mathcal{P}_n$. Note that
$\Delta L^n_h(\mathbf{x})$ vanishes on $j_n$-$\mathcal{F}^k_n$, and
it increases as we go away from $j_n$-$\mathcal{F}^k_n$ and into the
interior of $\mathcal{P}_n$. [The last sentence, as well as the rest
of the discussion leading to (\ref{eq:LagIndexSimp6noTheta}), would
receive a trivial modification if $j_n=r_G$ (corresponding to
constant $L_h$ over $\mathcal{P}_n$).] Therefore as $\beta\to 0$,
the integral in (\ref{eq:LagIndexSimp3noTheta}) localizes around
$j_n$-$\mathcal{F}^k_n$.

To further simplify (\ref{eq:LagIndexSimp3noTheta}), we now adopt a
set of new coordinates---affinely related to $x_i$ and with unit
Jacobian---that are convenient on $\mathcal{P}_n$. We pick a point
on $j_n$-$\mathcal{F}^k_n$ as the new origin, and parameterize
$j_n$-$\mathcal{F}^k_n$ with $\bar{x}_1,...,\bar{x}_{j_n}$. We take
$x_{\mathrm{in}}$ to parameterize a direction perpendicular to all
the $\bar{x}$s, and to increase as we go away from
$j_n$-$\mathcal{F}^k_n$ and into the interior of $\mathcal{P}_n$.
Finally, we pick $\tilde{x}_1,...,\tilde{x}_{r_G-j_n-1}$ to
parameterize the perpendicular directions to $x_{\mathrm{in}}$ and
the $\bar{x}$s. Note that, because $\Delta L^n_h$ is linear on
$\mathcal{P}_n$, it does not depend on the $\bar{x}$s; they
parameterize its flat directions. By re-scaling
$\bar{x},x_{\mathrm{in}},\tilde{x}\mapsto
\frac{\beta}{2\pi}\bar{x},\frac{\beta}{2\pi}x_{\mathrm{in}},\frac{\beta}{2\pi}\tilde{x}$,
we can absorb the $(\frac{2\pi}{\beta})^{r_G}$ factor in
(\ref{eq:LagIndexSimp3noTheta}) into the integral, and write the
result as
\begin{equation}
\begin{split}
I_n=\int_{\frac{2\pi}{\beta}\mathcal{P}_n} \mathrm{d}^{j_n}\bar{x}\
\mathrm{d}x_{\mathrm{in}}\ \mathrm{d}^{r_G-j_n-1}\tilde{x}\
e^{-2\pi(\frac{b+b^{-1}}{2})\Delta
L^n_h(x_{\mathrm{in}},\mathbf{\tilde{x}})}.\label{eq:ithIntegralSimpl1}
\end{split}
\end{equation}
To eliminate $\beta$ from the exponent, we have used the fact that
$\Delta L^n_h$ depends homogenously on the new coordinates. We are
also denoting the re-scaled polytope schematically by
$\frac{2\pi}{\beta}\mathcal{P}_n$.

Instead of integrating over all of $\frac{2\pi}{\beta}\mathcal{P}_n$
though, we can restrict to $x_{\mathrm{in}}<\epsilon/\beta$ with
some small $\epsilon>0$. The reason is that the integrand of
(\ref{eq:ithIntegralSimpl1}) is exponentially suppressed (as
$\beta\to 0$) for $x_{\mathrm{in}}>\epsilon/\beta$. We take
$\epsilon>0$ to be small enough such that a hyperplane at
$x_{\mathrm{in}}=\epsilon/\beta$, and parallel to
$j_n$-$\mathcal{F}^k_n$, cuts off a prismatoid
$P^n_{\epsilon/\beta}$ from $\frac{2\pi}{\beta}\mathcal{P}_n$. After
restricting the integral in (\ref{eq:ithIntegralSimpl1}) to
$P^n_{\epsilon/\beta}$, the integration over the $\bar{x}$s is easy
to perform. The only potential difficulty is that the range of the
$\bar{x}$ coordinates may depend on $x_{\mathrm{in}}$ and the
$\tilde{x}s$. But since we are dealing with a prismatoid, the
dependence is linear, and by the time the range is modified
significantly (compared to its $O(1/\beta)$ size on the re-scaled
$j$-face $\frac{2\pi}{\beta}(j_n$-$\mathcal{F}^k_n)$), the integrand
is exponentially suppressed. Therefore we can neglect the dependence
of the range of the $\bar{x}$s on the other coordinates in
(\ref{eq:ithIntegralSimpl1}). The integral then simplifies to
\begin{equation}
\begin{split}
I_n\approx\left(\frac{2\pi}{\beta}\right)^{j_n}\
\mathrm{vol}(j_n\text{-}\mathcal{F}^k_n)\int_{\hat{P}^n_{\epsilon/\beta}}
\mathrm{d}x_{\mathrm{in}}\ \mathrm{d}^{r_G-j_n-1}\tilde{x}\
e^{-2\pi(\frac{b+b^{-1}}{2})\Delta
L^n_h(x_{\mathrm{in}},\mathbf{\tilde{x}})},\label{eq:ithIntegralSimpl2}
\end{split}
\end{equation}
where $\hat{P}^n_{\epsilon/\beta}$ is the pyramid obtained by
restricting $P^n_{\epsilon/\beta}$ to
$\bar{x}_1=...=\bar{x}_{j_n}=0$. The logarithms of the two sides of
(\ref{eq:ithIntegralSimpl2}) differ by $O(\beta)$, with the error
mainly arising from our neglect of the possible dependence of the
range of the $\bar{x}$ coordinates in (\ref{eq:ithIntegralSimpl1})
on $x_{\mathrm{in}}$ and the $\tilde{x}s$. (Recall that the other
error, arising from restricting the integral in
(\ref{eq:ithIntegralSimpl1}) to $P^n_{\epsilon/\beta}$, is
exponentially small.)

We now take $\epsilon\to\infty$ in (\ref{eq:ithIntegralSimpl2}).
This introduces an exponentially small error, as the integrand is
exponentially suppressed (as $\beta\to 0$) for
$x_{\mathrm{in}}>\epsilon/\beta$. The resulting integral is strictly
positive, because it is the integral of a strictly positive
function. We denote by $A_n$ the result of the integral multiplied
by $\mathrm{vol}(j_n$-$\mathcal{F}^k_n)$. Then $I_n$ can be
approximated as
\begin{equation}
\begin{split}
I_n\approx e^{-\frac{4\pi^2}{\beta}(\frac{b+b^{-1}}{2})L_{h\
\mathrm{min}}^n}\left(\frac{2\pi}{\beta}\right)^{j_n}
A_n.\label{eq:LagIndexSimp4noTheta}
\end{split}
\end{equation}

We are now in a position to argue $J_n\approx I_n$. If we had
integrated over $\mathcal{P}'_n$, then we would end up with an
expression similar to (\ref{eq:LagIndexSimp4noTheta}), in which
$L_{h\ \mathrm{min}}^n$ would be replaced with the minimum of $L_h$
over $\mathcal{P}'_n$; but since $L_h$ is piecewise linear, the
difference between the new minimum and $L_{h\ \mathrm{min}}^n$ would
be $O(\beta)$, which translates to an $O(\beta^0)$ multiplicative
difference between $J_n$ and $I_n$. Other sources of difference
between $J_n$ and $I_n$ similarly introduce negligible error; more
precisely, we have $\ln I_n=\ln J_n+O(\beta^0)$.

The dominant contribution to $\mathcal{I}(b,\beta)$ comes, of
course, from the terms/polytopes whose $L_{h\ \mathrm{min}}^n$ is
smallest. If these terms are labeled by $n=n_\ast^1,n_\ast^2,...$,
we can introduce $\mathfrak{h}_{qu}$ and
$\mathrm{dim}\mathfrak{h}_{qu}$ via
\begin{equation}
\mathfrak{h}_{qu}:=\bigcup_{n_\ast}
j_{n_\ast}\text{-}\mathcal{F}^k_{n_\ast},\quad
\mathrm{dim}\mathfrak{h}_{qu}:=\mathrm{max}(j_{n_\ast}).
\end{equation}
Put colloquially, if $\mathfrak{h}_{qu}$ has multiple connected
components, by $\mathrm{dim}\mathfrak{h}_{qu}$ we mean the dimension
of the component(s) with greatest dimension, while if a connected
component consists of several intersecting flat elements inside
$\mathfrak{h}_{cl}$, by its dimension we mean the dimension of the
flat element(s) of maximal dimension.

Our final estimate for the contribution to $\mathcal{I}(b,\beta)$
from $\cup_n\mathcal{P}'_n$ is thus
\begin{equation}
\begin{split}
Be^{-\mathcal{E}^{DK}_0(b,\beta)-\frac{4\pi^2}{\beta}(\frac{b+b^{-1}}{2})L_{h\
\mathrm{min}}}\left(\frac{2\pi}{\beta}\right)^{\mathrm{dim}\mathfrak{h}_{qu}},
\label{eq:LagIndexSimp5noTheta}
\end{split}
\end{equation}
where $L_{h\ \mathrm{min}}:=L_{h\ \mathrm{min}}^{n_\ast}$, and $B$
is some positive real number.\\

We are left with determining the contribution to
$\mathcal{I}(b,\beta)$ coming from $\mathcal{S}^{(\beta)}_g$. Over
$\mathcal{P}_n\cap\mathcal{S}^{(\beta)}_g$, the simple estimate
$W(\mathbf{x};b,\beta)= O(1)$ (which follows from the fact that
$W(\mathbf{x};b,\beta)$ is uniformly bounded on
$\mathcal{S}^{(\beta)}_g$) suffices for our purposes; we thus learn
that the contribution that the integral
(\ref{eq:LagIndexSimp2noTheta}) receives from
$\mathcal{P}_n\cap\mathcal{S}^{(\beta)}_g$ is not only positive, but
also
\begin{equation}
\begin{split}
O\left(\int_{\frac{2\pi}{\beta}(\mathcal{P}_n\cap\mathcal{S}^{(\beta)}_g)}
\mathrm{d}^{j_n}\bar{x}\ \mathrm{d}x_{\mathrm{in}}\
\mathrm{d}^{r_G-j_n-1}\tilde{x}\ e^{-2\pi(\frac{b+b^{-1}}{2})\Delta
L^n_h(x_{\mathrm{in}},\mathbf{\tilde{x}})}\right).\label{eq:ithIntegralSimplSg}
\end{split}
\end{equation}

Now, the argument of the $O$ above is nothing but the difference
between $I_n$ and $J_n$, which we already argued to be negligible.
Thus the contribution to $\mathcal{I}(b,\beta)$ coming from
$\mathcal{S}^{(\beta)}_g$ is negligible.\\

Using the explicit expression (\ref{eq:EdkEquiv}) for
$\mathcal{E}^{DK}_0(b,\beta)$, and noting that
(\ref{eq:LagIndexSimp5noTheta}) is an accurate estimate for
$\mathcal{I}(b,\beta)$ up to a multiplicative factor of order
$\beta^0$, we arrive at our main result:
\begin{equation}
\boxed{\begin{split} \ln \mathcal{I}(b,\beta)=
-\frac{\pi^2}{3\beta}(\frac{b+b^{-1}}{2})(\mathrm{Tr}R+12L_{h\
\mathrm{min}})
+\mathrm{dim}\mathfrak{h}_{qu}\ln(\frac{2\pi}{\beta})+O(\beta^0).
\label{eq:LagIndexSimp6noTheta}
\end{split}}
\end{equation}

\subsection{Connection with the $S^3$ partition function}\label{sec:S3}

In this subsection we comment on the connection between the
asymptotics of the index of a 4d SCFT, and the divergence of the
$S^3$ partition function $Z_{S^3}$ of the dimensionally reduced
daughter of the 4d theory.

We will show below that the degree of divergence of $Z_{S^3}$ (as
the cut-off of the matrix-integral computing it is taken to
infinity) is determined by the behavior of the Rains function $L_h$
near the origin of $\mathfrak{h}_{cl}$. In particular
\begin{itemize}
\item if the origin is an isolated local minimum of $L_h$, then
$Z_{S^3}$ is finite;
\item if the origin is part of an extended locus where $L_h$ is
locally minimized, then $Z_{S^3}$ is power-law divergent;
\item if the origin is not a local minimum of $L_h$, then $Z_{S^3}$
is exponentially divergent.
\end{itemize}

Note that it is the \emph{local} behavior of $L_h$ near the origin
that determines the degree of divergence of $Z_{S^3}$. On the other
hand, according to (\ref{eq:LagIndexSimp6noTheta}), the asymptotics
of the 4d index is determined by the \emph{global} properties of
$L_h$. Therefore, at least until theorems relating the local and
global properties of $L_h$ are established, the asymptotics of the
4d index is not as tightly connected to the divergence of $Z_{S^3}$
as one may have wished.

For instance, \emph{we can not say} (in absence of theorems of the
kind discussed in the previous paragraph) that `the
Di~Pietro-Komargodski asymptotics applies to the index if $Z_{S^3}$
is finite'; it may happen that in a (non-chiral unitary Lagrangian)
4d SCFT (with $r_\chi\in]0,2[$) \emph{the origin is an isolated
local, but not global, minimum} of $L_h$; that the origin is an
isolated local minimum would imply that $Z_{S^3}$ is finite; that
$L_h$ is minimized somewhere else would imply---according to
(\ref{eq:LagIndexSimp6noTheta})---that the Di~Pietro-Komargodski
formula receives a modification.

However, \emph{we can say with certainty} that (in a non-chiral
unitary 4d Lagrangian SCFT with $r_\chi\in]0,2[$) `if $Z_{S^3}$ is
exponentially divergent, then the Di~Pietro-Komargodski formula
receives a modification'; this is simply because if $L_h$ is not
locally minimized at the origin, it is certainly not globally
minimized
there either.\\

We now demonstrate the three propositions itemized above.

The starting point is the observation that the function
$\vartheta(x)$ featuring in $L_h$ simplifies if its argument is
``small enough'':
\begin{equation}
\vartheta(x)= |x|-x^2 \quad\quad\text{for
$x\in[-1,1]$}.\label{eq:varthetaSimp}
\end{equation}
Using the above simplification in the expression (\ref{eq:LhDef})
for $L_h$, we learn that for small enough $|\mathbf{x}|$ the Rains
function simplifies to the following \emph{homogenous}
function\footnote{Interestingly, on a discrete subset of its domain
(corresponding to the cocharacter lattice of the gauge group $G$),
the function $\tilde{L}_{S^3}$ coincides (up to normalization) with
the $S^2\times S^1$ Casimir energy $\epsilon_0$
\cite{Imamura:2011mp} associated to monopole sectors of the 3d
$\mathcal{N}=2$ theory obtained from dimensional reduction of the 4d
$\mathcal{N}=1$ gauge theory. In the context of 3d $\mathcal{N}=4$
theories, a different connection between $\tilde{L}_{S^3}$ and 3d
monopoles was discussed in \cite{Kapustin:2010f}.}:
\begin{equation}
\begin{split}
\tilde{L}_{S^3}(\mathbf{x})=\frac{1}{2}\sum_{\chi}(1-r_\chi)\sum_{\rho^{\chi}
\in\Delta_\chi}|\langle\rho^{\chi}\cdot
\mathbf{x}\rangle|-\sum_{\alpha_+}|\langle\alpha_+\cdot
\mathbf{x}\rangle|.\label{eq:hypURains}
\end{split}
\end{equation}
Note that there is no quadratic term in $\tilde{L}_{S^3}$, thanks to
the cancelation of the U($1$)$_R$-gauge-gauge anomaly.

Next, we consider (recall $\omega:=i(b+b^{-1})/2$)
\begin{equation}
Z_{S^3}(b;\Lambda):=\frac{1}{|W|}\int_{\Lambda} \mathrm{d}^{r_G}x\
\frac{\prod_\chi \prod_{\rho^{\chi}
\in\Delta_\chi}\Gamma_h(r_\chi\omega +\langle\rho^{\chi}\cdot
\mathbf{x}\rangle)}{\prod_{\alpha_+}\Gamma_h(\pm\langle\alpha_+\cdot
\mathbf{x}\rangle)}, \label{eq:LagZ3d}
\end{equation}
which is the matrix-integral computing the squashed-three-sphere
partition function of the dimensionally reduced daughter (c.f.
Eq.~(5.23) of \cite{Aharony:2013a}), assuming the same R-charge
assignments as those directly descending from the parent 4d theory.
We are keeping the cut-off $\Lambda$ explicit, emphasizing that the
integration is over the hypercube $|x_i|<\Lambda$.

To study the convergence/divergence of $Z_{S^3}(b;\Lambda)$ as
$\Lambda$ is taken to infinity, we use the estimate
(\ref{eq:hyperbolicGammaAsy}) for the hyperbolic gamma functions in
the integrand of (\ref{eq:LagZ3d}). We find that the integrand of
$Z_{S^3}(b;\Lambda)$ can be estimated, as $|\mathbf{x}|\to\infty$,
by
\begin{equation}
\frac{\prod_\chi \prod_{\rho^{\chi}
\in\Delta_\chi}\Gamma_h(r_\chi\omega +\langle\rho^{\chi}\cdot
\mathbf{x}\rangle)}{\prod_{\alpha_+}\Gamma_h(\pm\langle\alpha_+\cdot
\mathbf{x}\rangle)}\approx e^{-2\pi
(\frac{b+b^{-1}}{2})\tilde{L}_{S^3}(\mathbf{x})},
\label{eq:LagZ3dIntegrandAsy}
\end{equation}
with $\tilde{L}_{S^3}$ the homogeneous function defined above.

Note that whether the integrand of $Z_{S^3}$ decays or grows at
large $|\mathbf{x}|$, is determined by the behavior
$\tilde{L}_{S^3}(\mathbf{x})$, and does not depend on $b$ (recall
that we take $b>0$).

Here comes the crucial point: since $\tilde{L}_{S^3}(\mathbf{x})$ is
homogenous, its sign at large $|\mathbf{x}|$ is the same as its sign
at small $|\mathbf{x}|$. Since at small enough $|\mathbf{x}|$, the
two functions $\tilde{L}_{S^3}(\mathbf{x})$ and $L_{h}(\mathbf{x})$
coincide, the large-$|\mathbf{x}|$ behavior of the integrand of
$Z_{S^3}$ is connected to the behavior of the Rains function near
the origin of $\mathfrak{h}_{cl}$. Therefore,
\begin{itemize}
\item if the origin is an isolated local minimum of $L_h$, then
$L_h$, and hence $\tilde{L}_{S^3}$, is positive near the origin, and
since $\tilde{L}_{S^3}(\mathbf{x})$ is homogeneous, it is positive
also for large $|\mathbf{x}|$, leading in combination with
(\ref{eq:LagZ3dIntegrandAsy}) to the conclusion that the integrand
of $Z_{S^3}$ decays exponentially at large $|\mathbf{x}|$, and
implying that $Z_{S^3}$ is finite as $\Lambda\to\infty$;
\item if the origin is part of an extended locus where $L_h$ is
locally minimized, then $\tilde{L}_{S^3}(\mathbf{x})$ has flat
directions near the origin, and hence at large $|\mathbf{x}|$, and
therefore the integrand of $Z_{S^3}$ does not decay in certain
directions, leading to the conclusion that $Z_{S^3}$ is power-law
divergent in $\Lambda$ as $\Lambda\to\infty$;
\item if the origin is not a local minimum of $L_h$,
then $L_h$, and hence $\tilde{L}_{S^3}$, is negative somewhere near
the origin, and since $\tilde{L}_{S^3}(\mathbf{x})$ is homogeneous,
it is negative also for large $|\mathbf{x}|$ in certain directions,
leading in combination with (\ref{eq:LagZ3dIntegrandAsy}) to the
conclusion that the integrand of $Z_{S^3}$ grows exponentially at
large $|\mathbf{x}|$ in certain directions, and implying that
$Z_{S^3}$ is exponentially divergent in $\Lambda$ as
$\Lambda\to\infty$.\\
\end{itemize}

\section{Illustrative examples}

\subsection{$A_k$ SQCD theories with $N_f>\frac{2N}{k+1}$}

Take now the example of $A_k$ SQCD with SU($N$) gauge group. This
theory has a chiral multiplet with R-charge $r_a=\frac{2}{k+1}$ in
the adjoint, $N_f$ flavors in the fundamental with R-charge
$r_f=1-\frac{2}{k+1}\frac{N}{N_f}$, and $N_f$ flavors in the
anti-fundamental with R-charge $r_{\bar{f}}=r_f$. For $r_f$ to be
positive we must have $N_f>2N/(k+1)$.

We also assume that we are in the right range of parameters, so we
are inside the conformal window of this theory.

The superconformal index of this theory is (c.f. \cite{Dolan:2008})
\begin{equation}
\begin{split}
\mathcal{I}_{A_k}(b,\beta)=&\frac{(p;p)^{N-1}(q;q)^{N-1}}{N!}\Gamma^{N-1}((pq)^{r_a/2})\int
\mathrm{d}^{N-1}x\\ &\left(\prod_{1\le i<j\le
N}\frac{\Gamma((pq)^{r_a/2}(z_i/z_j)^{\pm 1})}{\Gamma((z_i/z_j)^{\pm
1})}\right) \prod_{i=1}^{N}\Gamma^{N_f}((pq)^{r_f/2} z_i^{\pm
1}),\label{eq:SQCDindex}
\end{split}
\end{equation}
with $\prod_{i=1}^{N} z_i=1$.

\begin{figure}[t]
\centering
    \includegraphics[scale=.7]{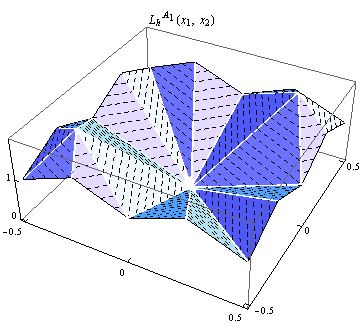}
\caption{The Rains function of the $A_1$ SU($3$) theory---also known
as SU($3$) SQCD. \label{fig:A1}}
\end{figure}

The Rains function of the theory is
\begin{equation}
\begin{split}
L_h^{A_k}(x_1,\dots,x_{N-1})&=N_f
(1-r_f)\sum_{i=1}^{N}\vartheta(x_i)+(1-r_a)\sum_{1\le i<j\le
N}\vartheta(x_i-x_j)-\sum_{1\le i<j\le N}\vartheta(x_i-x_j)\\
&=\frac{2}{k+1}(N\sum_{i}\vartheta(x_i)-\sum_{1\le i<j\le
N}\vartheta(x_i-x_j)).\label{eq:AkRains}
\end{split}
\end{equation}
The $x_N$ in the above expression is constrained by $\sum_{i=1}^{N}
x_i\in\mathbb{Z}$, although since $\vartheta(x)$ is periodic with
period one we can simply replace $x_N\to -x_1-\dots-x_{N-1}$. For
$k=1$ and $N=3$, the resulting function is illustrated in
Figure~\ref{fig:A1}.

We recommend that the reader convince herself that the Rains
function in (\ref{eq:AkRains}) can be easily written down by
examining the integrand of (\ref{eq:SQCDindex}). Whenever the index
of a theory is available in the literature, a similar examination of
the integrand quickly yields the theory's $L_h$ function.

Using Rains's generalized triangle inequality (\ref{eq:RainsGTI}),
in the special case where $d_i=0$, we find that the above function
is minimized when all $x_i$ are zero. This establishes that the
integrand of (\ref{eq:SQCDindex}) is localized around $x_i=0$, and
is exponentially suppressed everywhere else, as $\beta\to 0$.
Therefore $L^{A_k}_{h\ \mathrm{min}}=0$ and
$\mathrm{dim}\mathfrak{h}^{A_k}_{qu}=0$. We thus arrive at
\begin{equation}
\begin{split} \ln \mathcal{I}_{A_k}(b,\beta)=
-\frac{\pi^2}{3\beta}(\frac{b+b^{-1}}{2})(\mathrm{Tr}R) +O(\beta^0).
\label{eq:AkAsyFin}
\end{split}
\end{equation}

\subsubsection*{Much more precise asymptotics}

A more careful study shows \cite{Ardehali:2015c} (see
appendix~\ref{sec:app2} for the definition of the symbol $\sim$)
\begin{equation}
\begin{split} \ln \mathcal{I}_{A_k}(b,\beta)\sim
-\frac{\pi^2}{3\beta}(\frac{b+b^{-1}}{2})(\mathrm{Tr}R)+\ln
Z^{A_k}_{S^3}(b) +\beta E_{\mathrm{susy}}(b), \label{eq:AkAsyFin2}
\end{split}
\end{equation}
where
\begin{equation}
\begin{split}
Z^{A_k}_{S^3}(b)=\frac{\Gamma_h^{N-1}(r_a \omega)}{N!} \int
\mathrm{d}^{N-1}x \left(\prod_{1\le i<j\le N}\frac{\Gamma_h(r_a
\omega\pm (x_i-x_j) )}{\Gamma_h(\pm (x_i-x_j) )}\right)
\prod_{i=1}^{N}\Gamma_h^{N_f}(r_f \omega\pm x_i),\label{eq:SQCD3dZ}
\end{split}
\end{equation}
with the integral over $-\infty<x_i<\infty$.\\

\subsection{SO($2N+1$) SQCD with $N_f>2N-1$}

Consider the SO($n$) SQCD theories with $N_f$ chiral matter
multiplets of R-charge $r=1-\frac{n-2}{N_f}$ in the vector
representation. For the R-charges to be greater than zero, and the
gauge group to be semi-simple, we must have $0<n-2<N_f$.

We also assume that we are in the right range of parameters, so we
are inside the conformal window of this theory.

We perform the analysis for odd $n$; the analysis for even $n$ is
completely analogous, and the result is similar. The index of
SO($2N+1$) SQCD is given by (c.f. \cite{Dolan:2008})
\begin{equation}
\begin{split}
\mathcal{I}_{SO(2N+1)}(b,\beta)=&\frac{(p;p)^N(q;q)^N}{2^N
N!}\Gamma^{N_f}((pq)^{r/2})\\
&\times\int \mathrm{d}^N x
\frac{\prod_{j=1}^{N}\Gamma^{N_f}((pq)^{r/2}z_j^{\pm1})}{\prod_{j=1}^{N}
\Gamma(z_j^{\pm1})\prod_{i<j}(\Gamma((z_i z_j)^{\pm1})\Gamma((z_i/
z_j)^{\pm1}))}.\label{eq:SONindex}
\end{split}
\end{equation}

\begin{figure}[t]
\centering
    \includegraphics[scale=.7]{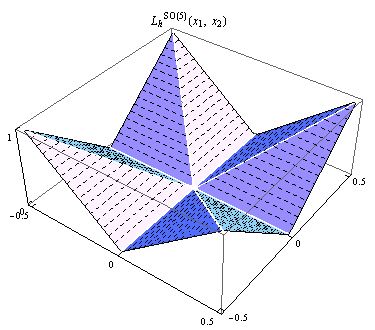}
\caption{The Rains function of the SO($5$) SQCD. \label{fig:SO5}}
\end{figure}

The Rains function of the theory is
\begin{equation}
\begin{split}
L_h^{SO(2N+1)}(\mathbf{x})=(2N-2)\sum_{j=1}^{N}\vartheta(x_j)
-\sum_{1\le i<j\le N}\vartheta(x_i+x_j)-\sum_{1\le i<j\le
N}\vartheta(x_i-x_j).\label{eq:soNRains}
\end{split}
\end{equation}
For the case $N=2$, corresponding to the SO($5$) theory, this
function is illustrated in Figure~\ref{fig:SO5}.

To find the minima of the above function, we need the following
result. For $-1/2\le x_i\le 1/2$
\begin{equation}
\begin{split}
(2N-2)\sum_{1\le j\le N}\vartheta(x_j)&-\sum_{1\le i<j\le
N}\vartheta(x_i+x_j)-\sum_{1\le i<j\le
N}\vartheta(x_i-x_j)=2\sum_{1\le i<j\le
N}\mathrm{min}(|x_i|,|x_j|)\\
&=2(N-1)\mathrm{min}(|x_i|)+2(N-2)\mathrm{min}_2(|x_i|)+\cdots
+2\mathrm{min}_{N-1}(|x_i|),\label{eq:SONvarthetaId}
\end{split}
\end{equation}
where $\mathrm{min}(|x_i|)$ stands for the smallest of
$|x_1|,\dots,|x_N|$, while $\mathrm{min}_2(|x_i|)$ stands for the
next to smallest element, and so on. To prove
(\ref{eq:SONvarthetaId}), one can first verify it for $N=2$, and
then use induction for $N>2$.

Applying (\ref{eq:SONvarthetaId}) we find that the Rains function in
(\ref{eq:soNRains}) is minimized to zero when one (and only one) of
the $x_j$ is nonzero, and the rest are zero. This follows from the
fact that $\mathrm{max}(|x_i|)$ does not show up on the RHS of
(\ref{eq:SONvarthetaId}). Therefore, unlike for the theories of the
previous subsection, here the matrix-integral is not localized
around the origin of the $x_i$ space, but \emph{localized around the
axes}. Equation (\ref{eq:LagIndexSimp6noTheta}) thus simplifies to
\begin{equation}
\begin{split}
\ln \mathcal{I}_{SO(2N+1)}(b,\beta)= -\mathcal{E}^{DK}_0(b,\beta)+
\ln\left(\frac{2\pi}{\beta}\right)+O(1)\quad\quad (\text{as
$\beta\to 0$}).\label{eq:LagIndexSimp1Coul2SON}
\end{split}
\end{equation}\\

The discussion in subsection~\ref{sec:S3} implies that the
three-sphere partition function $Z_{S^3}$ of the dimensionally
reduced daughter of this theory diverges as $Z_{S^3}\approx\Lambda$
(as the cut-off $\Lambda$ of the corresponding matrix-integral is
taken to infinity); this power-law divergence is closely related to
the (generically) subleading logarithmic term on the RHS of
(\ref{eq:LagIndexSimp1Coul2SON}). See subsection~3.2 of
\cite{Ardehali:2015c} for a more detailed discussion of the relation
between the power-law divergence of $Z_{S^3}$ and the subleading
asymptotics of the index.

\subsubsection*{Much more precise asymptotics for the SO($3$) theory with $N_f=2$ when $b=1$}

Luckily, for the special case of $N=1,N_f=2,b=1$, the asymptotic
expansion in (\ref{eq:LagIndexSimp1Coul2SON}) can be completed to
all orders, with the result reading \cite{Ardehali:2015c} (see
appendix~\ref{sec:app2} for the definition of the symbol $\sim$ used
below)
\begin{equation}
\begin{split}
\ln \mathcal{I}_{SO(3)}(\beta)\sim
\ln(\frac{\pi}{2\beta}-\frac{1}{2\pi})+\frac{3}{8}\beta\quad
(\text{as $\beta\to 0$}).\label{eq:SO3indexAsy3}
\end{split}
\end{equation}\\

\subsection{SU($N$) $\mathcal{N}=4$ SYM}

The SU($N$) $\mathcal{N}=4$ theory has the following index
\cite{Spiridonov:2010sv}:
\begin{equation}
\begin{split}
\mathcal{I}_{\mathcal{N}=4}(b,\beta)=&\frac{(p;p)^{N-1}(q;q)^{N-1}}{
N!}\Gamma^{3(N-1)}((pq)^{1/3})\\
&\times \int \mathrm{d}^{N-1} x \prod_{1\le i<j\le
N}\frac{\Gamma^{3}((pq)^{1/3}(z_i/z_j)^{\pm1})}{\Gamma((z_i/z_j)^{\pm1})},\label{eq:N=4index}
\end{split}
\end{equation}
with $\prod_{i=1}^{N}z_i=1$.

Recall that for the $A_k$ SQCD theories the integrand of the
matrix-integral was everywhere exponentially smaller than in the
origin of the $x_i$ space; in other words, the integral localized at
a point. We will shortly find that for the $\mathcal{N}=4$ theory
the matrix-integral does not localize at all.

The Rains function of the theory is
\begin{equation}
\begin{split}
L_h^{\mathcal{N}=4}=3(1-\frac{2}{3})\sum_{1\le i<j\le
N}\vartheta(x_i-x_j)-\sum_{1\le i<j\le N}\vartheta(x_i-x_j)=0.
\end{split}
\end{equation}
In other words, there is no effective potential, and the
matrix-integral does not localize:
$\mathfrak{h}_{qu}=\mathfrak{h}_{cl}$.
Eq.~(\ref{eq:LagIndexSimp6noTheta}) thus dictates
\begin{equation}
\begin{split}
\ln \mathcal{I}_{\mathcal{N}=4}(b,\beta)=
(N-1)\ln(\frac{2\pi}{\beta})+O(\beta^0).\label{eq:N=4indexAsy1}
\end{split}
\end{equation}
There is no $O(1/\beta)$ term on the RHS, because $\mathrm{Tr}R=0$
for the $\mathcal{N}=4$ theory (and also $L^{\mathcal{N}=4}_{h\
\mathrm{min}}=0$).\\

The discussion in subsection~\ref{sec:S3} implies that the
three-sphere partition function $Z_{S^3}$ of the dimensionally
reduced daughter of this theory diverges as
$Z_{S^3}\approx\Lambda^{N-1}$ (as the cut-off $\Lambda$ of the
corresponding matrix-integral is taken to infinity); this power-law
divergence is closely related to the logarithmic term on the RHS of
(\ref{eq:N=4indexAsy1}). See subsection~3.2 of \cite{Ardehali:2015c}
for more details.

\subsubsection*{More precise asymptotics}

A more careful treatment shows that \cite{Ardehali:2015c}
\begin{equation}
\begin{split}
\ln \mathcal{I}_{\mathcal{N}=4}(b,\beta)=
(N-1)\ln(\frac{2\pi}{\beta})+3(N-1)\ln\Gamma_h(\frac{2}{3}\omega)-\ln
N!+o(1)\quad (\text{as $\beta\to 0$}).\label{eq:N=4indexAsy}
\end{split}
\end{equation}\\

\subsection{The $\mathbb{Z}_2$ orbifold theory}

We now study a quiver gauge theory, to illustrate how easily Rains's
method generalizes to theories with more than one simple factor in
their gauge group.

Consider the $\mathbb{Z}_2$ orbifold of the $\mathcal{N}=4$ SYM with
SU($N$) gauge group. The theory consists of two SU($N$) gauge
groups, with one chiral multiplet in the adjoint of each, and one
doublet of bifundamental chiral multiplets from each gauge group to
the other. All the chiral multiplets have R-charge $r=2/3$.

The superconformal index is given by (c.f. \cite{Gadde:2010holo})
\begin{equation}
\begin{split}
\mathcal{I}_{\mathbb{Z}_2}(b,\beta)=\ &(\prod_{k=1,2}[
\frac{(p;p)^{N-1}(q;q)^{N-1}}{N!}\Gamma^{N-1}((pq)^{1/3})\int
\mathrm{d}^{N-1}x^{(k)}\\ &\left(\prod_{1\le i<j\le
N}\frac{\Gamma((pq)^{1/3}(z^{(k)}_i/z^{(k)}_j)^{\pm
1})}{\Gamma((z^{(k)}_i/z^{(k)}_j)^{\pm 1})}\right)]) \times
\prod_{i,j=1}^{N}\left(\Gamma((pq)^{1/3} (z^{(1)}_i/z^{(2)}_j)^{\pm
1})\right),\label{eq:Z2orbZ}
\end{split}
\end{equation}
with $\prod_{i=1}^{N} z^{(1)}_i=\prod_{i=1}^{N} z^{(2)}_i=1$.

\begin{figure}[t]
\centering
    \includegraphics[scale=.7]{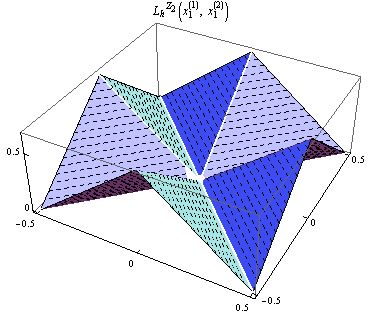}
\caption{The Rains function of the SU($2$)$\times$SU($2$) orbifold
theory. \label{fig:z2Lh}}
\end{figure}

The Rains function of the theory is
\begin{equation}
\begin{split}
L_h^{\mathbb{Z}_2}(\mathbf{x}^{(1)},\mathbf{x}^{(2)})&=-\frac{2}{3}\sum_{1\le
i<j\le N}\vartheta(x^{(1)}_i-x^{(1)}_j)-\frac{2}{3}\sum_{1\le i<j\le
N}\vartheta(x^{(2)}_i-x^{(2)}_j)+\frac{2}{3}\sum_{i,j=1}^{N}\vartheta(x^{(1)}_i-x^{(2)}_j).
\end{split}
\end{equation}
For the case $N=2$, corresponding to the SU($2$)$\times$SU($2$)
theory, this function is illustrated in Figure~\ref{fig:z2Lh}.

The generalized triangle inequality (\ref{eq:RainsGTI}) applies with
$c=x^{(1)},d=x^{(2)}$, and implies that $L_h^{\mathbb{Z}_2}$ is
positive semi-definite. It moreover shows that $L_h^{\mathbb{Z}_2}$
vanishes if the $x^{(1)}_i,x^{(2)}_j$ can be permuted such that
either of (\ref{cond:gti1}) or (\ref{cond:gti2}) holds. For
simplicity we consider all $x^{(1)}_i$ to be positive and very
small, except for $x^{(1)}_N=-x^{(1)}_1-\dots-x^{(1)}_{N-1}$ being
negative and very small, and similarly for $x^{(2)}_j$. Assuming
either (\ref{cond:gti1})  or (\ref{cond:gti2}), we conclude that
$x^{(1)}_i=x^{(2)}_i$. Based on this result, and also the $N=2$ case
whose Rains function is displayed in Figure~\ref{fig:z2Lh}, we
conjecture that for the $\mathbb{Z}_2$ orbifold theory
$\mathrm{dim}\mathfrak{h}_{qu}=N-1$, and thereby
\begin{equation}
\begin{split}
\ln \mathcal{I}_{\mathbb{Z}_2}(b,\beta)=
-\mathcal{E}^{DK}_0(b,\beta)+(N-1)
\ln\left(\frac{2\pi}{\beta}\right)+O(1)\quad\quad (\text{as
$\beta\to 0$}).\label{eq:Z2indexAsy}
\end{split}
\end{equation}\\

The discussion in subsection~\ref{sec:S3} implies that the
three-sphere partition function $Z_{S^3}$ of the dimensionally
reduced daughter of this theory diverges as
$Z_{S^3}\approx\Lambda^{N-1}$ (as the cut-off $\Lambda$ of the
corresponding matrix-integral is taken to infinity); this power-law
divergence is related to the subleading logarithmic term on the RHS
of (\ref{eq:Z2indexAsy}). See subsection~3.2 of
\cite{Ardehali:2015c} for more details.

\subsection{The SU($2$) ISS model}\label{sec:iss}

There are two famous interacting Lagrangian SCFTs with $c<a$. The
first is the Intriligator-Seiberg-Shenker (ISS) model of dynamical
SUSY breaking \cite{Intriligator:1994}. The theory is formulated in
the UV as an SU($2$) vector multiplet with a single chiral multiplet
in the four-dimensional representation of the gauge group. Although
originally suspected to confine (and to break supersymmetry upon
addition of a tree-level superpotential) \cite{Intriligator:1994},
the theory is currently believed to flow to an interacting SCFT in
the IR \cite{Intriligator:2005,Poppitz:2009}, where the chiral
multiplet has R-charge $3/5$. The IR SCFT has $c-a=-7/80$.

The index of this theory is (c.f. \cite{Vartanov:2010})
\begin{equation}
\begin{split}
\mathcal{I}_{ISS}(b,\beta)=\frac{(p;p)(q;q)}{2}\int\mathrm{d}x
\frac{\Gamma((pq)^{3/10}z^{\pm 1})\Gamma((pq)^{3/10}z^{\pm
3})}{\Gamma(z^{\pm 2})}.\label{eq:ISSindex}
\end{split}
\end{equation}

\begin{figure}[t]
\centering
    \includegraphics[scale=1.1]{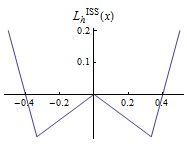}
\caption{The Rains function of the SU($2$) ISS theory.
\label{fig:ISS}}
\end{figure}

The Rains function of the theory is
\begin{equation}
\begin{split}
L^{ISS}_h(x)=\frac{2}{5}\vartheta(x)
+\frac{2}{5}\vartheta(3x)-\vartheta(2x).
\end{split}
\end{equation}
This function is plotted in Figure~\ref{fig:ISS}.

A direct examination reveals that $L^{ISS}_h(x)$ is minimized at
$x=\pm 1/3$, and $L^{ISS}_h(\pm 1/3)=-2/15$. The asymptotics of
$\mathcal{I}_{ISS}$ is hence given according to
(\ref{eq:LagIndexSimp6noTheta}) by
\begin{equation}
\begin{split}
\ln \mathcal{I}_{ISS}(b,\beta)=
\frac{\pi^2}{15\beta}(\frac{b+b^{-1}}{2})+O(\beta^0).\label{eq:AsyISS1}
\end{split}
\end{equation}
In other words we have $(c-a)_{\mathrm{shifted}}=c-a+1/10=1/80$.\\

The discussion in subsection~\ref{sec:S3} implies that the
three-sphere partition function of the dimensionally reduced
daughter of this theory is exponentially divergent; this severe
divergence is related to the modification that the
Di~Pietro-Komargodski formula receives in this case.

\subsubsection*{Much more precise asymptotics}

A more careful study shows \cite{Ardehali:2015c} (see
appendix~\ref{sec:app2} for the definition of the symbol $\sim$)
\begin{equation}
\begin{split}
\ln \mathcal{I}_{ISS}(b,\beta)\sim
\frac{16\pi^2}{3\beta}(c-a)_{\mathrm{shifted}}(\frac{b+b^{-1}}{2})+
\ln Y^{ISS}_{S^3}(b)+\beta E_{\mathrm{susy}}(b),\quad (\text{as
$\beta\to 0$})\label{eq:ISSindexAsy3}
\end{split}
\end{equation}
with
\begin{equation}
Y^{ISS}_{S^3}(b)=\int_{-\infty}^{\infty}\mathrm{d}x'
e^{-\frac{4\pi}{5}(b+b^{-1})x'}\times\Gamma_h(3x'+(3/5)\omega)\Gamma_h(-3x'+(3/5)\omega),\label{eq:Yiss}
\end{equation}
and $(c-a)_{\mathrm{shifted}}=(c-a)+1/10=1/80$. A numerical
evaluation using
\begin{equation}
\begin{split}
\ln \Gamma_h(ix;i,i)=(x-1)\ln (1-e^{-2\pi i x})-\frac{1}{2\pi
i}Li_2(e^{-2\pi i
x})+\frac{i\pi}{2}(x-1)^2-\frac{i\pi}{12},\label{eq:hypGammab=1}
\end{split}
\end{equation}
yields
$Y^{ISS}_{S^3}(b=1)\approx .423$.\\

\subsection{The SO($2N+1$) BCI model with $1<N<5$}\label{sec:BCI}

The second famous example of interacting SCFTs with $c<a$ is
provided by the ``misleading'' SO($n$) theory of Brodie, Cho, and
Intriligator \cite{Brodie:1998}. This is an $\mathcal{N}=1$ SO($n$)
gauge theory with a single chiral multiplet in the two-index
symmetric traceless tensor representation of the gauge group. The
theory is asymptotically free if $n\ge 5$. For $5\le n<11$ the
corresponding interacting IR SCFT has $c-a=-(n-1)/16$ (for greater
values of $n$ the R-symmetry of the IR SCFT is believed to mix with
an emergent accidental symmetry, and thus more care is called for;
c.f. \cite{Intriligator:2010}).

For the SO($2N+1$) theory (with $1<N<5$) we have  (c.f.
\cite{Vartanov:2010})
\begin{equation}
\begin{split}
\mathcal{I}_{BCI}(b,\beta)=&\frac{(p;p)^N(q;q)^N}{2^N
N!}\Gamma^N((pq)^{2/(2N+3)})\int\mathrm{d}^N x\\
&\prod_{i<j}\frac{\Gamma((pq)^{2/(2N+3)}z_i^{\pm1}
z_j^{\pm1})}{\Gamma(z_i^{\pm1}
z_j^{\pm1})}\prod_{j=1}^N\frac{\Gamma((pq)^{2/(2N+3)}z_j^{\pm1}
,(pq)^{2/(2N+3)}z_j^{\pm2})}{\Gamma(
z_j^{\pm1})}.\label{eq:BCIindex}
\end{split}
\end{equation}

\begin{figure}[t]
\centering
    \includegraphics[scale=.8]{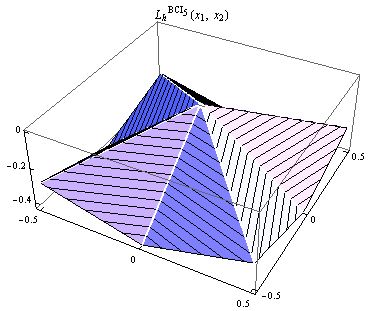}
\caption{The Rains function of the SO($5$) BCI theory.
\label{fig:BCI}}
\end{figure}

The Rains function of the theory is
\begin{equation}
\begin{split}
L_h^{BCI}(x)=\frac{4}{2N+3}\left((\frac{2N-1}{4})\sum_{j}\vartheta(2x_j)
-\sum_j\vartheta(x_j)-\sum_{i<j}\vartheta(x_i+x_j)-\sum_{i<j}\vartheta(x_i-x_j)\right).\label{eq:BCIRains}
\end{split}
\end{equation}
For $N=2$, corresponding to the SO($5$) theory, this function is
plotted in Figure~\ref{fig:BCI}.

To find the minima of the above function, we need the following
result, valid for $-1/2\le x_i\le 1/2$:
\begin{equation}
\begin{split}
&(\frac{2N-1}{4})\sum_{1\le j\le N}\vartheta(2x_j)-\sum_{1\le j\le
N}\vartheta(x_j)-\sum_{1\le i<j\le N}\vartheta(x_i+x_j)-\sum_{1\le
i<j\le N}\vartheta(x_i-x_j)=\\
&-\frac{3}{2}\sum_{ i<j}\mathrm{max}(|x_i|,|x_j|) +\frac{1}{2}\sum_{
i<j}\mathrm{min}(|x_i|,|x_j|)=\sum_{j}(-\frac{3N}{2}+2j-\frac{1}{2})\mathrm{min}_{N-j+1}(|x_i|),\label{eq:BCIvarthetaId}
\end{split}
\end{equation}
with $\mathrm{min}_N(|x_i|):=\mathrm{max}(|x_i|)$. The proof of
(\ref{eq:BCIvarthetaId}) is similar to that of
(\ref{eq:SONvarthetaId}).

Note that the coefficient of the $j$th term on the RHS of
(\ref{eq:BCIvarthetaId}) is negative if $j<\frac{3N+1}{4}$, and
positive otherwise. This implies that the Rains function
(\ref{eq:BCIRains}) is minimized when $\lfloor
\frac{3N+1}{4}\rfloor$ of the $|x_i|$ are maximized (i.e. $x_{i}=\pm
1/2$), and the rest of the $|x_i|$ are minimized (i.e. $x_{i}=0$).
Consequently, the minimum of the Rains function is
\begin{equation}
\begin{split}
L_{h\ \mathrm{min}}^{BCI}=-\frac{1}{2N+3}\sum_{1\le j\le \lfloor
\frac{3N+1}{4}\rfloor}(3N+1-4j).\label{eq:BCIRainsMin}
\end{split}
\end{equation}
This is less than zero for any $N>1$. Therefore the
Di~Pietro-Komargodski formula needs to be modified in the SO($2N+1$)
BCI model with $1<N<5$.

The discussion in subsection~\ref{sec:S3} implies that the
three-sphere partition function of the dimensionally reduced
daughter of this model (with $1<N<5$) is exponentially divergent;
this severe divergence is related to the modification that the
Di~Pietro-Komargodski formula receives in this case. See
subsection~3.3 of \cite{Ardehali:2015c}
for more details.\\

Consider now the concrete case of the SO($5$) theory corresponding
to $N=2$. This theory has $c-a=-1/4$. From
Eq.~(\ref{eq:BCIRainsMin}) we have in this case $L_{h\
min}^{BCI}(x)=-3/7$. The asymptotics of $\mathcal{I}$ is therefore
given according to (\ref{eq:LagIndexSimp6noTheta}) by
\begin{equation}
\begin{split}
\ln \mathcal{I}_{BCI_5}(b,\beta)=
\frac{8\pi^2}{21\beta}(\frac{b+b^{-1}}{2})+O(\beta^0).\label{eq:AsyBCI1}
\end{split}
\end{equation}
In other words $(c-a)_{\mathrm{shifted}}=c-a+9/28=1/14$.

\subsubsection*{Much more precise asymptotics for the SO($5$) BCI theory}

A more careful treatment shows \cite{Ardehali:2015c}
\begin{equation}
\begin{split}
\ln \mathcal{I}_{BCI_5}(b,\beta)\sim
\frac{16\pi^2}{3\beta}(c-a)_{\mathrm{shifted}}(\frac{b+b^{-1}}{2})+
\ln Y^{BCI_5}_{S^3}(b)+\beta E_{\mathrm{susy}}(b),\quad (\text{as
$\beta\to 0$})\label{eq:BCIindexAsy4}
\end{split}
\end{equation}
with
\begin{equation}
\begin{split}
Y^{BCI_5}_{S^3}(b)=&\frac{1}{2}\int_{-\infty}^{\infty}\mathrm{d}
x_1' \ \Gamma_h((4/7)\omega\pm
2x_1')\times\\
&\frac{\Gamma_h^2((4/7)\omega)}{2}\int_{-\infty}^{\infty} \mathrm{d}
x_2\ \frac{\Gamma_h((4/7)\omega\pm x_2)\Gamma_h((4/7)\omega\pm
2x_2)} {\Gamma_h(\pm x_2)},\label{eq:Ybci}
\end{split}
\end{equation}
and $(c-a)_{\mathrm{shifted}}=(c-a)+9/28=1/14$.  A numerical
evaluation using (\ref{eq:hypGammab=1}) yields
$Y^{BCI_5}_{S^3}(b=1)\approx .026$.\\

\subsection{Puncture-less SU($2$) class-$\mathcal{S}$ theories}\label{sec:classS}

\begin{figure}[t]
\centering
    \includegraphics[scale=2.5]{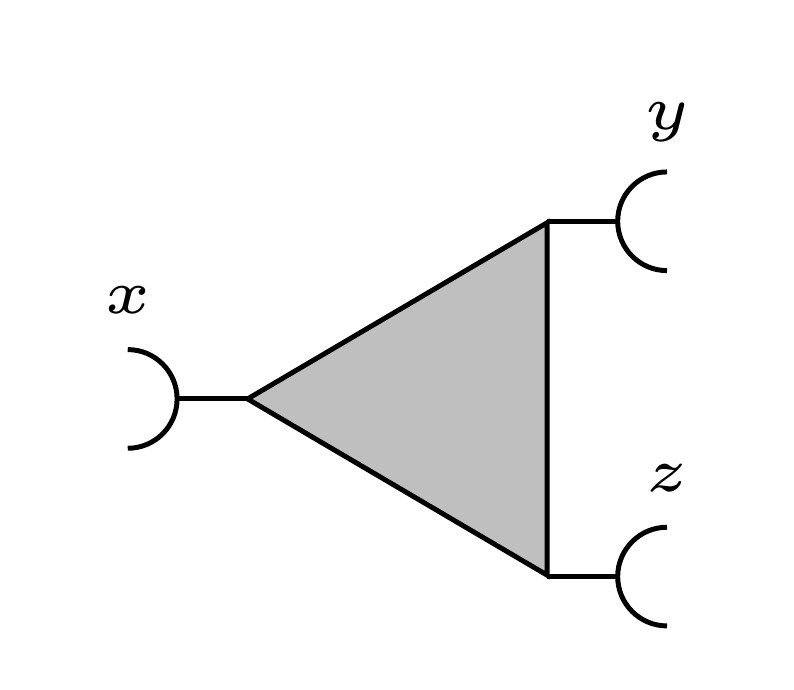}
\caption{The building block of the puncture-less Lagrangian
class-$\mathcal{S}$ theories. \label{fig:classSb}}
\end{figure}

\begin{figure}[t]
\centering
    \includegraphics[scale=1.8]{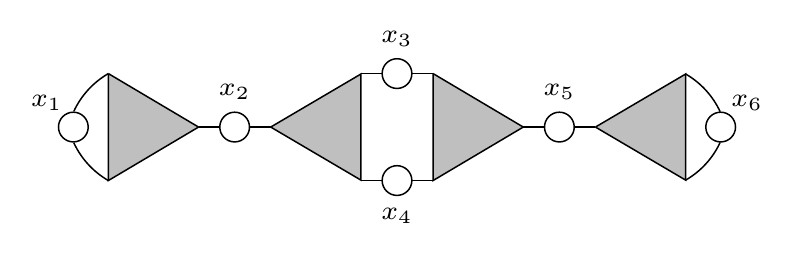}
\caption{The quiver diagram of the $g=3$ class-$\mathcal{S}$ theory.
\label{fig:g3S}}
\end{figure}

An interesting class of Lagrangian $\mathcal{N}=2$ SCFTs arise from
quiver gauge theories associated to Riemann surfaces of genus
$g\ge2$, without punctures (see e.g. \cite{Beem:2012ca} for a
discussion of the indices of these theories). These quivers can be
constructed from fundamental blocks of the kind shown in
Figure~\ref{fig:classSb}. The triangle in Figure~\ref{fig:classSb}
represents eight chiral multiplets of R-charge $2/3$ transforming in
the tri-fundamental representation of the three SU($2$) gauge
groups\footnote{We focus on class-$\mathcal{S}$ theories constructed
from $T_2$, and leave the study of higher-rank theories constructed
from $T_{N>2}$ to future work.} represented by the (semi-circular)
nodes; more precisely, when two semi-circular nodes are connected
together to form a circle, they represent an $\mathcal{N}=2$ SU($2$)
vector multiplet. A class-$\mathcal{S}$ theory of genus $g$ arises
when $2g-2$ of these blocks are glued back-to-back (and
forth-to-forth) along a straight line, with the leftmost and the
rightmost blocks having two of their half-circular nodes glued
together; see Figure~\ref{fig:g3S} for an example.

An $\mathcal{N}=2$ SU($2$) vector multiplet contributes to the Rains
function of the SCFT as
\begin{equation}
\begin{split}
L_h^{\mathcal{N}=2\
v}(x)=-\frac{2}{3}\vartheta(2x).\label{eq:SSNRains}
\end{split}
\end{equation}
A semi-circular node contributes half as much, and thus the three
semi-circular nodes in Figure~\ref{fig:classSb} contribute together
as
\begin{equation}
\begin{split}
L_h^{\mathrm{semi-nodes}}(x,y,z)=-\frac{1}{3}\left(\vartheta(2x)+\vartheta(2y)+\vartheta(2z)\right).\label{eq:3NRains}
\end{split}
\end{equation}

The eight chiral multiplets represented by the triangle in
Figure~\ref{fig:classSb} contribute to the Rains function of the
theory as
\begin{equation}
\begin{split}
L_h^{T_2}(x,y,z)=\frac{1}{3}\left(\vartheta(x+y+z)+\vartheta(x+y-z)+\vartheta(x-y+z)+\vartheta(-x+y+z)\right).\label{eq:T2Rains}
\end{split}
\end{equation}

Adding up (\ref{eq:3NRains}) and (\ref{eq:T2Rains}) we obtain the
contribution of a single block to the Rains function:
\begin{equation}
\begin{split}
L_h^{\mathrm{block}}(x,y,z)=&\frac{1}{3}[\vartheta(x+y+z)+\vartheta(x+y-z)+\vartheta(x-y+z)+\vartheta(-x+y+z)\\
&-\vartheta(2x)-\vartheta(2y)-\vartheta(2z)].\label{eq:T2Rains}
\end{split}
\end{equation}

With the Rains function of the block at hand, we can now write down
the Rains function of genus $g$ class-$\mathcal{S}$ theories. For
example, the Rains function of the $g=2$ theory is given by
\begin{equation}
\begin{split}
L_h^{\mathcal{S}_{g=2}}(x_1,x_2,x_3)=L_h^{\mathrm{block}}(x_1,x_1,x_2)+L_h^{\mathrm{block}}(x_2,x_3,x_3),\label{eq:Sg=2Rains}
\end{split}
\end{equation}
and the Rains function of the $g=3$ theory (illustrated in
Figure~\ref{fig:g3S}) is obtained as
\begin{equation}
\begin{split}
L_h^{\mathcal{S}_{g=3}}(x_1,x_2,x_3,x_4,x_5,x_6)=&L_h^{\mathrm{block}}(x_1,x_1,x_2)+L_h^{\mathrm{block}}(x_2,x_3,x_4)\\
&+L_h^{\mathrm{block}}(x_3,x_4,x_5)+L_h^{\mathrm{block}}(x_5,x_6,x_6).\label{eq:Sg=3Rains}\\
\end{split}
\end{equation}

Importantly, Rains's GTI (\ref{eq:RainsGTI}), with $c_1=x+y,\
c_2=x-y,\ d_1=z,\ d_2=-z$, implies that
\begin{equation}
\begin{split}
L_h^{\mathrm{block}}(x,y,z)\ge0.
\end{split}
\end{equation}
It is not difficult to show that the equality holds in a
finite-volume subspace of the $x,y,z$ space; take for instance
$x,y,z\approx .1$ within $.01$ of each other, and use the fact that
for small argument $L_h$ reduces to $\tilde{L}_{S^3}$ to show that
$L_h$ vanishes in the domain just described.

Since the Rains function of a $g\ge2$ class-$\mathcal{S}$ theory is
the sum of several block Rains functions, the positive
semi-definiteness of $L_h^{\mathrm{block}}$ guarantees the positive
semi-definiteness of $L_h^{\mathcal{S}_{g\ge2}}(x_i)$; moreover,
taking all $x_i$ to be around $.1$ and within $.01$ of each other we
can easily conclude (as in the previous paragraph) that for the
genus $g$ theory $\mathrm{dim}\mathfrak{h}_{qu}=3(g-1)$. The
relation (\ref{eq:LagIndexSimp6noTheta}) thus yields
\begin{equation}
\begin{split}
\ln \mathcal{I}_{\mathcal{S}_{g\ge2}}(b,\beta)=
\frac{16\pi^2}{3\beta}(c-a)(\frac{b+b^{-1}}{2})+3(g-1)\ln(\frac{2\pi}{\beta})+O(\beta^0),\label{eq:AsyClassS}
\end{split}
\end{equation}
with $c-a=-(g-1)/24$.\\

\section{Applications}

\subsection{Supersymmetric dualities}\label{sec:dualApp}

Dual QFTs must have equal partition functions. As a trivial
corollary, the high-temperature asymptotics of the index of dual 4d
SCFTs must match.

Assume now that both sides of the duality are non-chiral 4d
Lagrangian SCFTs with a semi-simple gauge group. The relation
(\ref{eq:LagIndexSimp6noTheta}) then yields two quantities to be
matched between the theories: $L_{h\ \mathrm{min}}$ and
$\mathrm{dim}\mathfrak{h}_{qu}$. Comparison of $L_{h\ \mathrm{min}}$
can rule out for instance the confinement scenario for the SU($2$)
ISS model: on the gauge theory (UV) side, as discussed above, we
have $L_{h\ \mathrm{min}}=-2/15$, while on the mesonic (IR)
side\footnote{Following \cite{Vartanov:2010}, we are assuming that
an index can be consistently assigned to the proposed IR theory,
even though the IR chiral multiplet would have R-charge
$12/5\notin]0,2[$. This assignment requires an analytic continuation
of the kind discussed in \cite{Ardehali:2015c}; the small-$\beta$
asymptotics of the resulting function can be obtained as in
\cite{Ardehali:2015b}. See \cite{Gerchkovitz:2013} for an
alternative take on this problem.} we have no gauge group and thus
$L_{h}=0$.

As another example, consider the recent $E_7$ SQCD duality of
\cite{Kutasov:2014,Kutasov:2014ed}. In that case a direct
examination reveals that $L_{h\
\mathrm{min}}=\mathrm{dim}\mathfrak{h}_{qu}=0$, both on the electric
and the magnetic side. Their proposal hence passes both our tests.\\

The case of the interacting $\mathcal{N}=1$ SCFTs with $c<a$ (namely
the IR fixed points of the ISS model, and the BCI$_{2N+1}$ model
with $1<N<5$) is particularly interesting. A dual description for
these theories is currently lacking. Our results for $L_{h\
\mathrm{min}}$ and $\mathrm{dim}\mathfrak{h}_{qu}$ on the electric
side might help to test future proposals for magnetic duals of these
theories.\\

\subsection{Holography}\label{sec:holoApp}

In the present chapter we analyzed the high-temperature asymptotics
of the indices of various gauge theories at \emph{finite $N$}. The
finite-$N$ indices of holographic SCFTs are expected to encode
information about micro-states of the supersymmetric Giant~Gravitons
of the dual string theories \cite{Bourdier:2015}. Take for instance
the SU($N$) $\mathcal{N}=4$ SYM. One of the novel results of
\cite{Ardehali:2015c} is the following high-temperature asymptotics
for the superconformal index of this theory (see
Eqs.~(\ref{eq:N=4indexAsy1}) and (\ref{eq:N=4indexAsy}) above):
\begin{equation}
\mathcal{I}(b=1,\beta)=\sum_{\mathrm{operators}}(-1)^F
e^{-\beta(\Delta-\fft12r)}\approx (\frac{1}{\beta})^{N-1}.
\end{equation}
The above \emph{canonical} relation can be transformed to the
\emph{micro-canonical} ensemble to yield the asymptotic
(fermion-number weighted) degeneracy of the protected high-energy
operators in the $\mathcal{N}=4$ theory:
\begin{equation}
N(E)\approx E^{N-2},
\end{equation}
with $E=\Delta-r/2$. This result should presumably be reproduced by
geometric quantization of the $1/16$ BPS Giant~Gravitons of IIB
theory on AdS$_5\times S^5$, along the lines of \cite{Biswas:2006}.
It would be interesting to see if this expectation pans out.\\

\subsection{Quantum Coulomb branch dynamics on $R^3\times
S^1$}\label{sec:crossed}

Take a non-chiral 4d SCFT with a semi-simple gauge group, and with
$r_\chi\in]0,2[$. Its superconformal index $\mathcal{I}(b,\beta)$
can be computed by a path-integral on $S_b^3\times S_\beta^1$, with
$S_b^3$ the unit-radius squashed three-sphere. We now replace the
$S_b^3$ with the round three-sphere $S_{r_3}^3$ of arbitrary radius
$r_3>0$. The path-integral on the new space gives
$\mathcal{I}(\beta;r_3)=\mathcal{I}(b=1,\beta/r_3)$; i.e. the
resulting partition function only depends on the ratio $\beta/r_3$,
as the theory is conformal. Thus, as far as $\mathcal{I}(\beta;r_3)$
is concerned, shrinking the $S^1$ (i.e. the high-temperature limit)
is equivalent to decompactifying the $S^3$. We hence fix $\beta$,
and send $r_3$ to infinity. In this limit we expect the unlifted
zero-modes on $S_{r_3}^3\times S_\beta^1$ to roughly correspond to
the quantum zero-modes on $R^3\times S^1$. Therefore at high
temperatures the unlifted holonomies of the theory on
$S_{r_3}^3\times S_\beta^1$ should be in correspondence with (a real
section of) the quantum Coulomb branch of the 3d $\mathcal{N}=2$
theory obtained from compactifying the 4d theory on the circle of
$R^3\times S^1$. In particular, we expect
$\mathrm{dim}\mathfrak{h}_{qu}$ to be equal to the (complex-)
dimension of the quantum Coulomb branch of the 3d theory. (Recall
that the Coulomb branch of the circle-compactified theory living on
$R^3$ consists not just of the holonomies around the $S^1$, but also
of the dual 3d photons; hence our references above to ``a real
section'' and ``complex-dimension''.)

We do not expect to recover the $R^3\times S^1$ Higgs branch from
the zero-modes on $S_{r_3}^3\times S_\beta^1$: for any (arbitrarily
small) curvature on the $S^3$, curvature couplings presumably lift
the Higgs-type zero-modes on $S_{r_3}^3\times S_\beta^1$.

From the point of view of $R^3\times S^1$, picking one of the $R^3$
directions as time\footnote{The following discussion is in the
spirit of the arguments in \cite{Shaghoulian:2015}, though our
treatment is not as precise. We are approaching $R^3$ from $S^3$,
rather than from $T^3$ (as in \cite{Shaghoulian:2015}). While on
$T^3$ each of the circles can be picked as the time direction,
picking a time direction along the $S^3$ makes the spatial sections
time-dependent, rendering our arguments in the paragraph of this
footnote somewhat hand-wavy. I thank E.~Shaghoulian for several
helpful conversations related to the subject of the present
subsection.}, we can relate $\mathcal{E}^{DK}_0$ to the Casimir
energy associated to the spatial manifold $R^2\times S^1$: we
reintroduce $r_3$ in $\mathcal{E}^{DK}_0$ (by replacing its $\beta$
with $\beta/r_3$), set in it $b=1$, interpret $\tilde{\beta}:=2\pi
r_3$ as the circumference of the crossed channel thermal circle, and
write
\begin{equation}
\mathcal{E}_0^{DK}(\beta;r_3)=\tilde{\beta}E_0^{R^2\times
S^1}(\beta),\quad\text{with}\quad E_0^{R^2\times
S^1}(\beta)=\frac{\pi}{6\beta}\mathrm{Tr}R.\label{eq:dkEcross}
\end{equation}
Now $E_0^{R^2\times S^1}(\beta)$ admits an interpretation as the
(regularized) Casimir energy associated to the spatial $R^2\times
S_\beta^1$. Similarly, resurrecting the $r_3$ in $V^{\mathrm{eff}}$,
and setting in it $b=1$, we obtain what can be loosely regarded as
$\tilde{\beta}$ times the quantum effective potential on (a real
section of) the crossed channel Coulomb branch. From this
perspective, the two tests we advocated in
subsection~\ref{sec:dualApp} would not really be new, but would
correspond to the comparison of low-energy properties on $R^3\times
S^1$.

The discussion in the previous three paragraphs is rather intuitive,
and should be considered suggestive at best. It is desirable to have
it made more precise. Nevertheless, in the examples of the SU($N$),
Sp($2N$), and SO($2N+1$) SQCD theories, and the SU($N$)
$\mathcal{N}=4$ SYM, it turns out \cite{Ardehali:2015c} that (upon
quotienting by the Weyl group) $\mathfrak{h}_{qu}$ \emph{does indeed
resemble} (a real section of) the $R^3\times S^1$ quantum Coulomb
branch; see \cite{Aharony:2013a,Aharony:2013b} and
\cite{Seiberg:1997}. We therefore conjecture that the relation
between $\mathfrak{h}_{qu}$ and the unlifted Coulomb branch on
$R^3\times S^1$ continues to remain valid, at least for all the
theories with a positive semi-definite Rains function. In
particular, we predict that, when placed on $R^3\times S^1$, the
SU($N$) $A_k$ SQCD theories (in the appropriate range of their
parameters such that all their $r_\chi$ are in $]0,2[$) have no
quantum Coulomb branch, and the $\mathbb{Z}_2$ orbifold of the
SU($N$) $\mathcal{N}=4$ theory has an $(N-1)$-dimensional unlifted
Coulomb branch.\\

For theories whose Rains function is not positive semi-definite, on
the other hand, it seems like this connection with $R^3\times S^1$
fails. The Rains function of the SU($2$) ISS model does not have a
flat direction, and appears to suggest a Higgs vacuum for the theory
on $R^3\times S^1$. However, the study of \cite{Poppitz:2009}
indicates that this theory possesses an unlifted Coulomb branch on
$R^3\times S^1$, and in particular does not necessarily break the
gauge group at low energies. It would be nice to understand if this
conflict is only a manifestation of the sloppiness of our intuitive
arguments above, or it has a more interesting origin.\\

\chapter{Taking the large-$N$ limit first: the holographic Weyl anomaly from the index}

\section{The large-$N$ limit of the 4d superconformal index: the multi-trace index}

It is often the case that asymptotically at large $N$ a hierarchy
appears in the spectrum of local operators of an SCFT. This
hierarchy is expected to be reflected in the superconformal index.
Take for instance the U($N$) $\mathcal{N}=4$ SYM, which has the
following \emph{Schur index} \cite{Bourdier:2015} (see
\cite{Gadde:2011MC} for the definition of the Schur index; in the
present chapter we discuss the Schur index only as a toy model of
the superconformal index):
\begin{equation}
\mathcal{I}_{Schur}(\beta)=\frac{(q;q)}{(q^{1/2};q^{1/2})^2}\sum_{n=0}^{\infty}(-1)^n\left[\begin{pmatrix}
N+n\\ N
\end{pmatrix}+\begin{pmatrix}
N+n-1\\ N
\end{pmatrix}\right]q^{(n N+n^2)/2},\label{eq:N=4SchurN}
\end{equation}
with $q=e^{-\beta}$. In the above expression, $n$ clearly has an
interpretation as a soliton-counting number. Of course, these
solitons are naturally interpreted in the gravity dual to the U($N$)
$\mathcal{N}=4$ SYM. They presumably correspond to Giant Gravitons
of the IIB theory on AdS$_5\times S^5$ \cite{Bourdier:2015}.

The large-$N$ limit suppresses (energetically) all the $n\neq 0$
terms in (\ref{eq:N=4SchurN}), and yields
\begin{equation}
\mathcal{I}^{N\to\infty}_{Schur}(\beta)=\frac{(q;q)}{(q^{1/2};q^{1/2})^2}.\label{eq:N=4SchurLargeN}
\end{equation}
This is the ``multi-trace'' Schur index of the U($N$)
$\mathcal{N}=4$ SYM; it can be obtained by summing over multi-trace
operators of the gauge theory in the planar limit.\\

Another general point that our toy model can help illustrate is that
the high-temperature asymptotics of the multi-trace index may be
(and generically is) very different from the asymptotics of the
finite-$N$ index (which we focused on in the previous chapter). Our
toy model index (\ref{eq:N=4SchurN}) has the high-temperature
asymptotics (see \cite{Ardehali:2015c} for the similar asymptotic
analysis of the Schur index of the SU($N$) $\mathcal{N}=4$ SYM)
\begin{equation}
\ln \mathcal{I}_{Schur}(\beta)= N\ln
(\frac{2\pi}{\beta})+O(\beta).\label{eq:N=4SchurfiniteNasy}
\end{equation}

Taking the $N\to\infty$ limit before the $\beta\to 0$ limit, changes
the high-temperature asymptotics drastically. The high-temperature
asymptotics of the large-$N$ Schur index (in
(\ref{eq:N=4SchurLargeN})) is found as (see appendix~\ref{sec:app2}
for the definition of the symbol $\sim$ used below)
\begin{equation}
\ln
\mathcal{I}_{Schur}^{N\to\infty}(\beta)=\ln(q;q)-2\ln(q^{1/2};q^{1/2})\sim\frac{\pi^2}{2\beta}+\frac{1}{2}\ln
(\frac{\beta}{8\pi}),\quad\quad(\text{as $\beta\to
0$})\label{eq:N=4largeNasy}
\end{equation}
differing significantly from the asymptotics of the finite-$N$
index in (\ref{eq:N=4SchurfiniteNasy}).\\

The purpose of the above discussion was to help orient the reader
for the following analysis of toric quiver gauge theories with
SU($N$) nodes. Toric quiver theories are a much-studied subset of
supersymmetric gauge theories whose field content can be efficiently
summarized using quiver diagrams. The latter are directed graphs
with nodes representing vector multiplets and edges representing
chiral multiplets. The nodes at the ends of an edge represent vector
multiplets under which the chiral multiplet (represented by the
edge) is charged. The direction of the edge encodes further
information about the representation of the gauge group according to
which the chiral multiplet transforms. The toric condition puts
further constraints on the theory, thereby guaranteeing some nice
properties such as existence of a non-trivial IR fixed point with a
holographic dual describable by ``toric geometry'' (see for instance
\cite{Franco:2006}). A canonical example is the $\mathcal{N}=4$ SYM
with SU($N$) gauge group, which can be represented by one node
(standing for the SU($N$) vector multiplet), and three directed
edges (standing for the three chiral multiplets in the adjoint) that
both emanate from and end on that one node.

Similarly to the case of the Schur index discussed above, \emph{in
the large-$N$ limit the index of toric quivers simplifies to the
multi-trace index}. The multi-trace index is the one obtained by
summing over multi-trace operators in the SCFT. These operators
correspond to the multi-particle KK states in the gravity dual.

The superconformal index of these SCFTs is studied in
\cite{Gadde:2010holo,Eager:2012hx,Agarwal:2013,Ardehali:2015b}, and
its large-$N$ limit is found to be
\begin{equation}
\mathcal{I}_{quiver}^{N\to\infty}(b,\beta)=\frac{1}{\prod_{i=1}^{n_z}((pq)^{r_i/2};(pq)^{r_i/2})\prod_{adj}\Gamma((pq)^{R_{adj}/2};p,q)},\label{eq:toricQmt}
\end{equation}
with the first product in the denominator being over the $n_z$
extremal BPS mesons (with R-charge $r_i$), and the second product
over the chiral multiplets (with R-charge $R_{adj}$) in the adjoints
of various nodes. For example, the SU($N$) $\mathcal{N}=4$ SYM has
$n_z=3$, and $r_{1,2,3}=R_{adj}=2/3$, with three adjoint chiral
multiplets in total.\\

Just as in the case of the Schur index discussed above, the
high-temperature asymptotics of
$\mathcal{I}_{quiver}^{N\to\infty}(b,\beta)$ is quite different from
the asymptotics of the same quiver at finite $N$; the latter
asymptotics can be obtained (for non-chiral quivers) from the
results of the previous section. Finding the asymptotics of
$\mathcal{I}_{quiver}^{N\to\infty}(b,\beta)$, on the other hand,
requires separate calculations (see (\ref{eq:conjLargeNsq}) below).\\

\section{From the multi-trace index to the single-trace index}

The single-trace index is defined as the plethystic log
\cite{Benvenuti:2006} of the multi-trace index
\begin{equation}
I_{s.t.}(b,\beta)\equiv\sum_{n=1}^{\infty}\frac{\mu(n)}{n}\ln
\mathcal{I}^{N\to\infty}(b,n\beta),\label{eq:plLogDef}
\end{equation}
where $\mu(n)$ is the M\"{o}bius function. The adjective
``single-trace'' is particularly appropriate for theories that admit
a planar limit in which single-trace operators are weakly
interacting. For such cases if in the definition of the index in
(\ref{eq:equivIndexDef}) one restricts the trace to the
``single-trace states'' in the Hilbert space, one obtains the
single-trace index as defined above. In AdS/CFT, the weakly
interacting mesons of the large-$N$ SCFT at large 't~Hooft coupling
map to the KK supergravity modes in the bulk. Therefore the boundary
single-trace index is (according to AdS/CFT) equal to the bulk
single-particle index, with the latter receiving contributions only
from the bulk single-particle KK states.\\

The single-trace index of SU($N$) toric quiver SCFTs can be easily
computed by taking the plethystic logarithm of the two sides of
(\ref{eq:toricQmt}); the result is
\begin{equation}
I_{s.t.\
quiver}(b,\beta)=\sum_{i=1}^{n_z}\frac{(pq)^{r_i/2}}{1-(pq)^{r_i/2}}-\sum_{adj}\frac{(pq)^{R_{adj}/2}-(pq)^{1-R_{adj}/2}}{(1-p)(1-q)}.\label{eq:toricQst}
\end{equation}

In the next section we will see that in its high-temperature
asymptotics, the above index encodes the subleading Weyl anomaly of
the underlying SU($N$) toric quiver SCFT.\\

An interesting problem, which is not relevant to the main discussion
of the present chapter, is the connection between the small-$\beta$
asymptotics of $I_{s.t.}(b,\beta)$ and
$\mathcal{I}^{N\to\infty}(b,\beta)$; this problem is addressed in
appendix~\ref{sec:app4}.\\

\section{Asymptotics of the single-trace index and the holographic Weyl anomaly}

In this section we present holographic results implying that the
subleading central charges of a holographic SCFT are encoded in the
high-temperature asymptotics of its large-$N$ index.

We focus on SCFTs whose dual geometry is of the form
AdS$_5\times$SE$_5$, with SE$_5$ a Sasaki-Einstein 5-manifold. The
KK spectrum of the IIB theory on AdS$_5\times$SE$_5$ organizes
itself into representations of the 4d $\mathcal{N}=1$ superconformal
group SU$(2,2|1)$.

The shortened multiplets of SU$(2,2|1)$ are listed in
Table~\ref{tbl:IndexContributions}, along with their contributions
to the single-trace index. For convenience, we have introduced
$t\equiv 1/\sqrt{pq}$, and $y\equiv\sqrt{p/q}$. The chiral and SLII
multiplets (on the 2nd and the last row, respectively) contribute to
the right-handed index\footnote{The index defined in
(\ref{eq:equivIndexDef}) is the right-handed index. One can also
define the left-handed index in which one replaces $r$ with $-r$ and
swaps $j_1$ and $j_2$ in the definition of the index in
(\ref{eq:equivIndexDef}). The index in
Table~\ref{tbl:IndexContributions} is defined as $ I^+_{s.t.} \equiv
\frac{1}{2}( I^R_{s.t.}+  I^L_{s.t.})$, in terms of the left and
right single-trace indices. For toric quivers $I^+_{s.t.} =
I^R_{s.t.}= I^L_{s.t.}$.}, while the CP-conjugate multiplets, namely
the anti-chiral and SLI multiplets (on the 3nd and the 4th row,
respectively), contribute to the left-handed index. Conserved
multiplets (on the 1st row), which are CP self-conjugate, contribute
to both.

\begin{table*}[t]
\centering
\begin{tabular}{|l|l|l|c|}
\hline
Shortening condition&Representation& $(1-t^{-1} y)(1-t^{-1} y^{-1})I^+_{s.t.}$\\
\hline
$E_0=2+j_1+j_2$, $\fft32r=j_1-j_2$&$\mathcal D(E_0,j_1,j_2,r)$&$ \ft12(-1)^{2(j_1+j_2)+1} t^{-(2E_0+2j_2+2)/3}\chi_{j_1}(y) +(j_1\leftrightarrow j_2)$\\
\hline
$E_0=\fft32r$&$\mathcal D(E_0,j_1,0,r)$&$ \ft12(-1)^{2j_1} t^{-2E_0/3}\chi_{j_1}(y)$\\
\hline
$E_0=-\fft32r$&$\mathcal D(E_0,0,j_2,r)$&$ \ft12(-1)^{2j_2} t^{-2E_0/3}\chi_{j_2}(y)$\\
\hline
$E_0=2+2j_1-\fft32r$&$\mathcal D(E_0,j_1,j_2,r)$&$\ft12(-1)^{2(j_1+j_2)+1} t^{-(2E_0+2j_1+2)/3}\chi_{j_2}(y)$\\
\hline
$E_0=2+2j_2+\fft32r$&$\mathcal D(E_0,j_1,j_2,r)$&$\ft12(-1)^{2(j_1+j_2)+1} t^{-(2E_0+2j_2+2)/3}\chi_{j_1}(y)$\\
\hline
\end{tabular}
\caption{Contributions to the superconformal index from the various
shortened multiplets.\label{tbl:IndexContributions}}
\end{table*}

We begin with relating $c-a$ to the index. First consider the chiral
and SLII multiplets. The contribution to $c-a$ from a generic chiral
multiplet $\mathcal D(E_0,j_1,0;r)$ in the bulk KK spectrum is given
by the following holographically derived expression
\cite{Beccaria:2014,Ardehali:2014}
\begin{equation}
(c-a)\big|_{\text{chiral}}= -\fft1{192}(-1)^{2j_1}(2E_0-3)(2j_1+1)
\left(1-8j_1(j_1+1)\right). \label{eq:CHc-a}
\end{equation}
Similarly, a generic SLII multiplet $\mathcal D(E_0,j_1,j_2;r)$ in
the bulk spectrum contributes \cite{Beccaria:2014,Ardehali:2014}
\begin{equation}
(c-a)\big|_{\text{SLII}}=\fft1{192}(-1)^{2j_1+2j_2}(2E_0+2j_2-1)(2j_1+1)
\left(1-8j_1(j_1+1)\right).\label{eq:sl2c-a}
\end{equation}
It is now possible to see how these expressions may be obtained from
the contributions to the right-handed index given in
Table~\ref{tbl:IndexContributions}. Since the SU(2) character
$\chi_j(y)$ is given by
\begin{equation}
\chi_j(y)=\fft{y^{2j+1}-y^{-(2j+1)}}{y-y^{-1}},
\end{equation}
the differential operator $(6(y\partial_y)^2-1)$ acting on the
contributions to the index gives $(2j+1)[8j(j+1)-1]$ when $y$ is set
to one.  The operator $(t\partial_t+1)$ then produces the
$E_0$-dependent factors in (\ref{eq:CHc-a}) and (\ref{eq:sl2c-a}).

The CP conjugate multiplets (anti-chiral and SLI) contribute
similarly to (\ref{eq:CHc-a}) and (\ref{eq:sl2c-a}) with the
appropriate replacement of quantum numbers, and are accounted for in
the left-handed index.  Finally, since conserved multiplets
contribute as the sum of one SLI and one SLII multiplet, they are
implicitly included in both the left- and right-handed indices.  Our
key observation is that the contribution to $c-a$ has a uniform
expression for every single bulk multiplet.  Hence a single
differential operator acting on the index can yield the appropriate
contribution to $c-a$ regardless of the shortening condition.
Summing over all the bulk KK multiplets, one finally arrives at
\begin{eqnarray}\label{eq:c-aIndex}
c-a &= &\lim_{t\to1}-\frac{1}{32}\left(t\partial_t
+1\right)\left(6(y\partial_y)^2-1\right)\nn\\
&&\times\left[(1-t^{-1} y)(1-t^{-1} y^{-1}) I^+_{s.t.}(t,y)\right]
\Big|^{\mbox{\scriptsize{finite}}}_{y=1},
\end{eqnarray}
where the fugacities are set to one after acting with the
differential operator on the index. Note that the factor $(1-t^{-1}
y)(1-t^{-1} y^{-1})$ multiplying the single-trace index removes the
contribution from descendant states. The result obtained is often
divergent, as we are working in the large-$N$ limit, so the
prescription is that the finite term in an expansion about $t=1$
yields the value of $c-a$.

A few remarks are now in order.
\begin{itemize}
\item The $t\to1$ limit corresponds to the high-temperature limit
$\beta\to 0$. Therefore the prescription (\ref{eq:c-aIndex})
extracts $c-a$ from the high-temperature asymptotics of
$I^+_{s.t.}$.

\item The index in (\ref{eq:c-aIndex}) is the
single-particle supergravity index, which is---according to the
AdS/CFT conjecture---equal to the single-trace index of the SCFT.

\item In the prescription (\ref{eq:c-aIndex}), the index provides a natural
regulator for the Kaluza-Klein sums encountered in the holographic
$c-a$ calculations of
\cite{Ardehali:2013gra,Ardehali:2013xla,Ardehali:2013xya}.\\
\end{itemize}

Following a similar approach, but using holographic expressions for
the individual central charges, one arrives at
\cite{Beccaria:2014,Ardehali:2015a}
\begin{equation}
\begin{split}
\delta
a&=\fft1{32}(t\partial_t+1)(-\ft92t\partial_t(t\partial_t+2)+\ft92(y\partial_y)^2-3)\hat
I(t,y),\\
\delta
c&=\fft1{32}(t\partial_t+1)(-\ft92t\partial_t(t\partial_t+2)-\ft32(y\partial_y)^2-2)\hat
I(t,y), \label{eq:acIndex}
\end{split}
\end{equation}
where $\hat I=(1-yt^{-1})(1-y^{-1}t^{-1}) I^+_{s.t.}$ is the
single-trace index with descendants removed, and $\delta$ indicates
that we are referring to the $O(N^0)$ part of the central charges
(and not their leading $O(N^2)$ piece). The fugacities are set to
one after acting with the differential operators on $\hat I$; we are
thus again dealing with the high-temperature limit of the index.\\

In principle, a successful application of Eq. (\ref{eq:acIndex}) to
a holographic SCFT can be viewed as a one-loop test of AdS/CFT. This
can be easily done for arbitrary SU($N$) toric quiver SCFTs without
adjoint matter that are dual to smooth Sasaki-Einstein 5-manifolds.
The single-trace index of such a toric theory is \cite{Eager:2012hx}
\begin{equation}
I_{s.t.}=\sum_i \frac{1}{t^{r_i/3}-1},\label{eq:toricIndex}
\end{equation}
where $r_i$ are the $R$-charges of extremal BPS mesons. Applying
(\ref{eq:acIndex}) to (\ref{eq:toricIndex}) gives
\begin{equation}
\delta a = -\frac{27}{32(t-1)^2}\sum_{i=1}^{n_v} \frac{1}{r_i} -
\frac{1}{32}\sum_{i=1}^{n_v}r_i+\cdots \label{eq:polToric}
\end{equation}
in an expansion about $t=1$.  Noting that $\sum r_i=6(\mbox{\# nodes
in the quiver})$, and keeping only the finite part, we obtain
\begin{equation}
\delta a=-\fft3{16}(\mbox{\# nodes in the
quiver}).\label{eq:c-aToric}
\end{equation}
This matches the expected result for the $O(1)$ part of $a$ based on
the decoupling of a U($1$) at each node in the quiver; since there
are no adjoint matter fields in the quiver, there are no additional
$O(1)$ contributions to $a$ in the field theoretical computation
through $a=\frac{1}{32}(9\mathrm{Tr}R^3-3\mathrm{Tr}R)$.

The successful matching for the $O(1)$ part of $c$ can be deduced
from a similar application of the second relation in
Eq.~(\ref{eq:acIndex}) to (\ref{eq:toricIndex}).\\

The prescriptions in Eq.~(\ref{eq:acIndex}) can also be successfully
applied to the single-trace index of an arbitrary SU($N$) toric
quiver, as given in (\ref{eq:toricQst}). However, the result would
count as a successful test of AdS/CFT only up to the following two
assumptions: $i)$ a combinatorial conjecture \cite{Agarwal:2013}
that has gone into the derivation of (\ref{eq:toricQst}), which
although strongly supported in \cite{Agarwal:2013}, is not yet
proven; $ii)$ the assumption that the index (\ref{eq:toricQst})
(which is derived as the single-trace index of the SCFT) equals the
single-particle index of the gravity dual (which is what goes on the
RHS of the prescriptions in Eq.~(\ref{eq:acIndex})). The equality in
the assumption $ii$ is not yet proven \cite{Eager:2012hx} when the
toric quiver has adjoint matter or is dual to singular toric
SE$_5$.\\

Finally, the study of \cite{Ardehali:2015b} shows that the validity
of the prescriptions in Eq.~(\ref{eq:acIndex}) is guaranteed if the
high-temperature asymptotics of the single-trace index of SCFTs dual
to AdS$_5\times$SE$_5$ has the following form (see
appendix~\ref{sec:app2} for the definition of the symbol $\sim$ used
below):
\begin{equation}
\begin{split}
I_{s.t.}\sim&\,
\frac{2H}{\beta\left(b+b^{-1}\right)}+\frac{G\left(b+b^{-1}\right)}{2\beta}+C\\
&\,-\beta\left(\frac{4}{27}(b+b^{-1})^3\left(3\delta c-2\delta
a\right)+\frac{4}{3}(b+b^{-1})\left(\delta a-\delta
c\right)\right),\label{eq:StExpandedQuiverGH}
\end{split}
\end{equation}
with $G$,$H$,$C$ constants that are insignificant for the formulas
in Eq.~(\ref{eq:acIndex}), except that $H$ determines the pole terms
that according to the prescription of \cite{Ardehali:2015a} one
should drop. The above asymptotics was explicitly verified in
\cite{Ardehali:2015b} for the single-trace index of the SU($N$)
toric quivers, shown in (\ref{eq:toricQst}).

\chapter{Concluding remarks}

\section{Summary of the high-temperature content of the index}

\subsection{Finite-$N$ non-chiral theories}

We have shown that the high-temperature expansion of the
superconformal index of finite-rank non-chiral SCFTs (having all
their $r_\chi$ inside $]0,2[$) looks like
\begin{equation}
\ln\mathcal{I}(b,\beta)=
\frac{A(b)}{\beta}+B\ln(\frac{2\pi}{\beta})+C(b)+o(\beta^0),\quad\quad(\text{as
$\beta\to0$})\label{eq:highTexp}
\end{equation}
with
\begin{equation}
A(b)=\frac{16\pi^2}{3}(\frac{b+b^{-1}}{2})(c-a-\frac{3}{4}L_{h\
min}),
\end{equation}
\begin{equation}
B=\mathrm{dim}\mathfrak{h}_{qu},
\end{equation}
and $C(b)$ some real function of $b$ that we have not found a
general expression for.\\

Based on various examples that we have looked at, it seems like
whenever $L_{h\ min}=0$, the (complex-) dimension of the quantum
Coulomb branch of the theory on $R^3\times S^1$ coincides with
$\mathrm{dim}\mathfrak{h}_{qu}$. Thus we can say the following.
\begin{center}
\emph{For theories with positive semi-definite Rains function, the
high-temperature expansion of $\ln\mathcal{I}(b,\beta)$ encodes $i)$
in its order-$1/\beta$ term the difference of the central charges
$c-a$, and $ii)$ in its order-$\ln(1/\beta)$ term the (complex-)
dimension of the quantum Coulomb branch of the theory on $R^3\times
S^1$.}
\end{center}
Note that while we have proven item $i$ above, item $ii$ is only a
conjecture based on various examples studied in
\cite{Ardehali:2015c}.\\

Moving on to the subleading terms, the following statement was
demonstrated in \cite{Ardehali:2015c} for $C(b)$. (We define
$Z_{S^3}(b):=Z_{S^3}(b;\infty)$; see subsection~\ref{sec:S3} for the
definition of $Z_{S^3}(b;\Lambda)$.)
\begin{center}
\emph{For theories whose Rains function is minimized only at the
origin of $\mathfrak{h}_{cl}$ (hence have $L_{h\
min}=\mathrm{dim}\mathfrak{h}_{qu}=0$), the high-temperature
expansion of $\ln\mathcal{I}(b,\beta)$ encodes in its
order-$\beta^0$ term the logarithm of the squashed three-sphere
partition function $Z_{S^3}(b)$ of the dimensionally reduced theory;
in other words, for the said theories $C(b)=\ln Z_{S^3}(b)$.}
\end{center}
[The above statement was claimed in \cite{Ardehali:2015c} to hold
even for chiral theories; however, while for non-chiral theories it
is straightforward to show $Z_{S^3}(b)\neq0$, for chiral theories we
have not been able to show that $Z_{S^3}(b)$ is non-zero; we thus
emphasize that the above statement is demonstrated in
\cite{Ardehali:2015c} for chiral theories assuming
$Z_{S^3}(b)\neq0$.]\\

Although we have not been able to make general statements about the
$o(\beta^0)$ terms on the RHS of (\ref{eq:highTexp}), based on the
examples studied in \cite{Ardehali:2015c} it seems that
\begin{equation}
\ln\mathcal{I}(b,\beta)=
\frac{A(b)}{\beta}+B\ln(\frac{2\pi}{\beta})+C(b)+D(b)\beta
+O(\beta^2)\quad\quad(\text{as $\beta\to0$}).\label{eq:highTexp2}
\end{equation}
For theories whose Rains function is minimized on a set of isolated
points, the above asymptotics can actually be demonstrated (with
$B=0$, of course); it can moreover be shown that the error term is
not just $O(\beta^2)$, but beyond all orders (and of the type
$e^{-1/\beta}$) \cite{Ardehali:2015c}. Furthermore, in those
theories $D(b)$ coincides with the SUSY Casimir energy\footnote{The
SUSY Casimir energy relates the superconformal index
$\mathcal{I}(b,\beta)$ to its corresponding partition function
$Z^{\mathrm{SUSY}}(b,\beta)$ computed via path-integration on
$S_b^3\times S^1_\beta$ \cite{Ardehali:2015b,Assel:2015s}:
$Z^{\mathrm{SUSY}}(b,\beta)=e^{-\beta E_{\mathrm{susy}}(b)}
\mathcal{I}(b,\beta)$. } (encountered also in (\ref{eq:EcBequiv})
above)
\begin{equation}
\begin{split}
E_{\mathrm{susy}}(b)=\frac{2}{27}(b+b^{-1})^3(3c-2a)+\frac{2}{3}(b+b^{-1})(a-c).\label{eq:EcBequiv2}
\end{split}
\end{equation}
Therefore we can say the following \cite{Ardehali:2015c}.
\begin{center}
\emph{For theories whose Rains function is minimized on a set of
isolated points in $\mathfrak{h}_{cl}$, the high-temperature
expansion of $\ln\mathcal{I}(b,\beta)$ takes the form shown in
(\ref{eq:highTexp2}), with $B=0$, and with the error being not only
$O(\beta^2)$ but also exponentially small. Moreover, the
order-$\beta$ term encodes the SUSY Casimir energy; in other words,
for the said theories $D(b)=E_{\mathrm{susy}}(b)$.}
\end{center}
The above statement implies that (whenever the Rains function is
minimized on a set of isolated points) the central charges $a$ and
$c$---and hence the `t~Hooft anomalies $\mathrm{Tr}R$ and
$\mathrm{Tr}R^3$---are both encoded in the order-$\beta$ term in the
high-temperature expansion of $\ln\mathcal{I}(b,\beta)$. It can
actually be shown that introducing flavor fugacities $u_a=e^{i\beta
m_a}$ in the superconformal index, the relation
$D(b)=E_{\mathrm{susy}}(b)$ generalizes to
$D(b;m_a)=E_{\mathrm{susy}}(b;m_a)$, with $E_{\mathrm{susy}}(b;m_a)$
the equivariant SUSY Casimir energy (which encodes all the `t~Hooft
anomalies in the theory \cite{Bobev:2015}); thus (whenever the Rains
function is minimized on a set of isolated points) all the `t~Hooft
anomalies are encoded in the order-$\beta$ term in the
high-temperature expansion of $\ln\mathcal{I}(b,\beta;m_a)$. This
statement is related (but not equivalent) to some of the claims in
\cite{Spiridonov:2012sv}, which were made there in the context of
SU($N$) SQCD.

\subsection{Large-$N$ toric quivers}

It was shown in \cite{Ardehali:2015b} that for SU($N$) toric quiver
SCFTs (see appendix~\ref{sec:app2} for the definition of the symbol
$\sim$ used below)
\begin{equation}
\begin{aligned} \ln
\mathcal{I}^{N\rightarrow\infty}_{quiver}(b,\beta)\sim
&\,\frac{\pi^2}{6\beta(\frac{b+b^{-1}}{2})
}\sum_{i=1}^{n_z}\frac{1}{r_i}+\frac{16\pi^2
(\frac{b+b^{-1}}{2})}{3\beta}\sum_{adj}(\delta c_{adj}-\delta
a_{adj})+\frac{n_z}{2}\ln(\beta/2\pi)+\ln
Y_b \\
&\,+\beta\left(\frac{2}{27}(b+b^{-1})^3(3\delta c-2\delta
a)+\frac{2}{3}(b+b^{-1})(\delta a-\delta
c)\right),\label{eq:conjLargeNsq}
\end{aligned}
\end{equation}
where the notation is similar to that in (\ref{eq:toricQmt}), except
for $\ln Y_b=\frac{1}{2}\sum_{i=1}^{n_z}\ln (
r_i(\frac{b+b^{-1}}{2}))-\sum_{adj}\ln\Gamma_h(iR_{adj}(\frac{b+b^{-1}}{2}))$,
with $\Gamma_h(\ast)$ a special function explained in
appendix~\ref{sec:app1}.\\

Based on various specific examples, it was conjectured in
\cite{Ardehali:2014,Ardehali:2015b} that
\begin{equation}
\sum_{i=1}^{n_z}\frac{1}{r_i}=\frac{3}{16\pi^3}\left(19\mathrm{vol}(SE)+\frac{1}{8}\mathrm{Riem}^2(SE)\right),\label{eq:sumOneOverRi}
\end{equation}
where $SE$ denotes the Sasaki-Einstein 5-manifold dual to the quiver
gauge theory. The above conjecture was motivated by the finding in
\cite{Eager:2010} that one can ``hear the shape of the dual
geometry'' in the asymptotics of the Hilbert series of mesonic
operators in the SCFT. We note that the leading high-temperature
behavior of the index of the quivers is contained in the first two
terms of (\ref{eq:conjLargeNsq}). The first term, according to
(\ref{eq:sumOneOverRi}), is dictated by the geometry of the dual
internal manifold, while the second is given by the $O(1)$ part of
the contribution of adjoint matter to $c-a$.  The latter is hence
the only part of the finite-$N$ Di~Pietro-Komargodski formula
that escapes metamorphosis into ``geometry'' in the planar limit.\\

Interestingly, the order-$\beta$ term on the RHS of
(\ref{eq:conjLargeNsq}) is $\beta$ times $\delta
E_{\mathrm{susy}}(b)$, where by the latter we mean
$E_{\mathrm{susy}}(b)$ as in (\ref{eq:EcBequiv2}) but with the
central charges in it replaced with their $O(N^0)$ pieces. Therefore
the order-$\beta$ term in the high-temperature expansion of $\ln
\mathcal{I}^{N\rightarrow\infty}_{quiver}(b,\beta)$ is somewhat
similar to the corresponding term for the finite-$N$ non-chiral
theories whose Rains function is minimized on a set of isolated
points (see the previous subsection).

Note that the discussion below (\ref{eq:StExpandedQuiverGH})
(combined with relation (\ref{eq:largeNAsy})) implies that the
holographic computation of the subleading central charges can be
thought of as extracting $\delta c$ and $\delta a$ from the
order-$\beta$ term of the high-temperature expansion of $\ln
\mathcal{I}^{N\rightarrow\infty}(b,\beta)$. In the case of the toric
quivers, the holographic prescriptions in (\ref{eq:acIndex}) thus
extract $\delta c$ and $\delta a$ from $\delta
E_{\mathrm{susy}}(b)$.

\section{Future directions}
The main result of this dissertation is the high-temperature
asymptotics of the EHIs arising as the superconformal index of
unitary non-chiral 4d Lagrangian SCFTs. The most important extension
of our work would be to chiral SCFTs; the preliminary investigation
of \cite{Ardehali:2015c} seems to indicate that the extension would
not be straightforward.\\

A particularly interesting outcome of our work has been the
connection between the high-temperature asymptotics of the index,
and the Coulomb branch dynamics on $R^3\times S^1$; see
subsection~\ref{sec:crossed}. This is undoubtedly worth pursuing
more carefully. Even before aiming at establishing the connection in
a general context, it would be nice to validate it by examining the
Coulomb branch dynamics of the $A_k$ SQCD (with SU($N$) gauge group)
and the $Z_2$ orbifold theory (with SU($N$)$\times$SU($N$) gauge
group), to see if the (complex-) dimension of their quantum Coulomb
branch on $R^3\times S^1$ coincides with their
$\mathrm{dim}\mathfrak{h}_{qu}$, which we have found to be
respectively zero and $N-1$.
\clearpage

%

\appendix 

\addtocontents{toc}{\protect\contentsline{chapter}{\vspace*{-6mm}}{}{}} 
\addappheadtotoc 

\let\oldaddtoc\addtocontents
\renewcommand{\addtocontents}[2]{%
    \expandafter\ifstrequal\expandafter{#1}{toc}%
    {
    \oldaddtoc{loa}{#2}}%
    {
    \oldaddtoc{#1}{#2}}
}
\chapter{Useful special functions}\label{sec:app1}

The \textbf{Pochhammer symbol} ($|q|\in ]0,1[$)
\begin{equation}
(a;q):=\prod_{k=0}^{\infty}(1-a q^k),\label{eq:PochDef}
\end{equation}
is related to the more familiar Dedekind eta function via
\begin{equation}
\eta(\tau)=q^{1/24}(q;q), \label{eq:etaPoc}
\end{equation}
with $q=e^{2\pi i\tau}.$

The eta function has an SL$(2,\mathbb{Z})$ modular property that
will be useful for us: $\eta(-1/\tau)=\sqrt{-i\tau}\eta(\tau)$.

The Pochhammer symbol $(q;q)$ equals the inverse of the generating
function of integer partitions. It also appears in the index of 4d
SUSY gauge theories that contain vector multiplets.\\

The \textbf{elliptic gamma function} is defined as
($\mathrm{Im}(\tau),\mathrm{Im}(\sigma) >0$)
\begin{equation}
\Gamma(x;\sigma,\tau):=\prod_{j,k\ge
0}\frac{1-z^{-1}p^{j+1}q^{k+1}}{1-z p^{j}q^{k}},\label{eq:GammaDef}
\end{equation}
with $z:=e^{2\pi i x}$, $p:=e^{2\pi i \sigma}=e^{-\beta b}$, and
$q:=e^{2\pi i \tau}=e^{-\beta b^{-1}}$. The above expression gives a
meromorphic function of $x\in\mathbb{C}$. For generic choice of
$\tau$ and $\sigma$, the elliptic gamma has simple poles at
$x=l-m\sigma-n\tau$, with $m,n\in\mathbb{Z}^{\ge 0}$,
$l\in\mathbb{Z}$.

We sometimes write $\Gamma(x;\sigma,\tau)$ as $\Gamma(z;p,q)$, or
simply as $\Gamma(z)$. Also, the arguments of elliptic gamma
functions are frequently written with ``ambiguous'' signs (as in
$\Gamma(\pm x;\sigma,\tau)$); by that one means a multiplication of
several gamma functions each with a ``possible'' sign of the
argument (as in $\Gamma(+x;\sigma,\tau)\times
\Gamma(-x;\sigma,\tau)$). Similarly $\Gamma(z^{\pm
1}):=\Gamma(z;p,q)\times \Gamma(z^{-1};p,q)$.

The elliptic gamma function appears in the exact solution of some
important 2d integrable lattice models. It also features in the
index of 4d Lagrangian SUSY QFTs that contain chiral multiplets.\\

Following Rains \cite{Rains:2009}, we define the \textbf{hyperbolic
gamma function} by
\begin{equation}
\Gamma_h(x;\omega_1,\omega_2):=\exp
\left(\mathrm{PV}\int_{\mathbb{R}}\frac{e^{2\pi i x w}}{(e^{2\pi
i\omega_1 w}-1)(e^{2\pi i\omega_2
w}-1)}\frac{\mathrm{d}w}{w}\right).\label{eq:hyperbolicGamma}
\end{equation}
The above expression makes sense only for
$0<\mathrm{Im}(x)<2\mathrm{Im}(\omega)$, with
$\omega:=(\omega_1+\omega_2)/2$. In that domain, the function
defined by (\ref{eq:hyperbolicGamma}) satisfies
\begin{equation}
\Gamma_h(x+\omega_2;\omega_1,\omega_2)=2\sin (\frac{\pi
x}{\omega_1})\Gamma_h(x;\omega_1,\omega_2).\label{eq:hyperbolicGammaRecursion}
\end{equation}
This relation can then be used for an inductive meromorphic
continuation of the hyperbolic gamma function to all
$x\in\mathbb{C}$. For generic $\omega_1,\omega_2$ in the upper half
plane, the resulting meromorphic function
$\Gamma_h(x;\omega_1,\omega_2)$ has simple zeros at
$x=\omega_1\mathbb{Z}^{\ge1}+\omega_2\mathbb{Z}^{\ge1}$ and simple
poles at $x=\omega_1\mathbb{Z}^{\le0}+\omega_2\mathbb{Z}^{\le0}$.

For convenience, we will frequently write $\Gamma_h(x)$ instead of
$\Gamma_h(x;\omega_1,\omega_2)$, and $\Gamma_h(x\pm y)$ instead of
$\Gamma_h(x+y)\Gamma_h(x-y)$.

The hyperbolic gamma function has an important property that can be
easily derived from the definition (\ref{eq:hyperbolicGamma}):
\begin{equation}
\Gamma_h(-\mathrm{Re}(x)+i\mathrm{Im}(x);\omega_1,\omega_2)=(\Gamma_h(\mathrm{Re}(x)+i\mathrm{Im}(x);\omega_1,\omega_2))^\ast,\label{eq:hyperbolicGammaConj}
\end{equation}
with $\ast$ denoting complex conjugation.

We also define the non-compact quantum dilogarithm $\psi_b$ (c.f.
the function $e_b(x)$ in \cite{Faddeev:2001}; $\psi_b(x)=e_b(-i x)$)
via
\begin{equation}
\psi_b(x):=e^{-i\pi
x^2/2+i\pi(b^2+b^{-2})/24}\Gamma_h(ix+\omega;\omega_1,\omega_2),\label{eq:hyperbolicGammaPsi}
\end{equation}
where
\begin{equation}
\omega_1:=i b,\quad\omega_2:=i b^{-1},\quad\text{and}\quad
\omega:=(\omega_1+\omega_2)/2.
\end{equation}
For generic choice of $b$, the zeros of $\psi_b(x)^{\pm 1}$ are of
first order, and lie at $\pm((b+b^{-1})/2+b\mathbb{Z}^{\ge
0}+b^{-1}\mathbb{Z}^{\ge 0})$. Upon setting $b=1$ we get the
function $\psi(x)$ of \cite{Felder:1999}; i.e.
$\psi_{b=1}(x)=\psi(x)$.

An identity due to Narukawa \cite{Narukawa:2004} implies the
following important relation between $\psi_{b}(x)$ and the elliptic
gamma function (see also Appendix~A of \cite{Ardehali:2015b})
\begin{equation}
\begin{split}
\Gamma(z;\sigma,\tau)&=\frac{e^{2i\pi
Q_{-}(x;\sigma,\tau)}}{\psi_b(\frac{2\pi i
x}{\beta}+\frac{b+b^{-1}}{2})}\prod_{n=1}^{\infty}\frac{\psi_b(-\frac{2\pi
in}{\beta}-\frac{2\pi i
x}{\beta}-\frac{b+b^{-1}}{2})}{\psi_b(-\frac{2\pi
in}{\beta}+\frac{2\pi i
x}{\beta}+\frac{b+b^{-1}}{2})}\\
&=e^{2i\pi
Q_{+}(x;\sigma,\tau)}\psi_b(\text{\footnotesize{$-\frac{2\pi i
x}{\beta}-\frac{b+b^{-1}}{2}$}})\prod_{n=1}^{\infty}\frac{\psi_b(-\frac{2\pi
in}{\beta}-\frac{2\pi i
x}{\beta}-\frac{b+b^{-1}}{2})}{\psi_b(-\frac{2\pi
in}{\beta}+\frac{2\pi i
x}{\beta}+\frac{b+b^{-1}}{2})},\label{eq:GammaFVSq}
\end{split}
\end{equation}
where
\begin{equation}
\begin{split}
Q_{-}(x;\sigma,\tau)=&-\frac{x^3}{6\tau\sigma}+\frac{\tau+\sigma-1}{4\tau\sigma}x^2-\frac{\tau^2+\sigma^2+3\tau\sigma-3\tau-3\sigma+1}{12\tau\sigma}x\\
&-\frac{1}{24}(\tau+\sigma-1)(\tau^{-1}+\sigma^{-1}-1),\\
Q_{+}(x;\sigma,\tau)=&Q_{-}(x;\sigma,\tau)+(x-\frac{\tau+\sigma}{2})^2/2\tau\sigma-(\tau^2+\sigma^2)/24\tau\sigma,\label{eq:MmaDef}
\end{split}
\end{equation}


\chapter{Some asymptotic analysis}\label{sec:app2}

\indent We say $f(\beta)=O(g(\beta))$ as $\beta\to0$, if there exist
positive real numbers $C,\beta_0$ such that for all $\beta<\beta_0$
we have $|f(\beta)|< C|g(\beta)|$. We say $f(x,\beta)=O(g(x,\beta))$
\emph{uniformly} over $S$ as $\beta\to0$, if there exist positive
real numbers $C,\beta_0$ such that for all $\beta<\beta_0$ and all
$x\in S$ we have $|f(x,\beta)|< C|g(x,\beta)|$.

We will write $f(\beta)=o(g(\beta))$, if $f(\beta)/g(\beta)\to 0$ as
$\beta\to 0$.

We use the symbol $\sim$ when writing the all-orders asymptotics of
a function. For example, we have
\begin{equation}
\ln(\beta+e^{-1/\beta})\sim \ln\beta,\quad\quad(\text{as $\beta\to
0$})
\end{equation}
because we can write the LHS as the sum of $\ln\beta$ and
$\ln(1+e^{-1/\beta}/\beta)$, and the latter is beyond all-orders in
$\beta$.

More precisely, we say $f(\beta)\sim g(\beta)$ as $\beta\to 0$, if
we have $f(\beta)- g(\beta)=O(\beta^n)$ for any (arbitrarily large)
natural $n$.

We will write $f(\beta)\simeq g(\beta)$ if $\ln f(\beta)\sim \ln
g(\beta)$ (with an appropriate choice of branch for the logarithms).
By writing $f(x,\beta)\simeq g(x,\beta)$ we mean that $\ln
f(x,\beta)\sim \ln g(x,\beta)$ for all $x$ on which
$f(x,\beta),g(x,\beta)\neq 0$, and that $f(x,\beta)= g(x,\beta)=0$
for all $x$ on which
either $f(x,\beta)=0$ or $g(x,\beta)=0$.\\

With the above notations at hand, we can asymptotically analyze the
Pochhammer symbol as follows. The low-temperature ($T\to 0$, with
$q=e^{-1/T}$) behavior is trivial:
\begin{equation}
(q;q)\simeq 1\quad\quad(\text{as $1/\beta\to
0$}).\label{eq:PochLowT}
\end{equation}

The high-temperature ($\beta\to 0$, with $q=e^{-\beta}$) asymptotics
is nontrivial. It can be obtained using the SL($2,\mathbb{Z}$)
modular property of the eta function, which yields
\begin{equation}
\ln \eta(\tau=\frac{i\beta}{2\pi})\sim
-\frac{\pi^2}{6\beta}+\frac{1}{2}\ln(\frac{2\pi}{\beta})\quad\quad(\text{as
$\beta\to 0$}).
\end{equation}
The above relation, when combined with (\ref{eq:etaPoc}), implies
\begin{equation}
\ln(q;q)\sim
-\frac{\pi^2}{6\beta}+\frac{1}{2}\ln(\frac{2\pi}{\beta})+\frac{\beta}{24}\quad\quad(\text{as
$\beta\to
0$}).\label{eq:PochAsy}\\
\end{equation}

For the hyperbolic gamma function, Corollary~2.3 of
\cite{Rains:2009} implies that when $x\in\mathbb{R}$
\begin{equation}
\begin{split}
\ln\Gamma_h(x+r\omega;\omega_1,\omega_2)= -\frac{i\pi}{2} x|x|-i\pi
(r-1)\omega|x|+O(1), \quad\quad(\text{as
$|x|\to\infty$})\label{eq:hyperbolicGammaAsy}
\end{split}
\end{equation}
for any fixed real $r$, and fixed $b>0$.

From the asymptotics of the hyperbolic gamma function, it follows
that for fixed $\mathrm{Re}(x)$ and fixed $b>0$
\begin{equation}
\ln\psi_b(x)\sim 0,\quad\quad\quad (\text{as } \beta\to 0, \text{
for } \mathrm{Im}(x)= -1/\beta)\label{eq:psiAsy}
\end{equation}
with a transcendentally small error, of the type $e^{-1/\beta}$.

The above estimate can be combined with (\ref{eq:GammaFVSq}) to
yield the small-$\beta$ estimates
\begin{equation}
\begin{split}
\Gamma(x;\sigma,\tau)&\simeq \frac{e^{2i\pi
Q_{-}(x;\sigma,\tau)}}{\psi_b(\frac{2\pi i
x}{\beta}+\frac{b+b^{-1}}{2})},\quad\quad
(\text{for } -1<\mathrm{Re}(x)\le 0 )\\
&\simeq e^{2i\pi
Q_{+}(x;\sigma,\tau)}\psi_b(\text{\footnotesize{$-\frac{2\pi i
x}{\beta}-\frac{b+b^{-1}}{2}$}}),\quad\quad (\text{for }
0\le\mathrm{Re}(x)<1 ) \label{eq:GammaAsyPosNegZ}
\end{split}
\end{equation}
with the range of $\mathrm{Re}(x)$ explaining our subscript
notations for $Q_+$ and $Q_-$. As a result of
(\ref{eq:GammaAsyPosNegZ}) we have for $x\in\mathbb{R}$, as
$\beta\to0$:
\begin{equation}
\begin{split}
\Gamma(-x+(\frac{\tau+\sigma}{2})r;\sigma,\tau)&\simeq
\frac{e^{2i\pi
Q_{-}(-\{x\}+(\frac{\tau+\sigma}{2})r;\sigma,\tau)}}{\psi_b(-\frac{2\pi
i \{x\}}{\beta}-(r-1)\frac{b+b^{-1}}{2})},\\
\Gamma(x+(\frac{\tau+\sigma}{2})r;\sigma,\tau)&\simeq e^{2i\pi
Q_{+}(\{x\}+(\frac{\tau+\sigma}{2})r;\sigma,\tau)}\psi_b(\text{\footnotesize{$-\frac{2\pi
i \{x\}}{\beta}+(r-1)\frac{b+b^{-1}}{2}$}}),
\label{eq:GammaAsyPosNegZ2}
\end{split}
\end{equation}
with $\{x\}:=x-\lfloor x\rfloor$. The above estimates are first
obtained in the range $0\le x<1$, and then extended to $x\in
\mathbb{R}$ using the periodicity of the LHS under $x\to x+1$.\\

\chapter{Generalized triangle inequalities}\label{sec:app3}
Define  $\vartheta(x):=\{x\}(1-\{x\})$. The Lemma~3.2 of
\cite{Rains:2009} says that for any sequence of real numbers
$c_1,\dots, c_n$, $d_1,\dots, d_n$, the following inequality holds:
\begin{equation}
\sum_{1\le i,j\le n}\vartheta(c_i-d_j)-\sum_{1\le i<j\le
n}\vartheta(c_i-c_j)-\sum_{1\le i<j\le n}\vartheta(d_i-d_j)\ge
\vartheta(\sum_{1\le i\le n}(c_i-d_i)),\label{eq:RainsGTI}
\end{equation}
with equality iff the sequence can be permuted so that either
\begin{equation}
\{c_1\}\le\{d_1\}\le\{c_2\}\le\cdots\le
\{d_{n-1}\}\le\{c_n\}\le\{d_n\},\label{cond:gti1}
\end{equation}
or
\begin{equation}
\{d_1\}\le\{c_1\}\le\{d_2\}\le\cdots\le
\{c_{n-1}\}\le\{d_n\}\le\{c_n\}.\label{cond:gti2}
\end{equation}
The proof can be found in \cite{Rains:2009}.

Re-scaling with $c_i,d_i\mapsto vc_i,vd_i$, taking $v\to 0^+$, and
using the relation $\vartheta(vx)=v|x|-v^2 x^2$ (which holds for
small enough $v$), Rains obtains the following corollary of
(\ref{eq:RainsGTI}):
\begin{equation}
\sum_{1\le i,j\le n}|c_i-d_j|-\sum_{1\le i<j\le
n}|c_i-c_j|-\sum_{1\le i<j\le n}|d_i-d_j|\ge |\sum_{1\le i\le
n}(c_i-d_i)|,\label{eq:RainsGTIav}
\end{equation}
with equality iff the sequence can be permuted so that either
\begin{equation}
c_1\le d_1 \le c_2 \le\cdots\le  d_{n-1} \le c_n \le d_n,
\end{equation}
or
\begin{equation}
d_1 \le c_1 \le d_2 \le\cdots\le  c_{n-1} \le d_n \le c_n.
\end{equation}\\

The fact that the inequality (\ref{eq:RainsGTIav}) arise as a
corollary of (\ref{eq:RainsGTI}) justifies the name ``generalized
triangle inequality'' for the latter.\\

Various generalized triangle inequalities (GTIs) allow us to
analytically address the minimization problems for the piecewise
linear functions $L_h$ arising in Chapter~\ref{sec:finiteNchap}. In
several physically interesting cases, the required GTI is a
corollary of Rains's GTI shown in (\ref{eq:RainsGTI}) above.

\chapter{Proof of an ansatz in
\cite{Ardehali:2015b}}\label{sec:app4}

The universal property of large-$N$ SCFTs that allows a systematic
study of their high-temperature asymptotics is the \emph{large-$N$
factorization}. The factorization implies that the index of
large-$N$ theories is conveniently expressed in terms of the
single-trace index as
\begin{equation}
\ln\mathcal{I}^{N\to\infty}(\beta,b)=\sum_{n=1}^{\infty}\frac{1}{n}I_{s.t.}(n\beta,b).\label{eq:fullIndexStIndex}
\end{equation}

Let us now review a useful technique in asymptotic analysis, which
we will find useful when studying large-$N$ indices expressible as
in (\ref{eq:fullIndexStIndex}).\\

Say we are interested in the small-$\beta$ asymptotics of a real
function $F(\beta)$ that can be written in the form
\begin{equation}
F(\beta)=\sum_{m=1}^{\infty}f(m\beta),\label{eq:ZagierForm}
\end{equation}
with $f(\beta)$ a real function having the $\beta\to 0$ asymptotic
development
\begin{equation}
f(\beta)\sim \sum_{\lambda\ge -1}^{\infty}b_{\lambda}
\beta^{\lambda}.
\end{equation}
Assume moreover that $f(\beta)$ and all its derivatives decay faster
than $1/\beta^{1+\varepsilon}$ as $\beta\to \infty$, for some
$\varepsilon>0$. Then, according to Zagier \cite{Zagier:2006}, the
$\beta\to 0$ asymptotics of $F(\beta)$ is given by
\begin{equation}
F(\beta)\sim\frac{1}{\beta}\left(b_{-1}\ln(\frac{1}{\beta})+I^{\ast}_f\right)+\sum_{\lambda>-1}b_\lambda
\zeta(-\lambda)\ \beta^{\lambda},\label{eq:ZagierResult}
\end{equation}
with
$I^{\ast}_{f}:=\int_{0}^{\infty}(f(x)-b_{-1}e^{-x}/x)\mathrm{d}x$.\\

Equation (\ref{eq:fullIndexStIndex}) has a remarkable resemblance to
the sums to which Zagier's method applies. In fact, dividing both
sides of (\ref{eq:fullIndexStIndex}) by $\beta$, we arrive at
\begin{equation}
\frac{\ln\mathcal{I}^{N\to\infty}(\beta)}{\beta}=\sum_{n=1}^{\infty}f(n\beta),\label{eq:StIndexAlmostZagier}
\end{equation}
with $f(\beta)=I_{s.t.}(\beta)/\beta$.

In all the examples we are aware of, $I_{s.t.}(\beta)$ has a leading
asymptotics of the form $I_{-1}/\beta$. Therefore $f(\beta)$ defined
above has a leading asymptotics of the form $I_{-1}/\beta^2$, and
thus Zagier's formula (\ref{eq:ZagierResult}) does not immediately
apply to it. However, defining
$\tilde{f}(\beta):=f(\beta)-I_{-1}/\beta^2$, we obtain
\begin{equation}
\frac{\ln\mathcal{I}^{N\to\infty}(\beta)}{\beta}=\frac{\pi^2}{6\beta^2}I_{-1}+\sum_{n=1}^{\infty}\tilde{f}(n\beta),\label{eq:StIndexZagierForm}
\end{equation}
and now Zagier's method can be applied to find the asymptotics of
the sum on the RHS of the above relation. The result is
\begin{equation}
\frac{\ln\mathcal{I}^{N\to\infty}(\beta)}{\beta}\sim\frac{\pi^2}{6\beta^2}I_{-1}+\frac{1}{\beta}(I_0\ln(\frac{1}{\beta})+I^{\ast}_{\tilde{f}})+\sum_{m=0}^{\infty}\tilde{f}_m\zeta(-m)\beta^{m},\label{eq:StIndexZagierResult}
\end{equation}
where $I^{\ast}_{\tilde{f}}:=\int_0^{\infty}
(\frac{I_{s.t.}(x)-I_{-1}/x}{x}-I_0\frac{e^{-x}}{x})\mathrm{d}x$,
and $\tilde{f}_m$ is the coefficient of $\beta^m$ in the asymptotics
of $\tilde{f}(\beta)$. Also $I_{0}$ is the $\beta$-independent term
in the asymptotic expansion of $I_{s.t.}$.

Let $I_{n}$ be the coefficient of $\beta^n$ in the asymptotics of
the single-trace index $I_{s.t.}$. From
\begin{equation}
\tilde{f}(\beta)=I_{s.t.}(\beta)/\beta-I_{-1}/\beta^2,\label{eq:tildeF}
\end{equation}
we obtain $\tilde{f}_n=I_{n+1}$ for $n=0,1,\dots$. Therefore we can
write (\ref{eq:StIndexZagierResult}) as
\begin{equation}
\ln\mathcal{I}^{N\to\infty}(\beta)\sim\frac{\pi^2}{6\beta}I_{-1}+I_0\ln(\frac{1}{\beta})+I^{\ast}_{\tilde{f}}+\sum_{m=1}^{\infty}I_{m}\zeta(-m+1)\beta^{m}.\label{eq:largeNAsy}
\end{equation}
This relation is the main result of this appendix. It expresses the
all-orders small-$\beta$ asymptotics of
$\ln\mathcal{I}^{N\to\infty}$ in terms of data that can be found
from the single-trace index.\\

As an application of the result (\ref{eq:largeNAsy}), we derive the
ansatz given in \cite{Ardehali:2015b} for the asymptotics of the
index of the $A_k$ SQCD in the Veneziano limit.

The single-trace index of the $A_k$ SQCD in the Veneziano limit is
given by \cite{Ardehali:2014}
\begin{equation}
I^{A_k}_{s.t.} =
\frac{\tau^{-\frac{2}{k+1}}}{1-\tau^{-\frac{2}{k+1}}} +
\frac{\tau^{-\frac{4k}{k+1}}}{1-\tau^{-\frac{4k}{k+1}}} -
\frac{\tau^{-\frac{2k}{k+1}}}{1-\tau^{-\frac{2k}{k+1}}} - \frac{
\big(\tau^{-\frac{2}{k+1}} - \tau^{-\frac{2k}{k+1}}\big) - N_f^2
\frac{\big(\tau^{\frac{2N_c}{(k+1)N_f}} -
\tau^{-\frac{2N_c}{(k+1)N_f}}\big)^2}{\tau^2\big(1-\tau^{-\frac{2}{k+1}}\big)\big(1+\tau^{-\frac{2k}{k+1}}\big)}}{(1-p)(1-q)},\label{eq:IstAk}
\end{equation}
where $\tau:=(pq)^{-1/2}$.

Expanding $I^{A_k}_{s.t.}$ at high temperatures we find a series of
the form
\begin{equation}
I^{A_k}_{s.t.}(\beta) =\frac{I_{-1}}{\beta}+I_0+\sum_{m\
odd>0}I_m\beta^m.\label{eq:IstAkExp}
\end{equation}

Note that no positive even powers of $\beta$ show up in the
expansion. This is because $I^{A_k}_{s.t.}(\beta)$ is ``almost'' an
odd function of $\beta$: one can directly check from
(\ref{eq:IstAk}) that
$I^{A_k}_{s.t.}(\beta)+I^{A_k}_{s.t.}(-\beta)=-1$.


Plugging (\ref{eq:IstAkExp}) in (\ref{eq:largeNAsy}) we find that
\begin{equation}
\ln \mathcal{I}^{N\rightarrow\infty}_{A_k}(\beta)
\sim\frac{\pi^2}{\beta}I_{-1}+I_0\ln(\frac{1}{\beta})+I^{\ast}_{\tilde{f}}+I_1\beta.\label{eq:IAkAsy}
\end{equation}

Using the actual values
\begin{equation}
\begin{split}
I_{-1}&=\frac{2k^3+3k^2-1}{4k(1+k)}\left(\frac{1}{(\frac{b+b^{-1}}{2})}\right)+\frac{16kN_c^2-8k^2+8k}{4k(1+k)}\left(\frac{b+b^{-1}}{2}\right)\\
I_0&=-\frac{1}{2}\\
I_1&=-\left(\frac{4}{27}(b+b^{-1})^3(3c-2a)+\frac{4}{3}(b+b^{-1})(a-c)\right).\label{eq:AkCoeffs}
\end{split}
\end{equation}
we obtain
\begin{equation}
\begin{split}
\ln\mathcal{I}^{N_c\rightarrow\infty}_{A_k}(\beta,b)\sim&\frac{2k^3+3k^2-1}{4k(1+k)}\left(\frac{\pi^2}{6\beta(\frac{b+b^{-1}}{2})}\right)+\frac{16kN_c^2-8k^2+8k}{4k(1+k)}\left(\frac{\pi^2(\frac{b+b^{-1}}{2})}{6\beta}\right)\\
&-\frac{1}{2}\ln (\frac{1}{\beta})+I^{\ast}_{\tilde{f}}(b)
+\beta\left(\frac{2}{27}(b+b^{-1})^3(3c-2a)+\frac{2}{3}(b+b^{-1})(a-c)\right).\label{eq:AkAnsatz}
\end{split}
\end{equation}
This is the ansatz of \cite{Ardehali:2015b}, now rigorously derived,
and supplemented with the $\beta$-independent term
$I^{\ast}_{\tilde{f}}(b)$ which was left undetermined in that work.
\clearpage
\renewcommand{\addtocontents}[2]{\oldaddtoc{#1}{#2}}


\backmatter

\addcontentsline{toc}{chapter}{Bibliography} 
\begin{singlespace}




\end{singlespace}

\end{document}